\documentclass[longauth]{aa} 

\usepackage{graphicx}
\usepackage{txfonts}
\begin{document}


  \title{MICONIC: JWST/MIRI MRS observations of the nuclear
     and circumnuclear regions of Mrk~231}

   \titlerunning{MICONIC: JWST/MIRI MRS observations of Mrk~231}
   
   \author{A. Alonso Herrero\inst{1}
         \and
         L. Hermosa Muñoz\inst{1}
         \and
         A. Labiano\inst{2}
         \and
         P. Guillard\inst{3, 4}
          \and
         V. A. Buiten\inst{5}
          \and
          D. Dicken\inst{6}
          \and
          P. van der Werf\inst{5}
          \and
          J. \'Alvarez-M\'arquez\inst{7}
          \and
          T. B\"oker\inst{8}
          \and
          L. Colina\inst{7}
          \and
          A. Eckart\inst{9,10}
          \and
          M. Garc\'{\i}a-Mar\'{\i}n\inst{8}
          \and
          O. C. Jones\inst{6}
          \and
          L. Pantoni\inst{11}
          \and
          P. G. P\'erez-Gonz\'alez\inst{7}
          \and
          D. Rouan\inst{12} 
          \and
          M. J. Ward\inst{13}
          \and
          M. Baes\inst{11}
          \and
          G. \"Ostlin\inst{14}
          \and
          P. Royer\inst{15}
          \and
           G. S. Wright\inst{6}
          \and
          M. G\"udel\inst{16, 17}
          \and
          Th. Henning\inst{18}
          \and
          P.-O. Lagage\inst{19}
         \and
         E. F. van Dishoeck\inst{5}
} 

       \institute{Centro de Astrobiolog\'{\i}a (CAB), CSIC-INTA,
         Camino Bajo del Castillo s/n, E-28692 Villanueva de la Ca\~nada, Madrid,
     Spain\\
     \email{aalonso@cab.inta-csic.es}\label{inst1}
   \and
       {Telespazio UK for the European Space Agency (ESA), ESAC,
Camino Bajo del Castillo s/n, 28692 Villanueva de la Ca\~nada, Spain}\label{inst2}
\and
{Sorbonne Universit\'e, CNRS, UMR 7095, Institut d’Astrophysique de
  Paris, 98bis bd Arago, 75014 Paris, France}\label{inst3}
\and
{Institut Universitaire de France, Minist\`ere de l’Enseignement Sup\'erieur et de la Recherche, 1 rue Descartes, 75231 Paris Cedex 05,
  France}\label{inst4}
\and
{Leiden Observatory, Leiden University, PO Box 9513, 2300 RA Leiden, The Netherlands}\label{inst5}
\and
{UK Astronomy Technology Centre, Royal Observatory, Blackford Hill Edinburgh, EH9 3HJ, Scotland, UK}\label{inst6}
\and
{Centro de Astrobiolog\'{\i}a (CAB), CSIC-INTA, Ctra. de Ajalvir km 4,
  Torrejón de Ardoz, 28850, Madrid, Spain}\label{inst7}
\and
{European Space Agency, c/o Space Telescope Science Institute, 3700 San
  Martin Drive, Baltimore MD 21218, USA}\label{inst8}
\and
{I. Physikalisches Institut, Universit\"at zu K\"oln, Z\"ulpicher Str. 77, 50939, K\"oln, Germany}\label{inst9}
\and
{Max-Planck Institut f\"ur Radioastronomie (MPIfR), Auf dem H\"ugel
  69, 53121 Bonn, Germany}\label{inst10}
\and
{Sterrenkundig Observatorium, Universiteit Gent, Krijgslaan 281 S9,
  B-9000 Gent, Belgium}\label{inst11} 
\and
{LESIA, Observatoire de Paris, Universit\'e PSL, CNRS, Sorbonne
  Universit\'e, Sorbonne Paris Cite\'e, 5 place Jules Janssen, F-92195
  Meudon, France}\label{inst12}
\and
{Centre for Extragalactic Astronomy, Durham University, South Road,
  Durham DH1 3LE, UK}\label{inst13}
\and
{Department of Astronomy, Stockholm University, The Oskar Klein
  Centre, AlbaNova, SE-106 91 Stockholm, Sweden}\label{inst14}
\and
{Institute of Astronomy, KU Leuven, Celestijnenlaan 200D, 3001 Leuven,
  Belgium}\label{inst15}
\and
{Dept. of Astrophysics, University of Vienna,
  T\"urkenschanzstr. 17, A-1180 Vienna, Austria}\label{inst16} 
\and
{ETH Z\"urich, Institute for Particle Physics and Astrophysics, Wolfgang-Pauli-Str. 27, 8093 Z\"urich, Switzerland}\label{inst17}
\and
{Max Planck Institute for Astronomy, Konigstuhl 17, 69117 Heidelberg, Germany}\label{inst18}
\and
{Universite Paris-Saclay, Universite Paris Cite, CEA, CNRS, AIM,
  F-91191 Gif-sur-Yvette, France}\label{inst19}
}

   \date{Received 2024; accepted 2024}

  \abstract
{We present JWST/MIRI MRS spatially resolved $\sim 5-28\,\mu$m observations of the
  central $\sim 4-8\,$kpc of the ultraluminous infrared galaxy and broad
  absorption line quasar
  Mrk~231. These are part of the Mid-Infrared Characterization of Nearby Iconic galaxy Centers
(MICONIC) program of the  MIRI European Consortium
  guaranteed time observations.
  No high excitation lines (i.e., [Mg\,{\sc v}] at $5.61\,\mu$m or [Ne\,{\sc v}] at
  $14.32\,\mu$m) typically associated with the presence of an active
  galactic nucleus (AGN) are detected in the nuclear region of Mrk~231. This 
  is likely due to the intrinsically X-ray weak nature of its
  quasar.  Some intermediate ionization potential lines, for instance,
  [Ar\,{\sc iii}] at $8.99\,\mu$m and [S\,{\sc iv}] at 
  $10.51\,\mu$m, are not detected either,  
  even though they are clearly
  observed in a star-forming region $\sim 920\,$pc south-east of the
  AGN. Thus, the strong nuclear mid-infrared (mid-IR)  continuum is
  also in part
  hampering the detection of faint lines in the nuclear region. The
   nuclear [Ne\,{\sc iii}]/[Ne\,{\sc ii}] line 
  ratio is consistent with values observed in star-forming galaxies. Moreover, we
  resolve for the first time the nuclear starburst in the mid-IR 
    low-excitation line
  emission (size of $\sim 400\,$pc, FWHM). Several pieces of evidence
  also indicate that it is partly obscured even at these wavelengths. At
  the AGN position, the ionized and warm molecular gas emission lines have modest  
  widths ($W_{\rm 80}\sim 300\,{\rm km\,s}^{-1}$). There are, however,
  weak blueshifted wings reaching velocities  $v_{\rm 02}\simeq -400\,{\rm km\,s}^{-1}$ in
  [Ne\,{\sc ii}]. The nuclear
  starburst is at the center of a large ($\sim 8\,$kpc),
  massive rotating disk with widely-spread, low velocity
outflows. 
Given the high star formation rate of Mrk~231, we speculate
  that part of the nuclear outflows and the large-scale non-circular
  motions observed in the mid-IR are driven by its powerful nuclear starburst. } 


   \maketitle
%

\section{Introduction}\label{sec:introduction}

The InfraRed Astronomical Satellite (IRAS) discovered a large number of the important
population of infrared (IR) bright galaxies, mostly in the local
Universe. They were classified in terms of their IR ($8-1000\,\mu$m)
luminosities ($L_{\rm IR}$) as 
luminous and ultraluminous IR galaxies (LIRGs and ULIRGs) with
$L_{\rm IR} = 10^{11}-10^{12}\,L_\odot$ and 
$L_{\rm IR} = 10^{12}-10^{13}\,L_\odot$, respectively. Follow-up observations revealed 
that a  significant number of local ULIRGs are interacting and merger
systems, are powered by intense star formation (SF) activity and/or an active
galactic nucleus (AGN), and contain large amounts of molecular gas. It
was also proposed that the dust-enshrouded ULIRG phase precedes the
formation of quasars and eventually elliptical galaxies \citep[see][for a
review]{SandersMirabel1996}.
Sensitive mid-infrared (mid-IR) observations taken with the InfraRed
Space Observatory (ISO) and the {\it Spitzer} Space Telescope provided
diagnostic tools that allowed estimates of  
the AGN bolometric contribution to $L_{\rm IR}$ and  nuclear extinctions
of local ULIRGs, among many other properties \citep[see e.g.,][]{Genzel1998, 
  Veilleux2009}.

Mrk~231 (IRAS~12540+5708) is a  ULIRG ($\log [L_{\rm IR}/L_\odot]=
12.5$) and a broad absorption line (BAL) quasar
\citep[see e.g., ][and references therein]{Boksenberg1977,
  Lipari2005, Rupke2005}. It is 
also the nearest quasar, located at a 
redshift of $z=0.04217$ that,  
following \cite{Feruglio2015}, corresponds to a luminosity distance of
$D = 187.6\,$Mpc and an angular scale of 837\,pc/\arcsec. It was recognized, more than
fifty years ago, as  an extraordinarily high IR luminosity
source from ground-based $10\,\mu$m observations \citep{RiekeLow1972},
later confirmed by 
IRAS. Subsequent deep optical imaging
demonstrated that Mrk~231 is an advanced merger \citep{Sanders1987} or even a triple merger
\citep{Misquitta2024}. 

Emission from dust heated by both 
AGN and SF activity is responsible for its high IR
luminosity, although estimates of the AGN bolometric
contribution vary widely, from $\simeq 30\%$ to $\simeq 70\%$
\citep[see e.g.,][]{DownesSolomon1998, Lonsdale2003, Veilleux2009, Yamada2023}.
Mrk~231 is also one of the
best examples of a local quasar and ULIRG, where 
multi-phase (neutral, ionized, and molecular) and multi-scale outflows
 have been detected \citep{Lipari2009, Fischer2010,
  Rupke2011, Cicone2012, 
  Feruglio2010, Feruglio2015, Aalto2012, Aalto2015, Morganti2016, Veilleux2016,
  GonzalezAlfonso2018, Misquitta2024}. The velocities are typically of the order of
hundreds of km\,s$^{-1}$, although they reach velocities in excess 
of  $1000\,{\rm km\,s}^{-1}$ in the neutral and molecular gas.
These outflows are
likely driven by both the  AGN and the intense SF  taking place in
Mrk~231, although the high velocity outflow is likely AGN driven
\citep[see][and references therein]{Rupke2013}.  
\cite{Leighly2014} proposed a physical model for feedback
from the AGN component, based on spectra ranging 
from the ultraviolet to the near-IR, and using the Pa$\alpha$ line they
estimated a black hole
mass of $2.3\times10^8\,M_\odot$.

  \begin{figure*}

\sidecaption
    \includegraphics[trim=4.75cm 0cm 4.75cm 1cm, clip=true, width=12cm]{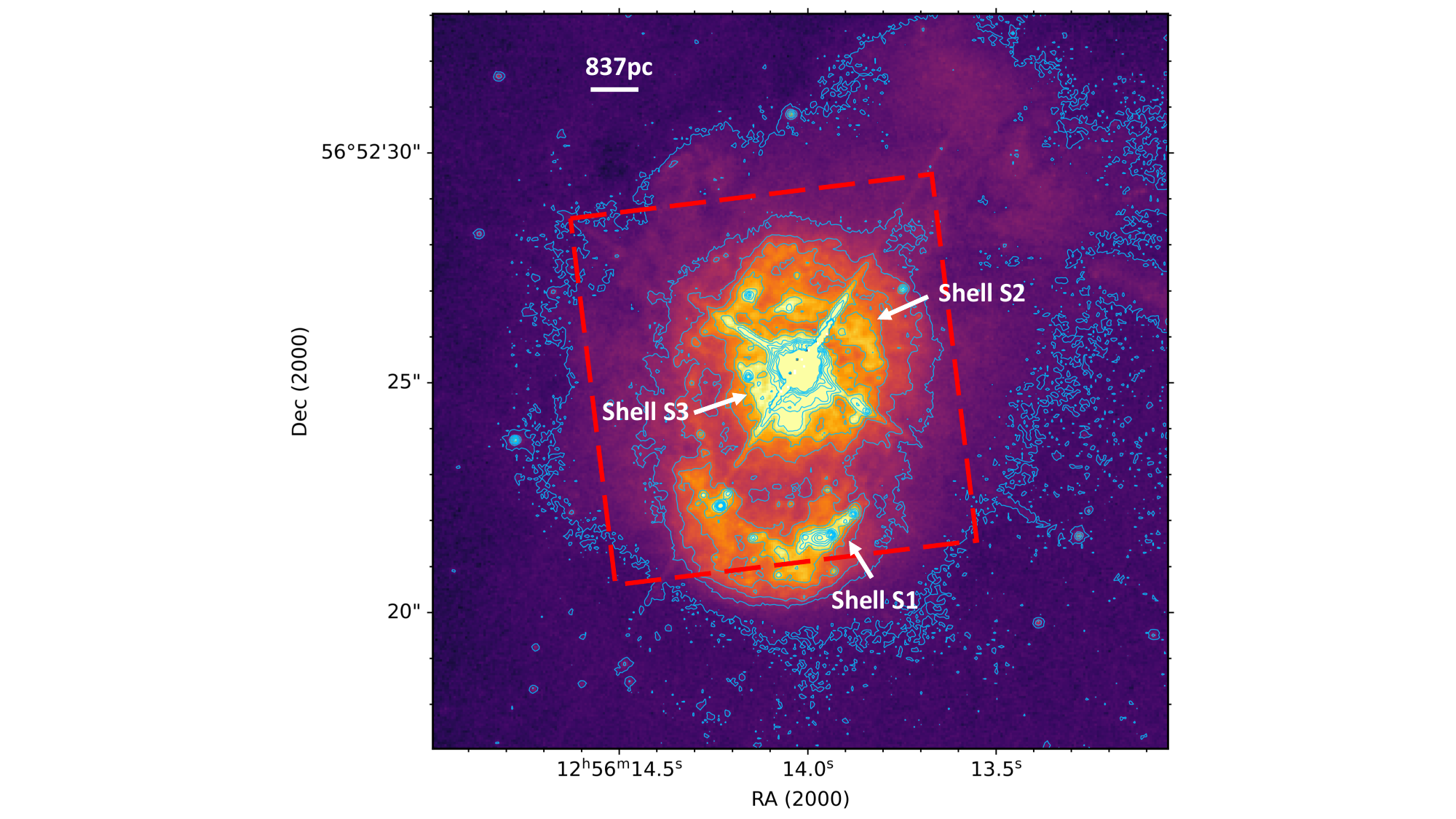}
     \caption{Archival HST/ACS image (color and contours, shown in a
       square root scale) of Mrk~231. The observation was taken with
       the optical F435W filter and it shows the central 
       $16\arcsec \times 16\arcsec$, which corresponds to 
       a region of $\simeq 13.4\,{\rm kpc} \times
       13.4\,{\rm kpc}$ in size. For reference, the overlaid dashed square represents
       the approximate FoV and 
       orientation of the MRS ch3 observations. We mark the
       positions of several expanding shells identified by \cite{Lipari2005} and
       discussed in the text.} 
     \label{fig:HSTimage}
\end{figure*}

Despite the presence of a quasar and the intense SF activity of Mrk~231, ISO and
{\it Spitzer} spectroscopic observations revealed just a handful of
mid-IR low-excitation emission lines and H$_2$ transitions accompanied
by a strong mid-IR continuum, but no high-excitation
lines. Additionally, only a few polycyclic
aromatic hydrocarbon (PAH) features with low equivalent widths (EW)
were detected
\citep[see e.g.,][]{Genzel1998,
  Armus2007, Veilleux2009}. Sub-arcsecond resolution mid-IR imaging,
spectroscopy, and polarimetry with ground-based $8-10\,$m-class
telescopes demonstrated that the emission is dominated by  a bright,
mostly unresolved, source 
(angular size of $0.1-0.4\arcsec\simeq 84-330\,$pc),  with a
moderately deep $9.7\,\mu$m  feature due to amorphous silicates 
\citep{Soifer2000, AlonsoHerrero2016a, AlonsoHerrero2016b, LopezRodriguez2017}.
The ground-based observations  also showed that the AGN accounts 
for a substantial fraction of the  mid-IR continuum 
emission of Mrk~231, with only a small contribution from SF
activity. The latter likely increases at longer wavelengths \citep[see
e.g.,][]{Efstathiou2022}.

We present integral field unit (IFU) spectroscopy of Mrk~231
obtained with the Mid-Infrared Spectrometer (MRS) of the Mid-InfraRed
Instrument 
\citep[MIRI,][]{Rieke2015, Wright2015, Wright2023} on board
of the {\it James Webb} Space 
  Telescope \citep[JWST,][]{Gardner2023}. The observations are part of the guaranteed time
  observations (GTO) program termed
  Mid-Infrared Characterization of Nearby Iconic galaxy Centers
(MICONIC) of the MIRI European
  Consortium, which includes Mrk~231, Arp~220,
  NGC~6240, 
  Centaurus A, and SBS0335-052, as well as the region surrounding
  SgrA$^*$ in our galaxy. These targets are also part of the
 GTO program for nearby 
  galaxies of the NIRSpec instrument team. The goal of this work is to
  carry out a spatially resolved study of the mid-IR emission of the
  nuclear and circumnuclear regions of Mrk~231.

  The paper is organized as
  follows. Section~\ref{sec:observations} describes the MRS
  observations, data reduction, and analysis, and  Sect.~\ref{sec:nuclearemission}
  the mid-IR nuclear 
  emission of Mrk~231. In Sects.~\ref{sec:ionizedgas} and
  ~\ref{sec:molecularhydrogen} 
  we analyze the spatially resolved observations of the ionized gas and
  warm molecular gas, 
  respectively. In  Sects.~\ref{sec:SFactivity} and \ref{sec:outflows}
  we use these new spatially resolved MRS observations to infer the SF
  activity and outflow properties of this ULIRG.  In
  Sect.~\ref{sec:conclusions}  we 
  present our summary.

  \section{MIRI MRS observations}\label{sec:observations}
\subsection{Data reduction}
We obtained  MIRI MRS observations of the central region of Mrk~231,
as part of JWST Cycle 1 program ID 1268 that also includes the NIRSpec
GTO observations of this target. The MIRI observations were taken
using a single pointing observed with a four-point dither, extended source pattern. The
on-source integration was 522\,s and made use of the FASTR1 readout
mode. We also took an MRS background observation using a two-point
dither, extended source pattern. The 
observations cover the full $\sim 5-28\,\mu$m spectral range.

We processed the MRS observations using version 1.12.3 of the JWST
Science Calibration Pipeline \citep{Bushouse2023}\footnote{See also
  https://jwst-docs.stsci.edu/jwst-science-calibration-pipeline-overview},
and the files from context 1135 of the Calibration References Data
System (CRDS). We followed the standard procedure for MRS data
\citep[see][]{Labiano2016,
  AlvarezMarquez2023} for stages 1 and 2 of the pipeline. The first
stage of the MRS pipeline \citep{Morrison2023, Dicken2022} performs the
detector-level and cosmic ray corrections, and transforms the ramps to
slope detector images. The default cosmic ray correction and
parameters did not leave any significant residuals in the data. The
second stage performs the specific corrections needed
for the MRS data \citep{Argyriou2023, Gasman2023, Patapis2024}. 

At the end of stage 2, our data showed some negative spikes in the
Level 2 science cubes. Investigation and discussion with STScI showed
that these were caused by bad or warm pixels (flagged as NaN) in the
detector, exactly atop the brightest part of the spectral traces. The
negative residuals were the result of the pipeline interpolating
these around using neighbouring pixels. At the time of our data
  reduction, we solved the problem by searching for isolated NaNs and
interpolating in
the spectral direction (D. Law, private communication) in the {\it
  cal.fits}. We note that currently the pipeline 
uses a pixel replacement routine to replace bad pixels flagged with a
bad-pixel map. We then ran pipeline stage 3  on
this modified datasets to combine the 4-dither pointings and
create the final  spectral cubes \citep{Law2023}. We  switched
off the  background corrections ({\it bkg\_subtract} and {\it
  master\_background}) and sky matching ({\it imatch}) steps
as they introduced artifacts. Although stage 3 is usually
ran just on the science data, we also created Level 3 background
cubes  with the same procedure as the science cubes. These can be used
to extract one-dimension (1D) background spectra to be compared to the science spectra.


We chose to produce the fully reconstructed science data cubes in the usual
orientation of north up, and east to the left. The resulting fields of
view (FoV) and pixel
sizes are (see  Fig.~\ref{fig:continuummaps}  for a few examples):  between
$5.1\arcsec \times  
5.6\arcsec$ and $5.1\arcsec \times 5.9\arcsec$ with a 0.13\arcsec
pixel size for channel 1 (ch1, observed wavelengths $4.9-7.65\,\mu$m), 
$6\arcsec \times 7\arcsec$ and $6.3\arcsec \times 7\arcsec$ with
0.17\arcsec \, for channel 2 (ch2, $7.51-11.7\,\mu$m), $7.8\arcsec
\times 8.6\arcsec$ and
$8.2\arcsec \times 8.6\arcsec$ with 0.2\arcsec \, for channel 3 (ch3,
$11.55-17.98\,\mu$m), 
and $10.1\arcsec\times 10.9\arcsec$ and $10.9\arcsec \times
10.9\arcsec$ with 0.35\arcsec \, for  
channel 4 (ch4, $17.7-27.9\,\mu$m). 
For reference, on the archival (proposal ID 10592)
{\it Hubble} Space Telescope (HST) image of Fig.~\ref{fig:HSTimage} taken
with the Advanced Camera for Surveys (ACS), we
overlaid the approximate FoV and 
orientation of the ch3 observations. These cover, apart from the
nuclear region, some of the expanding shells and super-bubbles identified by
\cite{Lipari2005}, which are believed to be associated with the
intense SF activity taking place in Mrk~231. We will
discuss this in detail in Sects.~\ref{sec:SFactivity} and \ref{sec:outflows}.

\subsection{Data  analysis}\label{subsec:analysis}

A bright nuclear point source dominates the mid-IR continuum emission
in all the MRS channels (see some continuum maps 
in Fig.~\ref{fig:continuummaps}), as already known from
ground-based mid-IR observations  at comparable angular resolutions
\citep{Soifer2000, AlonsoHerrero2016a, LopezRodriguez2017}. Indeed, the measured
continuum full width at half
maximum (FWHM) in all the MRS sub-channels
is consistent with the reported values for  point sources
\citep{Law2023, Argyriou2023}, ranging in Mrk~231 from $\sim 0.3\arcsec$
(251\,pc) at $5\,\mu$m to $\sim 0.8-0.9\arcsec$ at $23\,\mu$m (753\,pc).

We extracted a 1D nuclear spectrum for each of the MRS
sub-channels by  centering a relatively large aperture of 
1\arcsec \, ($\simeq 837\,$pc) radius at the
continuum peak. We chose it to  encompass most of the extended nuclear
line emission while avoiding emission from other circumnuclear star forming
regions  (see Sect.~\ref{sec:ionizedgas}).  
Additionally, the large aperture size  together with the four-point dither help
minimize the continuum wiggles that are due to 
the undersampling of the point spread 
function  \citep[see][]{Law2023}. We
applied the residual fringing correction to the nuclear 1D spectra  in
Fig.~\ref{fig:nuclearspectrum} with the {\sc rfc1d\_utils}
routine\footnote{
    https://github.com/spacetelescope/jwst/blob/master/jwst/residual\_fringe/utils.py.}  
\citep[Kavanagh et al in prep. and also][]{Gasman2023}.
We did not apply an aperture correction to the nuclear 1D spectra
because it contains both unresolved (continuum) and resolved (emission lines)
components. We note that the aperture radius is large enough that the
correction for the continuum unresolved emission would be less than
approximately $10\%$ for ch1 and ch2, 
but could reach $25\%$ at the longest wavelengths
\citep[see][]{Argyriou2023}.  For comparison we show the {\it
    Spitzer}/IRS spectra extracted as a 
  point source in Fig.~\ref{fig:SpitzerIRS}. The flux calibration of the
individual MRS channels was good 
and thus no correction factor was applied prior to the stitching of
the sub-channels, except for ch1A for which needed to be multiplied by a 1.02 factor
to match ch1B. Finally, given the extremely bright nuclear
flux of Mrk~231, we did not perform a background subtraction since its
contribution is negligible. 

The resulting rest-frame $\sim 4.7-23.5\,\mu$m nuclear spectrum is shown in
Fig.~\ref{fig:nuclearspectrum}. We did not include ch4C
because the red end is still affected by residual noise.
Fig.~\ref{fig:nuclearspectrum_channels}
shows the same
nuclear spectra but separately for each  
channel. We marked the positions of the pure rotational H$_2$  S(8)
to S(1) transitions and
 fine-structure lines as well as broad absorption features and
 PAHs features. We measured the fluxes
 and velocity dispersion of the emission lines from the nuclear 1D
 spectra before applying the residual fringing
       correction.  We fit a single Gaussian 
function and a local continuum, and then computed the FWHM after
correcting for the wavelength-dependent instrumental resolution
provided by \cite{Argyriou2023}. The uncertainties of the fluxes and
line widths were
calculated from the propagation of the errors of the fitted
parameters. The fluxes and line widths are listed in
Table~\ref{tab:linefluxes}.  We did not include the H$_2$  S(6) transition
because it is strongly affected by an H$_2$O absorption feature
(Fig.~\ref{fig:nuclearspectrum_channels}, top left). We note, however, that most
emission lines present weak wings, especially blueshifted. In
Sect.~\ref{subsec:outflows_nuclear} we will quantify them using a
non-parametric method.

To produce spatially resolved maps of the
fine-structure emission lines and bright rotational H$_2$ lines,
we  fitted a single Gaussian function and a linear continuum on a
spaxel-by-spaxel basis. We adopted a $3\times N_{\rm cont}$ detection
threshold at the peak of the
line over the continuum, where $N_{\rm cont}$ is the standard
deviation of the continuum adjacent to the line. With these fits, we
produced maps of the line intensity, the mean-velocity field using the
velocity corresponding to the peak of the line, and the velocity
dispersion after correcting from  instrumental resolution. We refer the reader 
to \cite{HermosaMunoz2024_NGC7172}
for a complete description of the
method used to construct the line maps. We discuss these maps in Sects.~\ref{sec:ionizedgas},
and \ref{sec:molecularhydrogen} for the ionized and molecular gas, respectively.

\section{Nuclear emission}\label{sec:nuclearemission}

\subsection{Continuum emission}
\begin{figure}
     \includegraphics[width=9cm]{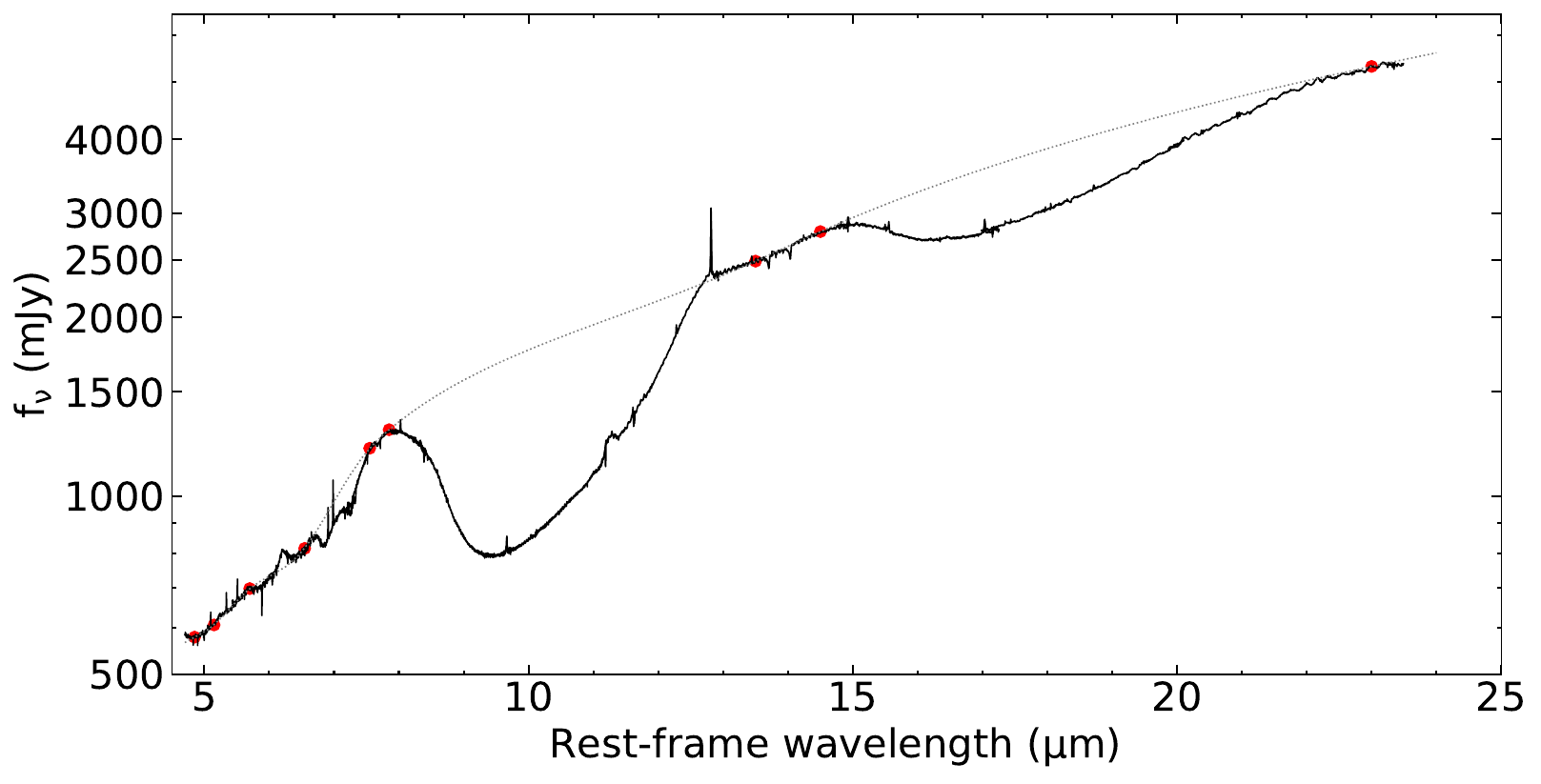}
     \caption{MRS nuclear spectrum (solid black line) of
       Mrk~231. It was extracted with a
       1\arcsec-radius aperture and covers the rest-frame $\sim
       4.7-23.5\,\mu$m spectral range. We applied the residual fringing
       correction to the extracted 1D spectra of each of the
       sub-channels. The dotted line is the fitted continuum using a
       cubic spline function and the red dots are
       the anchor points selected for the fit.}
     \label{fig:nuclearspectrum}
\end{figure}

The nuclear spectrum of Mrk~231 presents a strong and steeply rising mid-IR
continuum  (Fig.~\ref{fig:nuclearspectrum}). It is 
mostly produced by the bright point source, which
dominates the emission in all the MRS channels (see Fig.~\ref{fig:continuummaps}). The MRS
continuum fluxes in the $\simeq 8-12\,\mu$m spectral range are similar
to those reported from ground-based observations
\citep{Soifer2000, AlonsoHerrero2016a, LopezRodriguez2017} at $\simeq
0.3\arcsec$ resolution. 

The modeling
of the nuclear near- and mid-IR emission of Mrk~231 done by
\cite{LopezRodriguez2017} showed that most of the MRS continuum
emission would stem from the dusty molecular torus, although a
powerful nuclear starburst was also required to
reproduce the near and mid-IR emission (see also
Sects.~\ref{sec:ionizedgas} and \ref{sec:SFactivity}).
Indeed, using OH megamaser observations, \cite{Kloeckner2003} 
detected a 100\,pc-radius torus, which appears to be warped with
respect to the larger-scale disk. 
The expected size of the hot and warm dust
emission dominating the mid-IR continuum is expected to be smaller than the torus physical size since it would
mostly arise in its inner walls \citep[e.g., ][]{Hoenig2019,
  Nikutta2021, AlonsoHerrero2021}. Thus, the Mrk~231 torus would be
unresolved, even at the best MRS angular resolution \citep[$\simeq
0.3$\arcsec][]{Argyriou2023}.  

\subsection{Low excitation emission lines, rotational H$_2$ lines, and
  PAH features}\label{subsec:lines} 
\begin{figure*}
     \includegraphics[width=9cm]{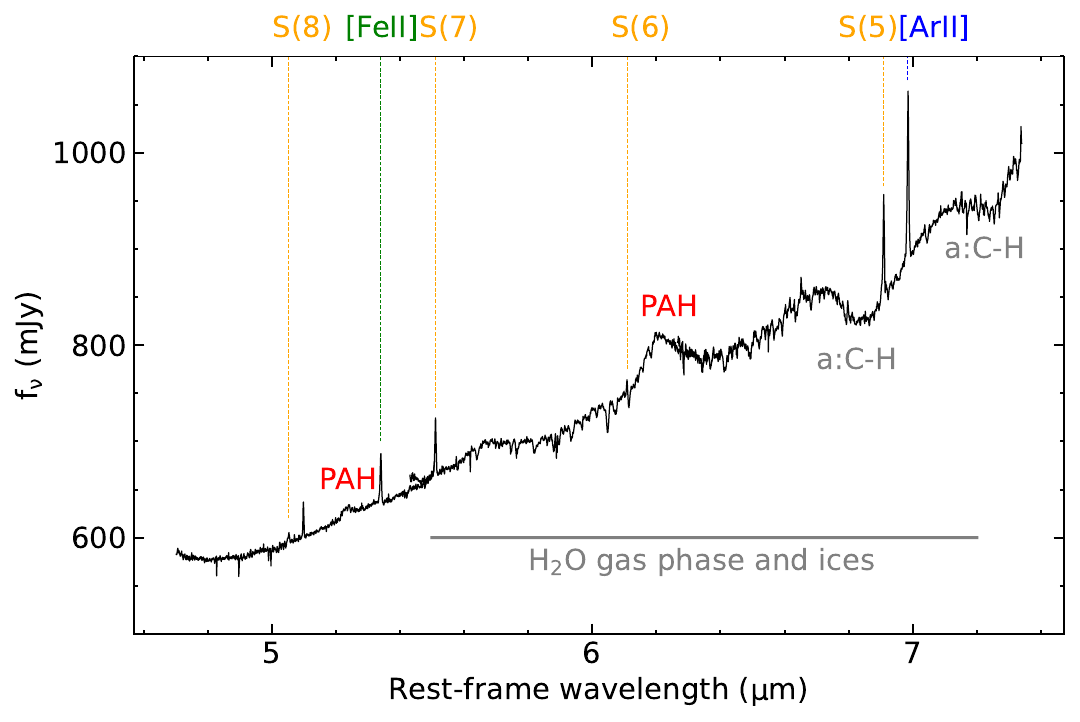}
     \includegraphics[width=9cm]{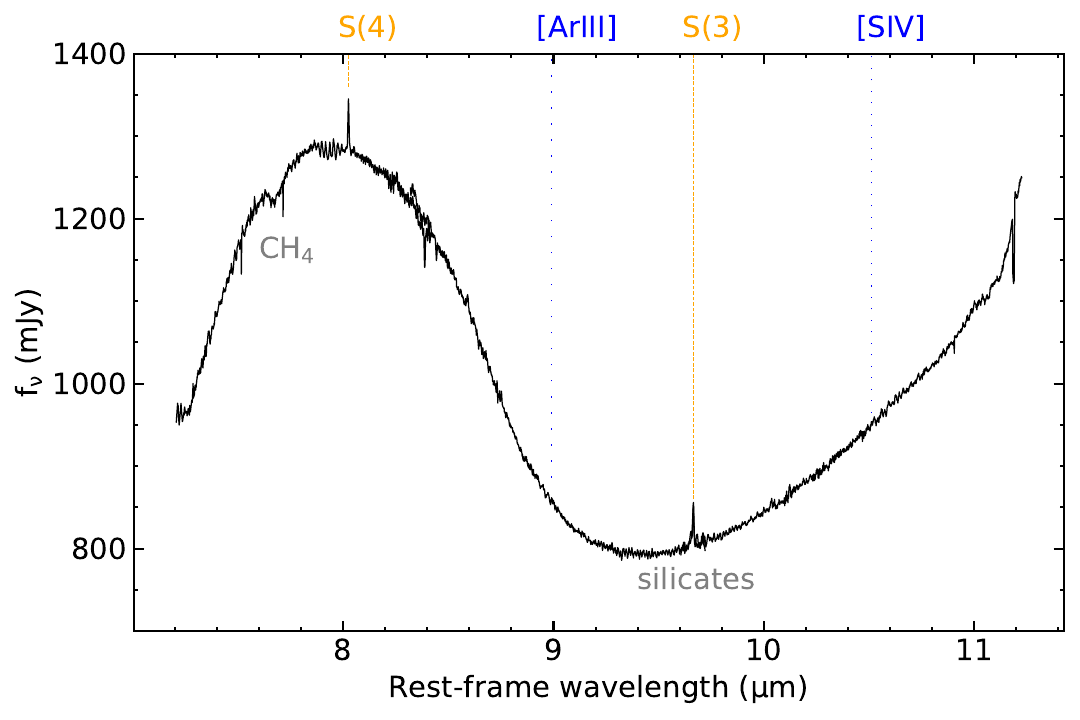}
     
     \includegraphics[width=9cm]{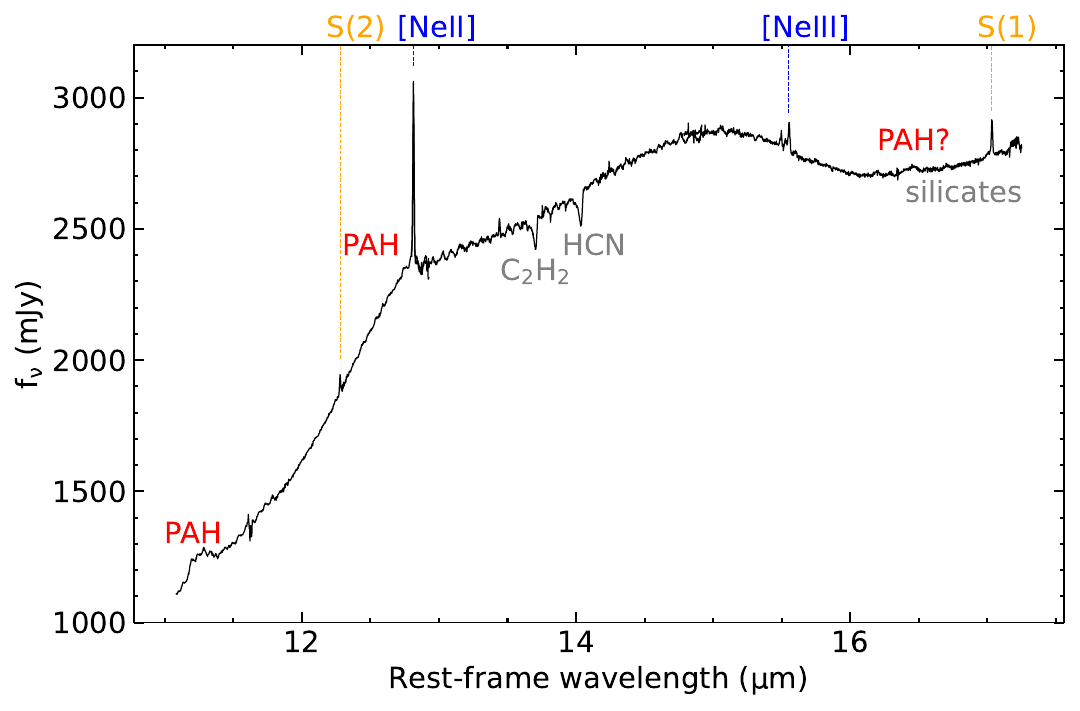}
     \includegraphics[width=9cm]{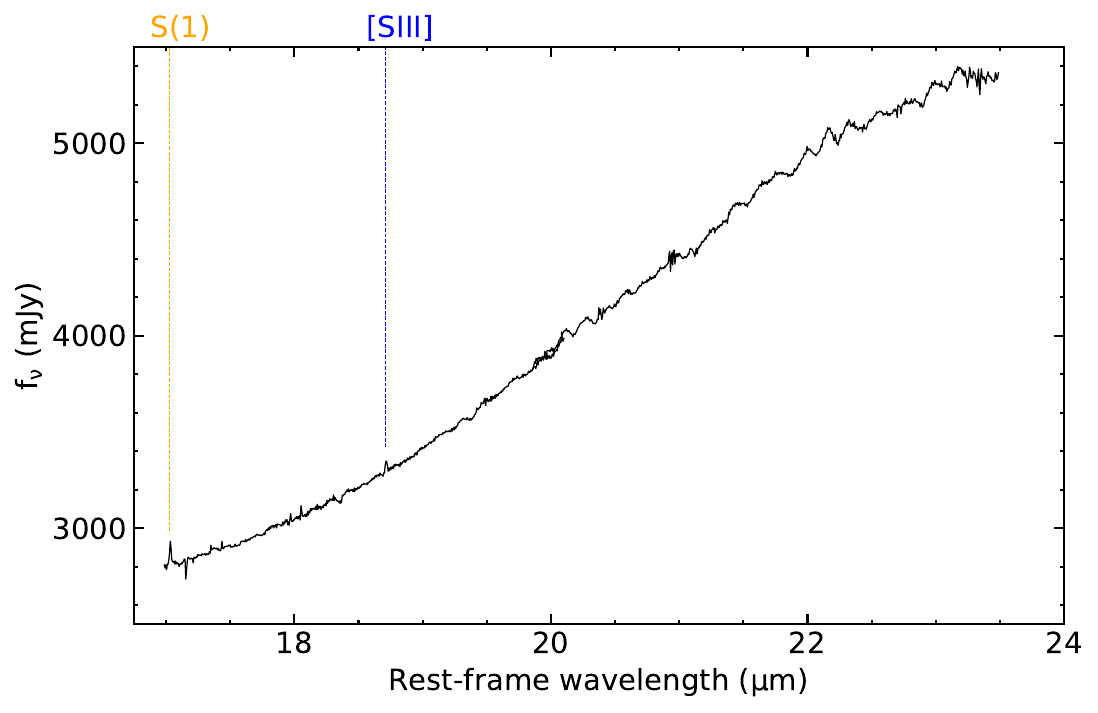}
     \caption{Nuclear spectrum shown for each of the
       MRS channels. As in
       Fig.~\ref{fig:nuclearspectrum}, the extraction aperture had
       a 1\arcsec-radius. Each panel shows the A, B, and
       C  sub-channels (ch1 {\it upper left}, ch2 {\it upper right},
       ch3 {\it lower left}, and ch4 {\it lower right}), except for ch4 for which we only
       show A and B (see text). The dashed lines mark the detected
       low-excitation fine-structure
       emission lines (blue and green) and rotational H$_2$ lines
       (orange). The loosely dotted lines are other relatively low
       IP lines not
       detected in the nuclear spectrum (see text for more details). Also
       labeled are the positions of PAHs, silicate features, and some molecular
       absorption bands. }
     \label{fig:nuclearspectrum_channels}
\end{figure*}

In contrast to other nearby AGN and (U)LIRGs observed with MRS
\citep[see e.g.,][and also van der Werf et al.  in prep. and Hermosa Mu\~noz et
al. in prep.]{PereiraSantaella2022, AlvarezMarquez2023, 
  GarciaBernete2022, GarciaBernete2024, Armus2023}, there are only a few  emission lines
with relatively low EW (Figs.~\ref{fig:nuclearspectrum} and
~\ref{fig:nuclearspectrum_channels}) in the nuclear spectrum. 
The series of pure rotational 0--0 S(8) to S(1) H$_2$ lines is clearly 
detected in the nuclear region of 
Mrk~231. It is also evident that only  fine-structure emission lines
with low ionization potentials (IPs), between 7.9\,eV and 41\,eV,
are present (Table~\ref{tab:linefluxes}).
On the other hand,  [Ar\,{\sc iii}] at
$8.99\,\mu$m and
[S\,{\sc iv}] at $10.51\,\mu$m (loosely dotted lines in the upper
right panel of Fig.~\ref{fig:nuclearspectrum_channels}) with IPs of 28
and 35\,eV, respectively,
are not detected in the nuclear region of Mrk~231. We note, however, that
these lines are present in a circumnuclear star-forming region (see
Sect.~\ref{subsec:starformingregions}).
When compared with  previous {\it
  Spitzer}/IRS spectroscopic observations \citep[see][and also
  Fig.~\ref{fig:SpitzerIRS}]{Armus2007,
  Veilleux2009}, [Fe\,{\sc ii}], [Ar\,{\sc
  ii}], [Ar\,{\sc  iii}], [Ne\,{\sc iii}], and  [S\,{\sc iii}], as well as the S(8) to S(4) H$_2$ lines
are new MRS detections in the (circum)nuclear region of Mrk~231.

\begin{table}
  \caption{MRS fluxes measured in the nuclear ($r=1\arcsec$)
    region of Mrk~231.}\label{tab:linefluxes}
  \begin{tabular}{lcccc}
    \hline
    Line & $\lambda_{\rm rest}$ & IP & flux & 
                                                              FWHM$_{\rm line}$\\
           & ($\mu$m) & (eV) & ($10^{-14}$ erg cm$^{-2}$ s$^{-1}$) & (km s$^{-1}$)\\
    \hline

    H$_2$ S(8) & 5.053 &... & $0.72\pm 0.10$ &  ...\\
       ${\rm [Fe\,II]}$ & 5.340   & 7.9 & $2.36\pm 0.13$ & $236\pm9$\\
    H$_2$ S(7) & 5.511 &... & $2.59\pm 0.14$ & $248\pm 10$ \\
    ${\rm [Mg V]}$ & 5.610 & 109 & $<1.1$ &  ...\\
    H$_2$ S(5) & 6.910 & ... &$3.21\pm 0.20$ & $208\pm 10$\\
    ${\rm [Ar\,II]}$ &  6.985 & 15.8 & $5.74\pm 0.12$ & $216\pm 4$ \\ 
    H$_2$ S(4) & 8.025 & ... &$1.64\pm 0.10$ & $209 \pm 10$ \\
    H$_2$ S(3) & 9.665 & ...&$0.73 \pm 0.09$ & ...\\
    H$_2$ S(2) & 12.279 & ...& $0.90 \pm 0.17$  & $206\pm 29$\\
    ${\rm [Ne\,II]}$ & 12.814 & 21.6 &  $11.14 \pm 0.41$ & $203 \pm 6$\\
    ${\rm [Ne\,V]}$ & 14.322 & 97.2&  $<1.3$ &  ...\\
   ${\rm[Ne\,III]}$ &15.555 & 41.6 & $1.37\pm 0.29$ & ...\\
    H$_2$ S(1) & 17.035 &... & $1.21\pm 0.23$ & $171 \pm 24$\\
    ${\rm [S\,III]}$& 18.713 & 23.3 & $0.29\pm0.11$ &...\\

    \hline
  \end{tabular}
  Notes.--- Fluxes and line widths are from fits with a single Gaussian. 
  FWHM$_{\rm line}$ are corrected for instrumental resolution and are
  only given for values with uncertainties of less than $15\%$. 
    The upper limits are $3\sigma_{\rm cont}$ (see text for details).
  \end{table}

The observed ratio of [Ne\,{\sc
  iii}]/[Ne\,{\sc ii}]$=0.12$  (Table~\ref{tab:linefluxes})
is well within the typical values of star-forming galaxies (average of
$0.17\pm0.07$), while local 
quasars present significant higher ratios 
(average of  $2.0 \pm 1.0$), both values  from {\it Spitzer}/IRS spectra
\citep[][]{PereiraSantaella2010},
Moreover, photoionization models  for AGN radiation
fields \citep[see][and references therein]{Feltre2023} predict  [Ne\,{\sc
iii}] to [Ne\,{\sc ii}] ratios above the observed ratio in Mrk~231. Only radiation fields with steep UV
continua, which are representative of low ionization nuclear emission-line regions
(LINERs), would produce low [Ne\,{\sc
iii}]/[Ne\,{\sc ii}] ratios \citep[see Fig.~6
of][]{Feltre2023}. However, the nuclear optical line emission of
Mrk~231 is classified as a 
Seyfert 1 galaxy \citep[see][and references therein]{Sanders1988}.

\cite{vanderWerf2010} modeled the CO rotational ladder  of Mrk~231 using a
combination of a photodissociation region (PDR) and an X-ray dominated
region (XDR), with the latter being responsible for the emission of
the high-J CO transitions. Models including an XDR produce
mostly [Ne\,{\sc   ii}], while the [Ne\,{\sc iii}] emission would be partly
suppressed \citep{PereiraSantaella2017}. Therefore,  the 
  relatively low 
  [Ne\,{\sc iii}]/[Ne\,{\sc ii}]  ratios in the nuclear region could be
  due to   AGN photoionization with a contribution from an XDR.
  \cite{vanderWerf2010} predicted a radial size of the X-ray illuminated
region of 160\,pc from the Mrk~231 using the
range of intrinsic X-ray luminosity given by \cite{Braito2004}. 
  We note that 
this putative XDR would have a  size smaller 
than that  of the resolved nuclear starburst 
obtained from the low-excitation [Ar\,{\sc ii}] and [Ne\,{\sc ii}]
emission lines (see
Sect.~\ref{subsec:linefluxmaps}). However, we have limited spatial
information 
about the nuclear distribution of the [Ne\,{\sc iii}] emission line (see
Sect.~\ref{subsec:linefluxmaps} and
Appendix~\ref{appendix:extramaps}). Thus, we cannot conclusively determine
the effects of a putative XDR of Mrk~231 on the observed nuclear mid-IR
line ratios.

  
Emission from the $6.2\,\mu$m and the $11.3\,\mu$m PAH features is
clearly seen in the nuclear region of Mrk~231, together with other fainter PAH
features typically detected
in PDRs
\citep[see e.g., in the Orion bar,][]{Chown2024}, such as
those at  $5.25$ and $12.7\,\mu$m, and tentatively at $16.45\,\mu$m
\citep[see][]{Donnan2023}.  The relatively low EW of the PAH features
for the 6.2 and $11.3\,\mu$m (see Sect.~\ref{subsec:CON} for the
measurements) reflects the 
dominance of an AGN produced mid-IR continuum, 
even though the extracted nuclear spectrum covers a relatively large physical
region (central $\simeq 1.6\,$kpc, in diameter).

\subsection{High-excitation emission lines}\label{subsec:lines_highexcitation}
\begin{figure}
\hspace{0.25cm}
     \includegraphics[width=8cm]{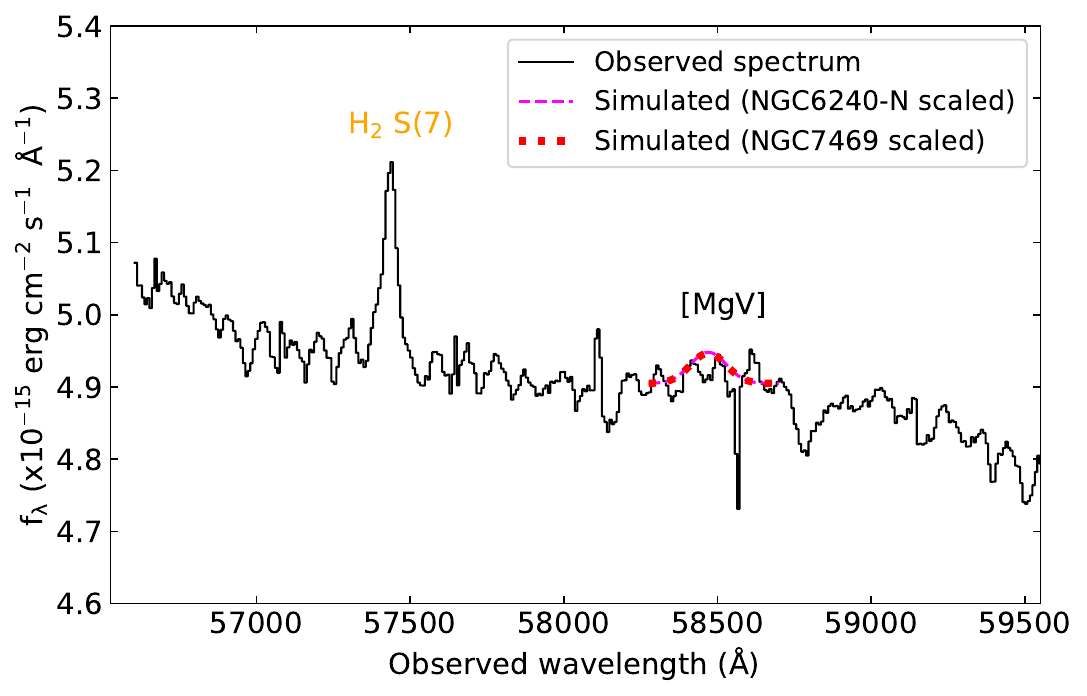}

\hspace{0.25cm}
     \includegraphics[width=8cm]{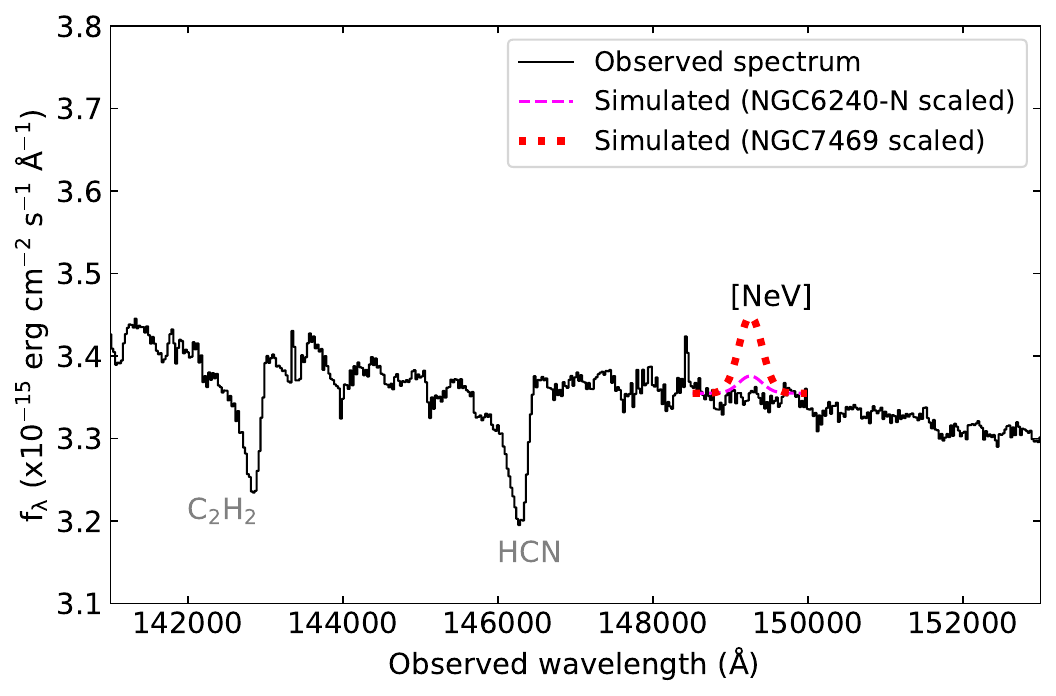}
     \caption{Zoomed-in spectral regions of the
       nuclear spectra  around the wavelengths of high-excitation
       lines [Mg\,{\sc v}] ({\it top}) and 
 [Ne\,{\sc v}] ({\it bottom}). The black lines are the observed
 spectra  extracted  with an
aperture radius of 0.5\arcsec (top) and an aperture 
radius of 1\arcsec (bottom), while the dotted and dashed lines are the
predicted high-excitation emission lines from MRS observations of
two local AGN assuming an intrinsic $2-10\,$keV luminosity of
$10^{44}\,{\rm erg\,s}^{-1}$  for Mrk~231 (see text for
details). The ch3-medium spectrum was corrected for residual
  fringing that was affecting mostly data at wavelengths longer than
  that of [Ne\,{\sc v}].}
     \label{fig:simulated_lines}
\end{figure}

Mid-IR high-excitation lines with ${\rm IP} \ge 95$\,eV (e.g.,
[Mg\,{\sc v}] at $5.61\,\mu$m and [Ne\,{\sc v}] at 
$14.32\,\mu$m) are generally associated with 
the presence of AGN  \citep{Sturm2002, PereiraSantaella2022, GarciaBernete2022,
  GarciaBernete2024, Armus2023, AlvarezMarquez2023}. Moreover, AGN
present a good correlation between 
the [Ne\,{\sc v}] at $14.32\,\mu$m luminosity, among several IR high
excitation lines, and the intrinsic 
$2-10\,$keV luminosity \citep{Spinoglio2022}. 
These and other
high-excitation lines remain undetected in 
the nuclear spectrum of Mrk~231 (see
Fig.~\ref{fig:nuclearspectrum_channels}), despite the higher 
sensitivity and angular resolution of the MRS observations
compared with previous observations \citep{Armus2007, Veilleux2009}.
 We can obtain an estimate of the line upper limits by measuring
  the standard deviation of the continuum ($\sigma_{\rm cont}$)
  adjacent to the lines (see Fig.~\ref{fig:simulated_lines}). For
  [Ne\,{\sc v}] we corrected the 1D spectrum for residual fringing
  which was affecting the data at wavelengths mostly redder than the line.
  We
  note that for both lines,  part of the continuum structure  near the
  lines is due to the
  presence of individual absorption features
(see
  Sect.~\ref{subsec:absorptionfeatures}). For the expected width of the  
  [Mg\,{\sc v}] and [Ne\,{\sc v}] lines of Mrk~231, we
took the FWHM$_{\rm line}$$\simeq 700\,{\rm km\,s}^{-1}$ value measured by
\cite{Rupke2011} using the optical H$\alpha$ line 
for the nuclear region, which they attributed to the nuclear
wind. This value is higher than the measured
FWHM$_{\rm line}$ of
the low excitation lines detected with MRS in Mrk~231 (Table~\ref{tab:linefluxes}),
but these are produced mainly 
by SF activity (see Sect.~\ref{sec:ionizedgas}).  However,
in NGC~7469 the broad components of [Ne\,{\sc v}] and [Mg\,{\sc v}]
have FWHM$_{\rm line}$$\simeq 726-814\,{\rm km\,s}^{-1}$
\citep{Armus2023}. 
The $3\sigma_{\rm cont}$ upper
  limits are $1.1\times 10^{-14}\,{\rm erg\,cm}^{-2}\,{\rm s}^{-1}$ and 
$1.3\times 10^{-14}\,{\rm erg\,cm}^{-2}\,{\rm s}^{-1}$ for [Mg\,{\sc v}]
and [Ne\,{\sc v}], respectively.

 We can also use MRS observations of two AGN in local LIRGs 
to predict
 the expected fluxes of [Mg\,{\sc v}] and
 [Ne\,{\sc v}] for Mrk~231 using their X-ray luminosities. 
The 
intrinsic $2-10\,$keV luminosity of Mrk~231 was estimated to be 
between $ 2 \times 10^{43}$ and $\simeq 10^{44}\,{\rm erg\,s}^{-1}$
\citep[][corrected for the distance assumed in this work]{Braito2004}, although other works
\citep[see e.g.,][]{Teng2014} inferred a lower value putting it in the
intrinsically X-ray weak quasar category.
Both the bright Seyfert 1 galaxy NGC~7469
\citep{Armus2023} and the AGN in the northern nucleus of NGC~6240 
(NGC~6240-N, Hermosa Mu\~noz et al.  in prep.) present similar $2-10\,$keV luminosities,
$3\times 10^{43}\,{\rm erg\,s}^{-1}$ 
\citep{Vasudevan2010} and $2\times 10^{43}\,{\rm
  erg\,s}^{-1}$ \citep{Puccetti2016}, respectively.

We used
the observed line fluxes of the 
high excitation lines in
NGC~7469 (summing the narrow and broad components) and NGC~6240-N,
scaled to the distance and a $2-10\,$keV luminosity of $10^{44}\,{\rm
  erg\,s}^{-1}$ for Mrk~231 to
predict the amplitudes of the two lines, assuming a single
Gaussian function. Again we took FWHM$_{\rm line}$$\simeq
  700\,{\rm km\,s}^{-1}$. Figure~\ref{fig:simulated_lines} shows the 
simulated lines over spectra extracted with an aperture radius of
0.5\arcsec \, for the [Mg\,{\sc v}] line and 1\arcsec \, for [Ne\,{\sc
  v}]. The smaller aperture for ch1 is intended to minimize the
continuum structure contribution near the line, while containing most of the
unresolved emission at these wavelengths. The simulated [Mg\,{\sc v}]
lines are consistent with the non-detection in the observed
spectrum, corresponding to a $\simeq 1.7\sigma_{\rm cont}$
  detection, approximately. The [Ne\,{\sc v}] line at $14.32\,\mu$m predicted from the
NGC~7469 MRS flux would have 
been detected if the $2-10\,$keV luminosity of Mrk~231 was 
$10^{44}\,{\rm  erg\,s}^{-1}$ but would be undetected if
Mrk~231 is an  intrinsically X-ray weak quasar \citep{Teng2014}.
On the other hand, the line flux
predicted from NGC~6240-N would be 
  buried within the noise and structure of the continuum emission.
Therefore, the lack of
detection of [Mg\,{\sc v}] and [Ne\,{\sc v}] in the nuclear region of
Mrk~231 is likely due a combination of
a strong mid-IR continuum, the presence of continuum structure due to
  absorption features, and its relatively weak X-ray emission
\citep[see also][]{Teng2014}.


\subsection{Absorption features}\label{subsec:absorptionfeatures}

The nuclear spectrum of Mrk~231 presents a $\sim$9.7$\,\mu$m
silicate feature  in moderate absorption, as
already known from  ground-based 
observations with similar angular resolution
\citep{AlonsoHerrero2016a, 
  AlonsoHerrero2016b, LopezRodriguez2017}. A
corresponding $\sim$18$\,\mu$m feature was previously observed with
{\it Spitzer}/IRS \citep{Armus2007, 
  Veilleux2009} at lower angular resolution. To measure the apparent optical
depths of the silicate features, we fitted the
continuum emission with a cubic spline function with several anchor
points (red dots in Fig.~\ref{fig:nuclearspectrum} and continuum fit plotted as
a dashed line). 
We calculated the apparent depths of 
the features as $S_{\rm feature} = {\rm ln} (f_{\rm feature}/f_{\rm
  cont})$ and obtained $S_{9.7\mu{\rm m}}=-0.75$ and 
$S_{18\mu{\rm m}}=-0.24$.  These values are consistent with past
observations, although they are subject to a precise extraction and
fit of the continuum emission. 

The high resolving power of MRS \citep{Labiano2021, Jones2023, 
  Argyriou2023} allows to identify numerous absorption
features in Mrk~231. These 
include the previously detected features  at $\sim 6.85$ and
$7.25\,\mu$m  (Fig.~\ref{fig:nuclearspectrum_channels}),
which are attributed to aliphatic hydrocarbons 
\citep[possibly a:C–H hydrogenated
amorphous carbon analogs, see][for more details]{Spoon2022}, and
at $\simeq 13.7\,\mu$m and $\simeq 14\,\mu$m 
produced  by acetylene (C$_2$H$_2$) and hydrogen cyanide (HCN), respectively 
\citep{Lahuis2007}. Additionally, a newly detected absorption
feature in Mrk~231 at $7.7\,\mu$m
is due to methane (CH$_4$) ice. All these 
absorption features are observed in the nuclear regions of
other local (U)LIRGs \citep{Rich2023, Buiten2024} and 
MICONIC targets (Buiten et al. in prep., Hermosa Mu\~noz et al.  in prep.).  

\begin{figure*}
  \includegraphics[width=19cm]{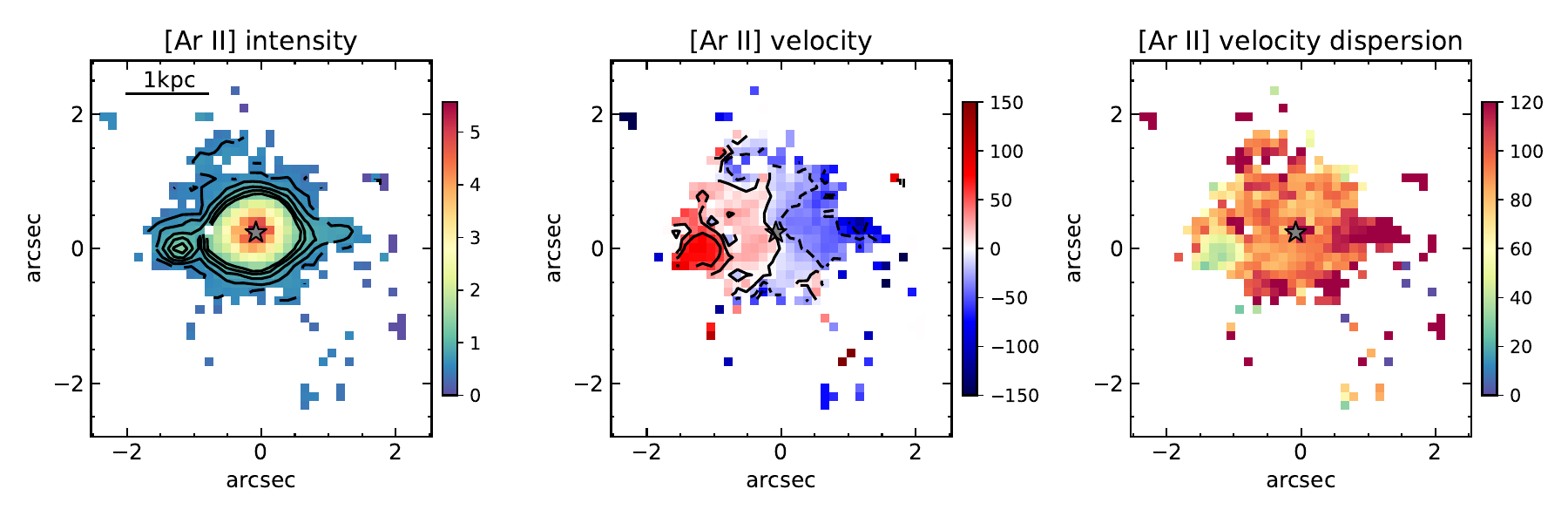}
  \vspace{-0.75cm}
     \caption{Maps of  [Ar\,{\sc ii}] at $\lambda_{\rm
         rest} = 6.99\,\mu$m. The maps were constructed by fitting a
       single Gaussian function and a
       local continuum to the emission line on a spaxel-by-spaxel basis.
       The panels show  the intensity and contours in units of
       $\sqrt{10^{-16}}$$\,{\rm erg\,cm}^{-2}\,{\rm s}^{-1}$ ({\it
         left panel}),
       the mean-velocity field in units of km\,s$^{-1}$ ({\it middle panel}), 
       and the velocity dispersion $\sigma$ (corrected for instrumental
       resolution)
       in units of km\,s$^{-1}$
       ({\it right panel}). The isovelocity contours are in a linear
       scale (solid lines positive values and dashed lines negative
       values). The star symbol marks the peak of the continuum 
       adjacent to the line, that is, the AGN position. The
       0,0 point on the axes refers to the center of this sub-channel
       array, after rotation. In 
       all maps, north is up and east to the left.}
     \label{fig:ArIImaps}
\end{figure*}     

\begin{figure*}
  \includegraphics[width=19cm]{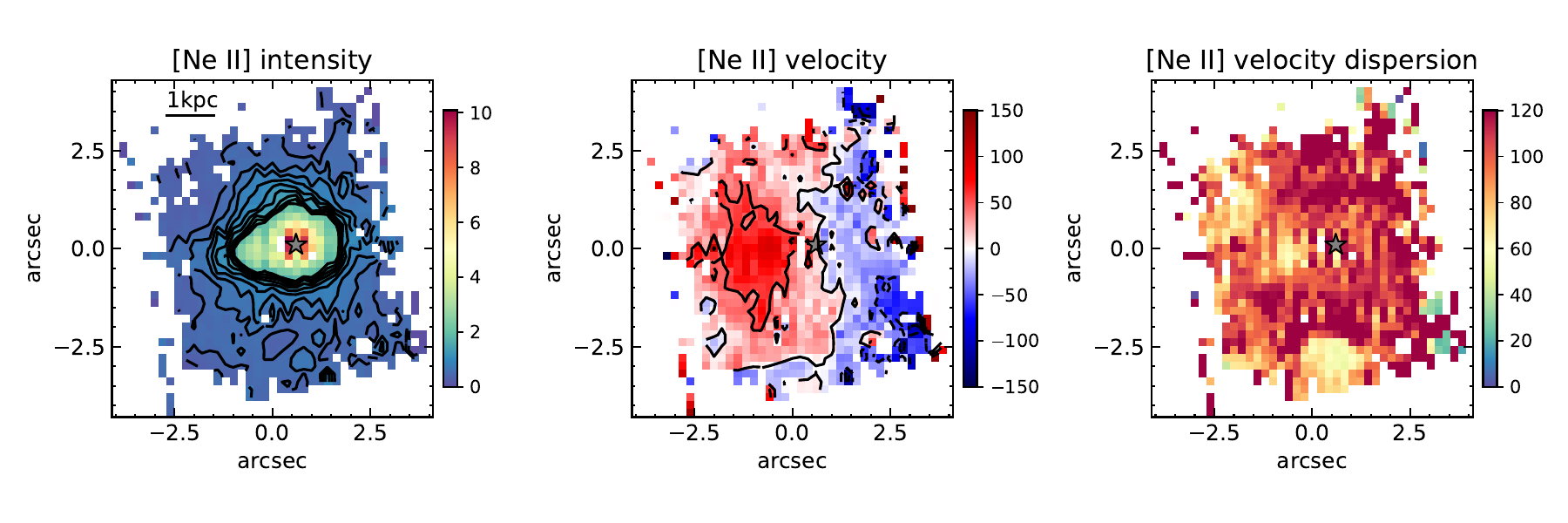}
  \vspace{-0.75cm}
     \caption{Maps of  [Ne\,{\sc ii}]  at $\lambda_{\rm
         rest} = 12.81\,\mu$m. Panels and 
       symbol are as in
       Fig.~\ref{fig:ArIImaps}. The
       0,0 point on the axes refers to the center of this sub-channel
       array, after rotation. In 
       all maps, north is up and east to the left.}
     \label{fig:NeIImaps}
\end{figure*}

\begin{figure}
\hspace{-1cm}
  \includegraphics[width=9cm]{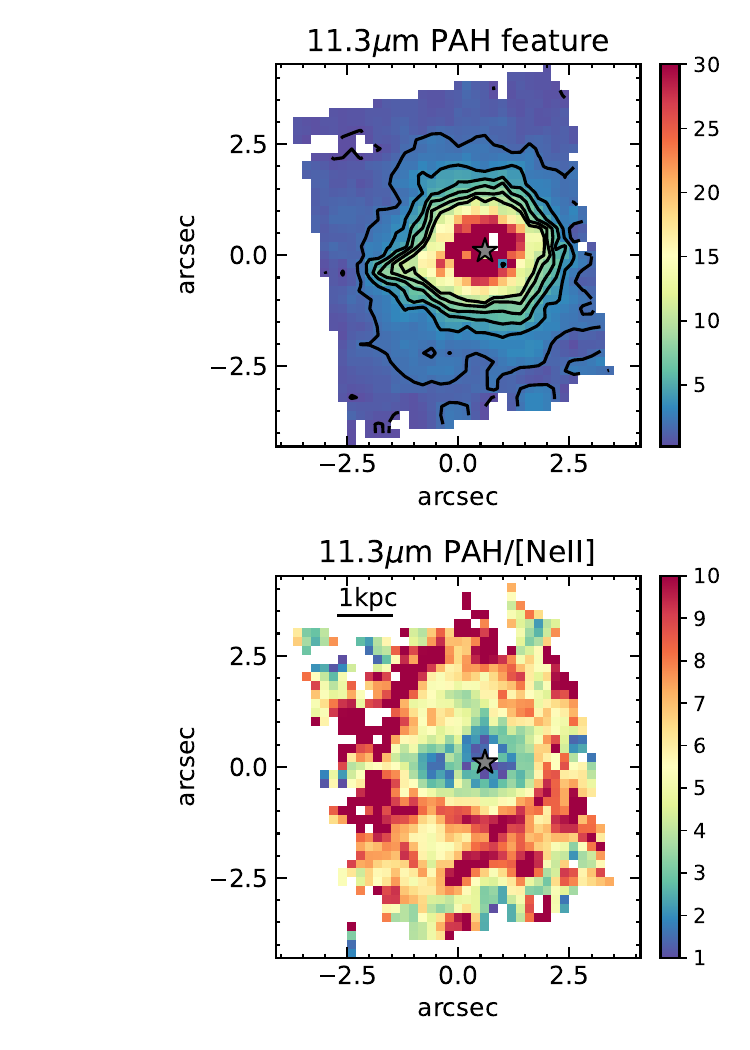}
\vspace{-0.5cm}
  \caption{Maps of the intensity of the
       $11.3\,\mu$m PAH feature ({\it top}) in units of
         $10^{-16}\,{\rm erg\,cm}^{-2}\,{\rm s}^{-1}$ and the $11.3\,\mu$m PAH to
       [Ne\,{\sc ii}] line ratio ({\it bottom}). The FoV and symbols are as in 
       Fig.~\ref{fig:NeIImaps}. The maps and contours are in a linear
       scale. }
     \label{fig:11.3micronPAHmap}
\end{figure}

Another remarkable finding is the large number of 
 H$_2$O  rovibrational absorption 
lines in the $5.5-7.2\,\mu$m range
(Fig.~\ref{fig:nuclearspectrum_channels}). These are associated with
the gas phase of this molecule \citep[][]{GonzalezAlfonso2024,
  GarciaBernete2024WATER}. Moreover, the H$_2$O features of Mrk~231  
appear to be similar to those of the embedded star-forming
regions in the LIRG system VV~114 \citep[][]{Buiten2024}.
A strong mid-IR continuum source behind the absorbing gas is required
to give rise to these features 
\citep{GonzalezAlfonso2024}, which in the case of Mrk~231 is produced
by dust heated by the 
AGN.
Using  {\it Herschel} observations
\cite{GonzalezAlfonso2010}  detected  far-IR water
vapour features in absorption and emission and
estimated the sizes  of the regions where these features are
formed to be at radii between 120
and 600\,pc from the Mrk~231 nucleus. The  latter  basically
corresponds to the extent of the nuclear starburst 
(see Sect.~\ref{sec:ionizedgas}). 
 
The nuclear spectrum of Mrk~231 also shows evidence of a broader and relatively shallow H$_2$O
feature at $\sim 6\,\mu$m, which is believed to be due to the presence of dirty
water ices in obscured
AGN and embedded nuclear regions of local ULIRGs
\citep[see][and references therein]{GarciaBernete2024}. For the measured nuclear
X-ray column density  
of Mrk~231 of $1.1 \times 10^{23}\,{\rm cm}^{-2}$ \citep{Teng2014},
according to \cite{GarciaBernete2024},
the expected apparent optical depth of 
this ice feature would be small, which is consistent with the
observations. A detailed modeling of this and other absorption features
will be carried out in a future paper.

\section{Extended ionized gas and PAH emission}\label{sec:ionizedgas}
In this section we discuss the properties of the extended ionized gas
and PAH
emission in Mrk~231. Figures~\ref{fig:ArIImaps} and
\ref{fig:NeIImaps}  present maps  of the line flux of the brightest
fine-structure lines in Mrk~231 ([Ar\,{\sc
  ii}] and [Ne\,{\sc ii}]
respectively) together with the  
mean-velocity field and velocity dispersion (left, middle, and right
panels). In Figs.~\ref{fig:NeIIImaps} and \ref{fig:SIIImaps}, we show similar 
maps for  [Fe\,{\sc ii}],  [Ne\,{\sc iii}], and [S\,{\sc 
  iii}].

\subsection{Line flux maps}\label{subsec:linefluxmaps}
The intensity maps of the
low-excitation emission lines exhibit not 
only bright
emission arising in the nuclear region but also an extended emission
component which includes 
several circumnuclear star-forming regions. 
This is in contrast to the bright point source present in the MRS continuum maps (see
Fig.~\ref{fig:continuummaps}). The location of a known star-forming
region at  approximately 1.1\arcsec \, (920\,pc) southeast (SE) of the AGN
is detected in the [Ne\,{\sc ii}] and [Ne\,{\sc iii}]
line maps (Figs.~\ref{fig:NeIImaps} and \ref{fig:NeIIImaps}),  and is
easily identified in the higher angular
resolution map of [Ar\,{\sc ii}] (Fig.~\ref{fig:ArIImaps}).
The larger FoV of the [Ne\,{\sc ii}]
maps also shows emission from another star-forming region located to
the northeast (approximately $1\arcsec$) of the AGN as well as some
regions at $\simeq 2.5-3\arcsec$ (2.1\,kpc)  to the south (see
Fig.~\ref{fig:NeIIsketch}).   These star-forming regions
were previously detected in Mrk~231 using optical IFU
observations \citep{Lipari2009, Rupke2011}. They are
related to the expanding shells S3 (at an approximate radial
distance from the AGN of 1.2\arcsec) and S1 (at 3.5\arcsec \, from the AGN), respectively
 \citep[see][and Fig.~\ref{fig:HSTimage}]{Lipari2005, Lipari2009} and
 associated bright optical knots 
detected with  HST \citep[][and also
Fig.~\ref{fig:HSTimage}]{Surace1998}.
We note that other works \citep[see 
e.g.,][]{Surace1998, Rupke2005, Rupke2011} described shell S1 as 
  southern arc of star-forming 
  regions and/or knots.

Interestingly, some of the bright 
circumnuclear star-forming regions in the MRS maps are
identified by their relatively 
low values of the velocity dispersion (Figs.~\ref{fig:ArIImaps} and
\ref{fig:NeIImaps}, right panels). As an example, the [Ar\,{\sc
  ii}] line from the region SE of the AGN has $\sigma \simeq
40\,{\rm km\,s}^{-1}$ (corrected for instrumental resolution). Additionally, the [Ne\,{\sc 
  ii}] intensity map shows more extended and diffuse emission\footnote{This is because [Ne\,{\sc ii}] is  brighter than
  [Ar\,{\sc ii}] and the ch3 pixels are larger than those of ch1, thus
  making it easier to detect extended diffuse emission in
  [Ne\,{\sc ii}].}, which is associated with
the overall interstellar medium of the galaxy, while the other more compact 
arisings in the star forming regions  (see
Sect.~\ref{subsec:extendedoutflows}) . 

From the [Ar\,{\sc ii}] and [Ne\,{\sc ii}] line flux maps, we measured
the size of the nuclear region responsible for the low-excitation line
emission.  From spatial profiles extracted with 1\arcsec, we
obtained nuclear sizes of 
$0.6-0.5\arcsec$ (FWHM, $502-418\,$pc) along the east-west (E-W) and
north-south (N-S) directions, respectively, for [Ar\,{\sc
  ii}] and slightly larger for [Ne\,{\sc ii}], $0.8-0.7\arcsec$
($660-586\,$pc, FWHM).  For the latter line we created an
  intensity map
  by integrating the line from a continuum-subtracted data cube. This is because
  the [Ne\,{\sc ii}] map created by fitting a single Gaussian
  (Fig.~\ref{fig:NeIImaps}, left) contains some NaN in the
  central region which hamper the fit of the spatial cuts. For comparison, the
derived sizes  from spatial profiles taken from MRS  continuum
  maps adjacent to
the lines are $0.4-0.3\arcsec$ along the E-W
and N-S directions
in ch1C and 0.6\arcsec \, for  ch3A.
This indicates that a
large fraction of the observed nuclear [Ar\,{\sc
  ii}] and [Ne\,{\sc ii}] emission is resolved.

It is likely that a contribution from unresolved emission is also
present in the low excitation emission lines.
The velocity channel maps of [Ne\,{\sc ii}] near the systemic velocity
(Fig.~\ref{fig:channelmaps}) show faint emission 
from the diffraction spikes 
associated with the unresolved source. Thus, a better estimate of the 
line emitting region can be obtained by subtracting in quadrature the
 FWHM of the core of the point spread
  function measured from the continuum images from the observed value. The
resulting  sizes for the [Ar\,{\sc ii}] and [Ne\,{\sc ii}]
  emissions are 
approximately 
$0.4-0.5\,$\arcsec$=335-420\,$pc (FWHM), respectively,  for
the nuclear line emission in both  the E-W and N-S
  directions. These sizes are 
   within the
range observed in other local ULIRGs using  
ALMA sub-millimeter continuum observations
\citep{PereiraSantaella2022} and similar to the values measured for the
  nuclear starburst of Mrk~231 with optical, near-IR, and radio
  observations \citep{Lipari2009, Davies2004, Carilli1998}. Therefore,
both the mid-IR sizes and observed
[Ne\,{\sc iii}]/[Ne\,{\sc ii}] ratio indicate that the MRS
observations have resolved  the nuclear starburst of Mrk~231.


\subsection{Extended PAH emission}

In galaxies, PAH molecules are excited by ultraviolet photons produced
by B stars rather than the more massive O stars 
\citep[see][]{Peeters2004}.  Thus, PAH emission is often considered a
tracer of the recent SF activity, while [Ne\,{\sc ii}] probes
youngest ages (a few million years).
We generated a map of the $11.3\,\mu$m PAH intensity
by integrating, on a spaxel-by-spaxel basis, the observed 
feature flux between $11.6\,\mu$m and $11.95\,\mu$m ($\lambda_{\rm
  rest}=11.13\,\mu$m and $\lambda_{\rm rest}=11.47\,\mu$m) and
subtracting a local continuum measured on both sides of the
feature. We note that for Mrk~231
this feature lies close to the edge of the MRS ch3A (see bottom left
panel of Fig.~\ref{fig:nuclearspectrum_channels}) and thus the local blue
continuum might be slightly overestimated. The resulting map
(Fig.~\ref{fig:11.3micronPAHmap}, top panel) exhibits PAH emission over the entire
ch3A FoV, with the brightest arising in the nuclear region. The
detection of the $11.3\,\mu$m PAH feature in the central few hundred
parsecs of Mrk~231 is in line 
with observations of the nuclear regions of other AGN with comparable X-ray
luminosities \citep[][]{GarciaBernete2022, GarciaBernete2024PAHs}. These
works showed that while
the  presence of an AGN can impact the properties of the PAH
molecules, it does not necessarily destroy them completely.

Although the overall morphology of the $11.3\,\mu$m PAH feature
emission is similar to  [Ne\,{\sc ii}] (Fig.~\ref{fig:NeIImaps}), some
differences are apparent in the $11.3\,\mu$m PAH
over [Ne\,{\sc ii}] line ratio 
map (Fig.~\ref{fig:11.3micronPAHmap}, bottom). The nuclear region and some of the
circumnuclear star-forming regions show
distinctly lower ratios than in the rest of the
extended emission. This behavior was observed in {\it Spitzer}/IRS
spatially resolved line ratio
maps of local LIRGs \citep{AlonsoHerrero2009, PereiraSantaella2010mapping}. While the
decreased $11.3\,\mu$m PAH/[Ne\,{\sc ii}] ratio in the nuclear
region of Mrk~231 and LIRGs might be in
part due to extinction due to silicate grains affecting the 
$11.3\,\mu$m PAH feature 
\citep{HernanCaballero2020} and/or the AGN effects, they
also reflect the fact that   
PAHs trace less massive and slightly older stellar populations than
other mid-IR low excitation 
emission lines \citep{Peeters2004}. Nevertheless, the
extended nature of the PAH and [Ne\,{\sc ii}] emissions reinforces the
conclusion that most of the mid-IR low excitation line
emission in Mrk~231 is produced by SF.

\subsection{Ionized gas kinematics}\label{subsec:kinematics_ionized}

The mean-velocity field of the [Ne\,{\sc ii}] line (middle
panel of Fig.~\ref{fig:NeIImaps}) shows evidence of
circular motions, where the velocities vary from approximately
$-100\,{\rm km\,s}^{-1}$ to $+80\,{\rm km\,s}^{-1}$. The isovelocity contours show
that the kinematic major axis of the galaxy lies close to the E-W
direction, as found by the kinematics of optical emission lines
\citep{Lipari2009} and the cold molecular gas \citep{DownesSolomon1998,
  Feruglio2015}.  Moreover, the [Ne\,{\sc ii}]  velocity field is similar to 
that derived from  the
optical H$\alpha$ line \citep{Lipari2009}. This includes the presence of
blueshifted motions to the  south of the AGN at the
location of star-forming regions associated with  shell S1 (see
Figs.~\ref{fig:HSTimage} and \ref{fig:NeIIsketch}), where clearing of
material and the possible  rupture of a super-bubble are taking place
\citep{Lipari2009}.

The higher angular resolution and spectral resolution view of the nuclear region
provided by the [Ar\,{\sc ii}] line (middle panel of
Fig.~\ref{fig:ArIImaps}) shows a rotational pattern 
as well as non-circular motions in the central 2\arcsec
along the kinematic minor axis (N-S direction). These
deviations from circular motions \citep[see the expected velocity
field for the rotating disk of Mrk~231 in the top left panel of
Fig.~ 5 in][]{Feruglio2015} are 
more pronounced in the warm molecular gas (see 
Sect.~\ref{subsec:kinematics_molecular}) as well as in the cold molecular gas
\citep{Feruglio2015}. There are also some deviations from pure
rotation in the [Ar\,{\sc ii}] and [Ne\,{\sc ii}] velocity fields at
the location
of the star-forming region $\simeq 1.1\arcsec$ SE of the AGN, which
were  also observed in H$\alpha$ \citep{Lipari2009, Rupke2011}. This
might be related to the location of this star-forming region in the
shell S3 \citep{Lipari2009}.

 \begin{figure*}
  \includegraphics[width=19cm]{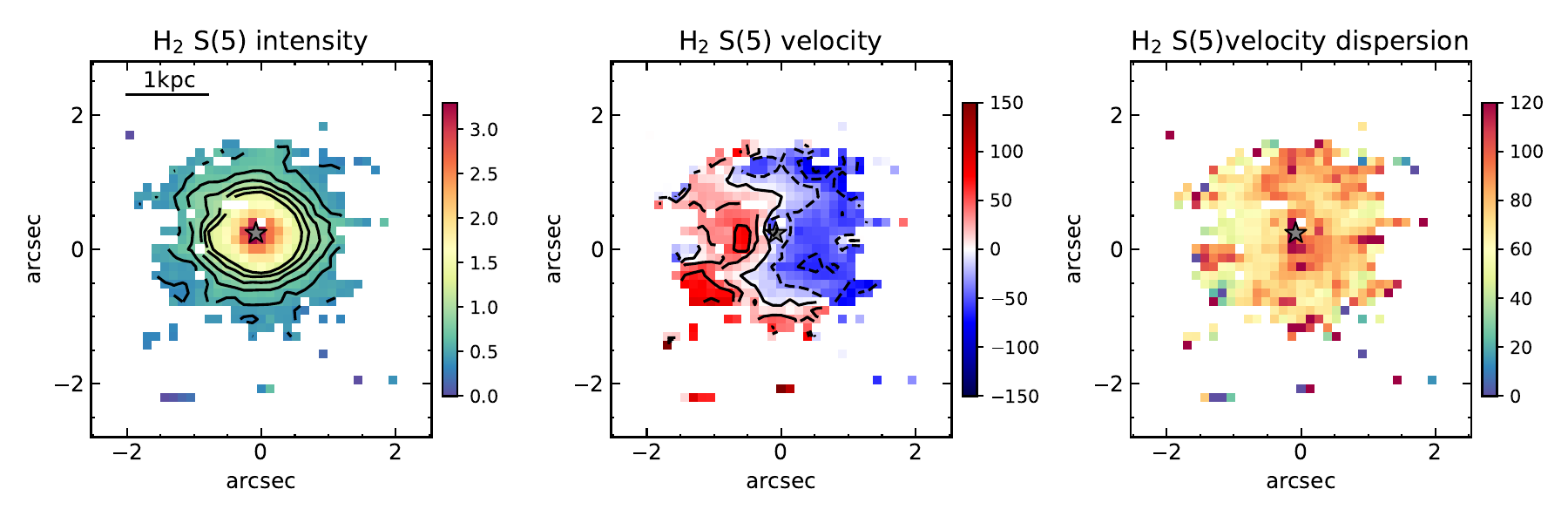}
  \vspace{-0.75cm}
     \caption{Maps of  H$_2$ S(5)  at $\lambda_{\rm
         rest} = 6.91\,\mu$m. Panels and 
       symbol are as in
       Fig.~\ref{fig:ArIImaps}. The
       0,0 point on the axes refers to the center of this sub-channel
       array, after rotation. In 
       all maps, north is up and east to the left.}
     \label{fig:H2S5maps}
\end{figure*}     

\begin{figure*}
  \includegraphics[width=19cm]{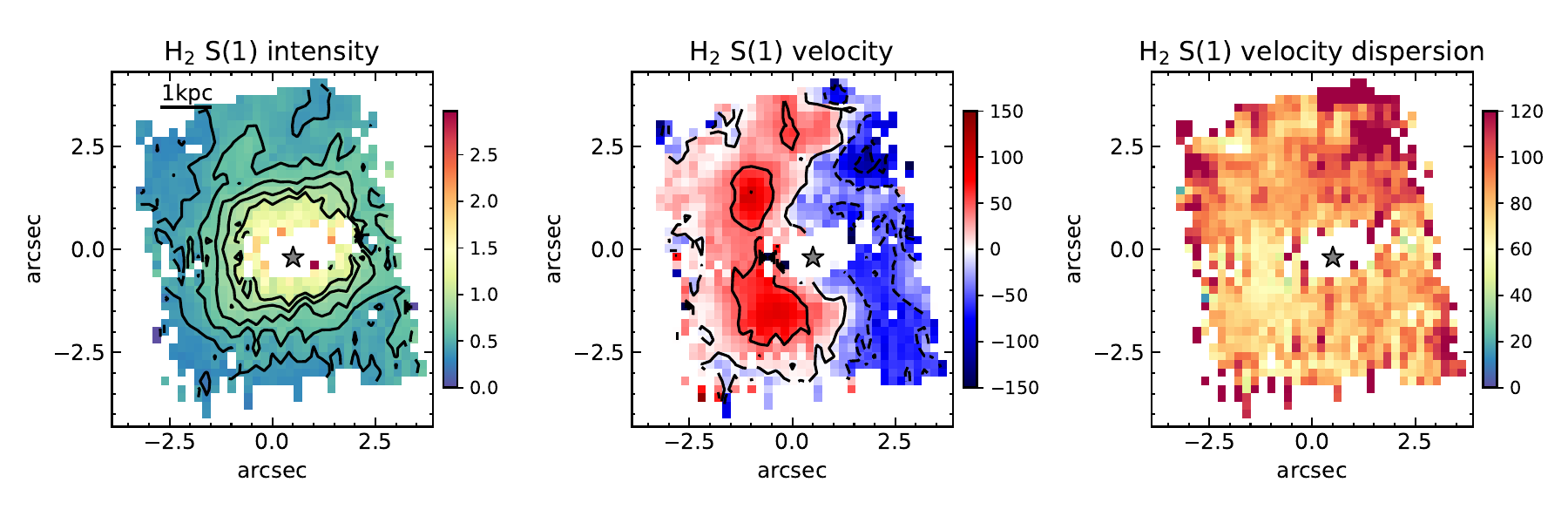}
  \vspace{-0.75cm}
     \caption{Maps of H$_2$ S(1)  at $\lambda_{\rm
         rest} =17.03\,\mu$m observed in ch3. Panels and 
       symbol are as in
       Fig.~\ref{fig:ArIImaps}. The
       0,0 point on the axes refers to the center of this sub-channel
       array, after rotation. In 
       all maps, north is up and east to the left.}
     \label{fig:H2S1maps}
\end{figure*}

The velocity dispersion maps of [Ar\,{\sc ii}] and [Ne\,{\sc ii}]
(right panels of Figs.~\ref{fig:ArIImaps} and \ref{fig:NeIImaps},
respectively) display moderate values ($\sigma\sim 60-120\,{\rm
  km\,s}^{-1}$)
These are higher than the values measured in the star-forming regions
discussed above, which have  $\sigma\sim 30-40\,{\rm km\,s}^{-1}$ for
the [Ar\,{\sc ii}] line or 
FWHM$_{\rm line}$ $\sim 71-94\,{\rm km\,s}^{-1}$. These are similar to those
measured  from
H$\alpha$ \citep{Rupke2011}. Although the lines  are relatively narrow, the
star-forming regions to the south might be located in
areas of Mrk~231 affected by starburst-driven winds, as we shall see in 
Sect.~\ref{sec:outflows}.

\section{Molecular hydrogen emission}\label{sec:molecularhydrogen}

In the nuclear region of Mrk~231 there is H$_2$ emission from
the S(1) to
the S(8) transitions (see Fig.~\ref{fig:nuclearspectrum}).
In this section we
discuss the morphology and kinematics of H$_2$ S(1) and S(5), 
compute the mass and temperature of the warm molecular gas observed in
the central region of Mrk~231. We note that the
S(1) transition is present in both the  ch3 and ch4 data cubes. In
Appendix~\ref{appendix:extramaps}, we show the ch3 S(2)  and  ch4 S(1) maps.

\subsection{Line flux maps}
The H$_2$ S(5) line displays clearly
extended emission in Mrk~231 (left panel of Fig.~\ref{fig:H2S5maps}), as found for the low-excitation emission
lines, with disk-like
morphology. The emission  confined to the approximately central 
$\simeq 2\arcsec$ region. It is resolved, with   sizes of $0.75-0.6\arcsec$ 
(FWHM $\simeq 0.6-0.5\,$kpc) as measured
  from spatial cuts along the E-W and N-S directions, respectively. These are slightly
larger than those of 
the  [Ar\,{\sc ii}] and [Ne\,{\sc ii}] emitting region. There is H$_2$ S(5) emission at
the location of the  star-forming region at 1.1\arcsec\, SE of
the AGN (Fig.~\ref{fig:ArIImaps}), but it does not show a
distinct peak there. The H$_2$ S(5) morphology resembles
 that of the cold molecular gas disk traced with the CO(1-0) and CO(2-1)
transitions as measured with millimetre interferometry
\citep{Bryant1996, DownesSolomon1998, Feruglio2015}. 

The H$_2$ S(1) line emission (Fig.~\ref{fig:H2S1maps}) 
fills almost entirely the ch3 FoV. The apparent
deficit of emission at the 
central spaxels is due to the low contrast of the line against the strong
continuum. The larger spaxel size of the ch4 H$_2$ S(1) map
(Fig.~\ref{fig:H2S1ch4maps}) alleviates this problem showing that
there is centrally peaked 
emission (FWHM$\simeq 1\,$kpc) at the AGN position, as well
as emission extending over  $\simeq 9\arcsec \times 9\arcsec$ 
(projected
size of $7.5\,{\rm  kpc} \times 7.5\,{\rm kpc}$). It is thus more
extended that the cold molecular gas disk
\citep[diameter of approximately
3\arcsec,][]{DownesSolomon1998}. However, it is likely that the large difference in
sizes derived for the cold and warm molecular gas is because the more 
diffuse and 
extended cold molecular gas emission was filtered in  the CO interferometric observations.
The H$_2$ S(1) morphology reveals additionally  a spiral-like shape to the
NE and SW, which appears to link with the central ($\sim
30\arcsec$) optical emission. The latter
shows tidal tails extending toward the north and south on larger scales 
\citep[$>1\arcmin$,  see Fig.~1 of][]{Sanders1988}. 



\begin{figure}

  \hspace{0.25cm}
  \includegraphics[width=8cm]{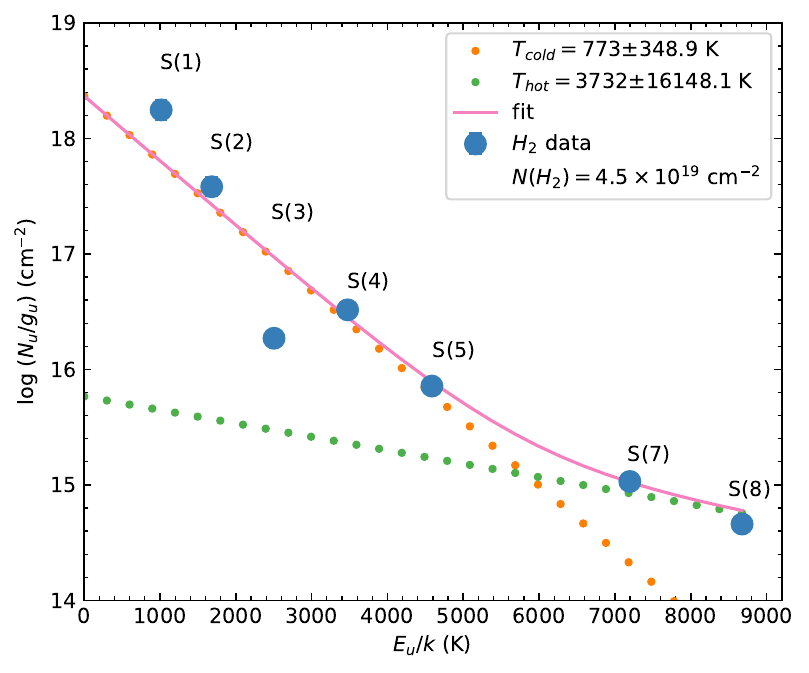}

  \hspace{0.25cm}
  \includegraphics[width=8cm]{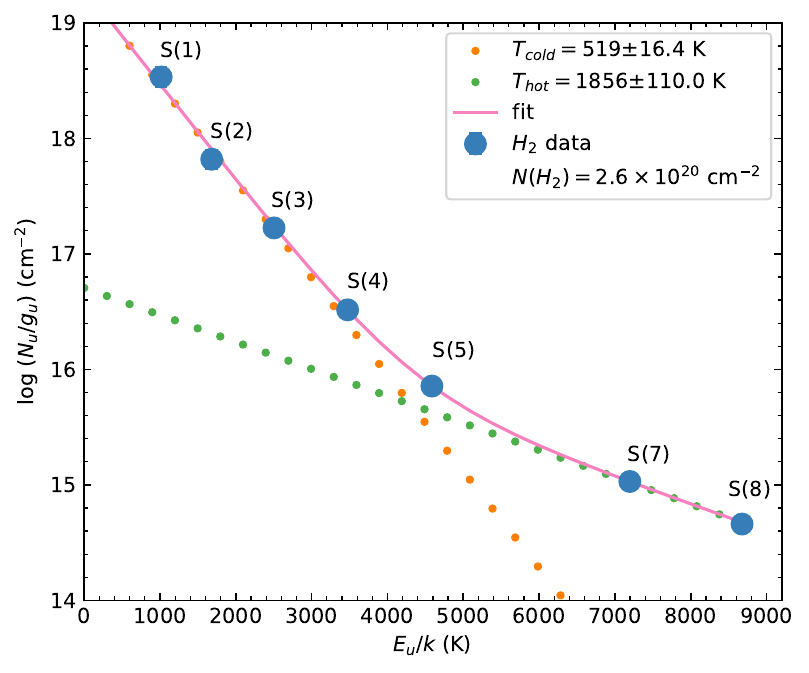}

  \caption{H$_2$ rotational diagrams for the nuclear region of
       Mrk~231. The fit was performed with a
       two-temperature model and assuming LTE conditions (see text for
       details). {\it Top}: using observed values and {\it bottom}: 
       using the S(3), S(2), and S(1) fluxes corrected for
       extinction (see text for details). }  
     \label{fig:H2diagram}
\end{figure}

\subsection{Warm molecular gas kinematics}\label{subsec:kinematics_molecular}
The H$_2$ S(5) mean-velocity field (middle panel of Fig.~\ref{fig:H2S5maps}) shows a clear
rotational signature consistent with a disk with a major axis oriented
approximately in the E-W direction. The observed velocities are 
in the $-80\,{\rm km\,s}^{-1}$ and $+100\,{\rm km\,s}^{-1}$ range. 
  There are, however, some 
  deviations from circular motions along the kinematic minor axis with
  a marked inverted S-shaped pattern. The H$_2$ S(5) velocity dispersions measured
  in the nuclear region are relatively
  low, $\sigma\sim 65\,{\rm km\,s}^{-1}$ (corrected for
instrumental resolution), except
along the kinematic minor axis mostly, 
where they reach slightly higher values. On larger scales, the H$_2$ S(1) line
present several regions with somewhat  higher velocity dispersions ($\sigma
> 85\,{\rm km\,s}^{-1}$, right
panel of Fig.~\ref{fig:H2S1maps}). The
  H$_2$ S(5) kinematics are remarkably similar to those of the 
interferometric CO(2-1) observations 
 obtained at a
$\simeq 0.5\arcsec$ resolution \citep[compare with Fig.~4 of][but note
their smaller FoV]{Feruglio2015}. They
 modeled their observations with an inner disk of $r=0.12\arcsec$
 radius,  which
 is warped
 with respect to the outer ring of radius
$r=0.24\arcsec$ with an E-W orientation and an inclination of
36$^{\rm o}$.

The kinematics of the H$_2$ S(1) and (2) lines (Figs.~\ref{fig:H2S1maps}, \ref{fig:H2S2maps}, and
Fig.~\ref{fig:H2S1ch4maps}) connect well  with those seen in the inner
region traced by the S(5) 
line. The mean-velocity fields show qualitatively the
same general E-W rotation pattern but with some 
interesting features indicating the presence of non-circular
motions. There are redshifted features to the NE and SE, and
blueshifted to the NW and SW. These are at projected
radial distances from the AGN of approximately 2\arcsec$\simeq$1.5\,kpc
and might be related to expanding shells S2 (at an approximate
radial distance from the AGN of 1.8\arcsec) and/or S3 (see
Fig.~\ref{fig:HSTimage}), where the ionized gas shows outflow
velocities  between $-500\,{\rm km\,s}^{-1}$ and $-230\,{\rm
  km\,s}^{-1}$ \citep{Lipari2005}. The enhanced H$_2$ velocity
dispersions observed in some of these regions could be the result of
the  interaction between the ionized
gas outflow and molecular gas in the galaxy. We analyse the
molecular warm outflows in Sect.~\ref{sec:outflows}. 

\subsection{Warm molecular gas mass and temperature}\label{subsec:H2rotdiagrams}

In this section we use the suite of 
 H$_2$ pure rotational transitions (Table~\ref{tab:linefluxes})
 detected in the nuclear region of Mrk~231
to estimate the warm molecular gas mass and temperatures. We note that
the H$_2$ ground S(0) transition is outside the ch4 spectral range,
and thus we can only obtain a lower limit for the mass. For this
analysis, we used the {\sc pdrtpy} routine 
\citep{Pound2023}, which fits a
two-temperature model. We 
assumed local thermal equilibrium (LTE) conditions and a constant ortho-to-para
ratio of 3 \citep[see][for more
details]{AlvarezMarquez2023}. 

Using the observed H$_2$ fluxes
the 
temperatures are not well constrained, as can be seen from the top
panel of Fig.~\ref{fig:H2diagram}. Both the S(3) and S(1)
transitions are not well fitted with this model because they
are located   within the
troughs of  the  $9.7\,\mu$m and $18\,\mu$m silicate feature absorptions,
respectively (see
 Fig.~\ref{fig:nuclearspectrum}).
We  tried values of the  extinction around $\tau_{9.7\mu{\rm m}}=2$, which are close to
the extinctions derived by \cite{Veilleux2009} and \cite{LopezRodriguez2017}
(see also Sect.~\ref{subsec:CON}), and the regular Milky Way
extinction law \citep[see][]{HernanCaballero2020} to derive
extinction-corrected S(1),  S(2),
and S(3) line fluxes.  A value of $\tau_{9.7\mu{\rm m}}=2.2$
  provided the smallest relative errors in the derived temperatures. As can be seen from
Fig.~\ref{fig:H2diagram} (bottom panel), the resulting fit to the
excitation diagram is smoother. We note that this was not achieved if
we corrected the fluxes with the lower extinction that would be
derived from the apparent optical depths of the
silicate features  (Sect.~\ref{sec:nuclearemission}). Moreover,
compared with the diagram without an extinction correction, the
temperatures are constrained significantly better, $T_{\rm 1} = 519\pm16\,$K and $T_{\rm  2} = 1856
\pm 110\,$K. For this fit, the derived total column density is higher,
$N_{\rm H2} = 2.6\times 
10^{20}\,{\rm cm}^{-2}$. We note that in order to get a better determination of the
extinction affecting the different emitting components in the nuclear region of Mrk~231, a
more detailed modeling, including 
NIRSpec observations, would be necessary \cite[see,
e.g.,][]{Donnan2024}.  

The corresponding warm (for the two temperatures fitted above)  H$_2$
mass is $M_{\rm H2}({\rm warm})=9\times
10^{6}\,{\rm M}_\odot$ for $r=1\arcsec$, from the fit with the
extinction-corrected fluxes.
Using {\it Spitzer}/IRS detected H$_2$ transitions measured over 
larger physical sizes, \cite{Petric2018} 
estimated masses of warm molecular gas of $\simeq 1.7\times 10^8\,{\rm M}_\odot$ for
the $T=106\,$K component and $\simeq 3.1\times 10^6\,{\rm M}_\odot$ for
that at $T=343\,$K.  Since {\it Spitzer} did not detect the S(4) to
S(8) transitions, it is not surprising that these estimates did not
include the higher temperature component detected in this work.
Also, as expected, the MRS-derived mass is intermediate between the masses of  
hot molecular gas of $2.5\times 10^4\,{\rm M}_\odot$  \citep[][for
our assumed distance]{Krabbe1997} derived from near-IR H$_2$ lines and
that of 
the cold molecular gas of $2.3\times
10^9\,{\rm M}_\odot$ 
\citep{DownesSolomon1998} derived from millimeter CO
transitions.

\begin{figure*}
\hspace{1cm}
  \includegraphics[width=16cm]{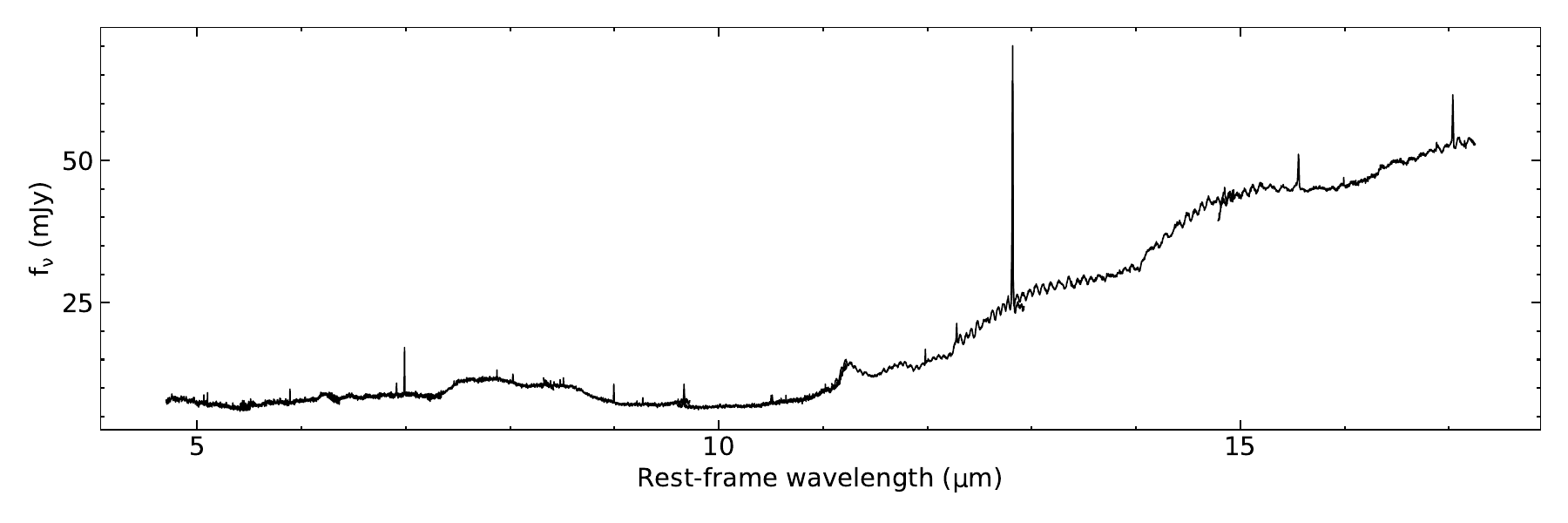}

\hspace{1cm}
\includegraphics[width=16.25cm]{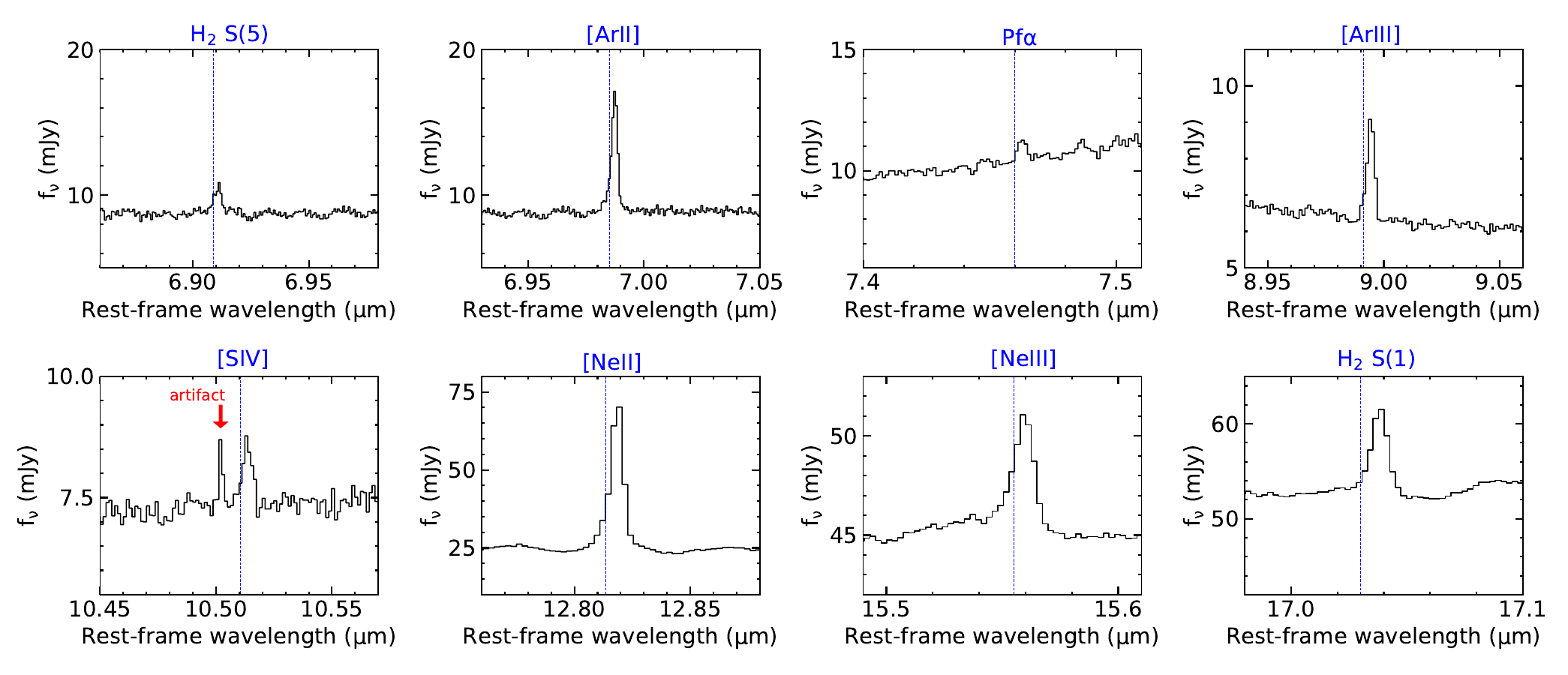}
\caption{MRS spectra of the star-forming region
       $1.1\arcsec$ SE of the
       AGN. {\it Top panel}: ch1 to ch3 spectra extracted
       with a $0.3\arcsec$ radius.
We applied  the residual fringing correction to the 1D spectra and
scaling factors to stitch together the spectra of the different
sub-channels. {\it Middle and
       bottom panels}: zoomed-in spectral regions of selected emission
     lines. In all panels, the
       rest-frame wavelengths were 
       computed with the redshift
       quoted in Sect.~\ref{sec:introduction} and appear redshifted with respect to the
systemic velocity (see mean-velocity maps of [Ar\,{\sc ii}] and [Ne\,{\sc
       ii}], middle panels of Fig.~\ref{fig:ArIImaps} and
     \ref{fig:NeIImaps}), due to rotation. }
     \label{fig:HIIregionspectrum}
\end{figure*}


\section{Star formation activity}\label{sec:SFactivity}

\subsection{An obscured nuclear starburst}\label{subsec:CON}
Several works in the literature inferred high values of the
extinction in Mrk~231 from the modeling of 
the observed nuclear and integrated spectral energy distributions 
\citep[i.e., $\tau_{{\rm 9.7}\mu{\rm m}}=2.23$, $A_V=36\pm 5\,$mag, and 
$\tau_V=162^{+52}_{-34}$,][respectively]{Veilleux2009, LopezRodriguez2017, Efstathiou2022}. 
Indeed, the  H$_2$ rotational diagram of the nuclear region derived
with our MRS observations 
revealed that the S(1) and S(3) lines are  affected by extinction
(Sect.~\ref{subsec:H2rotdiagrams}). In this section we
discuss additional pieces of evidence to support this. 

The comparison between the star
formation rate (SFR) derived from $L_{\rm IR}$ and some other
mid-IR indicators can 
provide extra evidence for the presence of high levels of
extinction. Since  there are no detections of 
mid-IR hydrogen
recombination lines in the nuclear region of Mrk~231, we can employ
the [Ne\,{\sc ii}] and [Ne\,{\sc iii}] lines  to estimate the
SFR. With 
the calibration from \cite{Zhuang2019}, which takes a Salpeter
initial mass function, and assuming solar metallicity, we obtained  a
SFR$\simeq 23\,{\rm
  M}_\odot\,{\rm yr}^{-1}$ for the central 1.6\,kpc. Integrating the
neon lines over the entire ch3 
FoV, we derived SFR$\simeq 40\,{\rm
  M}_\odot\,{\rm yr}^{-1}$. We note that if the metallicity of Mrk~231 is above solar, these
estimates should be taken as upper limits. The neon-based SFR value is 
below several literature estimates \citep{Carilli1998, Davies2004} and
in particular that 
of $\simeq 140\,{\rm M}_\odot\,{\rm
  yr}^{-1}$ from $L_{\rm IR}$, once the AGN bolometric contribution
\citep[$\simeq$70\%,][]{Veilleux2009, Veilleux2020} is subtracted.
This discrepancy thus could be resolved with a high value of the
nuclear extinction.

The PAH emission can also provide additional clues considering that
star-forming  galaxies show relatively constant ratios
between the  
EW of the 6.2, 11.3, and $12.7\,\mu$m PAH features. 
Using these, \cite{GarciaBernete2022CON} proposed a new method to identify 
deeply obscured nuclei since a decreased
mid-IR continuum within the $9.7\,\mu$m silicate feature boosts
the $11.3\,\mu$m PAH EW. This in turn produces 
lower EW(6.2$\mu$m PAH)/EW(11.3$\mu$m PAH) and 
EW(12.7$\mu$m PAH)/EW(11.3$\mu$m PAH) ratios. The main assumption here
is that PAH emission is
mostly produced in an extended component compared with the continuum
source, as is the case in Mrk~231.

We measured nuclear EWs of the 6.2$\mu$m, 11.3$\mu$m, and 12.7$\mu$m PAHs 
$0.007\pm 0.001\,\mu$m,  $0.012\pm 0.001\,\mu$m, and 
$0.0005\pm 0.0001\,\mu$m, respectively\footnote{The
uncertainties are driven by the continuum placement since the 6.2 and
$11.3\,\mu$m PAH features are close to the edge of their corresponding sub-channel
and the continuum red-wards the $12.7\,\mu$m feature is affected by the
[Ne\,{\sc ii}] line wing (see Sect.~\ref{subsec:outflows_nuclear}).}.
The observed EW ratios, within the uncertainties, place the nuclear
region of Mrk~231 within the  region 
defined in the EW(6.2$\mu$m
PAH)/EW(11.3$\mu$m PAH) versus
EW(12.7$\mu$m PAH)/EW(11.3$\mu$m PAH) diagram \citep[top panels of
Fig.~8 of][]{GarciaBernete2022CON} where deeply obscured nuclei are located. This is
consistent with the detection of the HCN–vib (3-2) transition and the
high nuclear column
density of $N_{\rm H}= 1.2 \times 10^{24}\,{\rm cm}^{-2}$ implied by
these observations, which would be responsible for obscuring the
  mid-IR  high-excitation emission lines \citep{Aalto2015}. 
  
The deep silicate absorption features that would be arising in the most
extinguished  part of the nuclear starburst are, at the  same time,
partly filled in by the AGN dust continuum 
emission. In other words, the combination of both emissions produces
the relatively low apparent
optical depths of the silicates observed in the nuclear MRS
spectrum (see Sect.~\ref{sec:nuclearemission}).  Therefore, the
extinctions derived from the apparent depths need to be taken as lower limits to
the true nuclear value.

\begin{figure*}
\vspace{-2.5cm}
     \includegraphics[width=5.75cm]{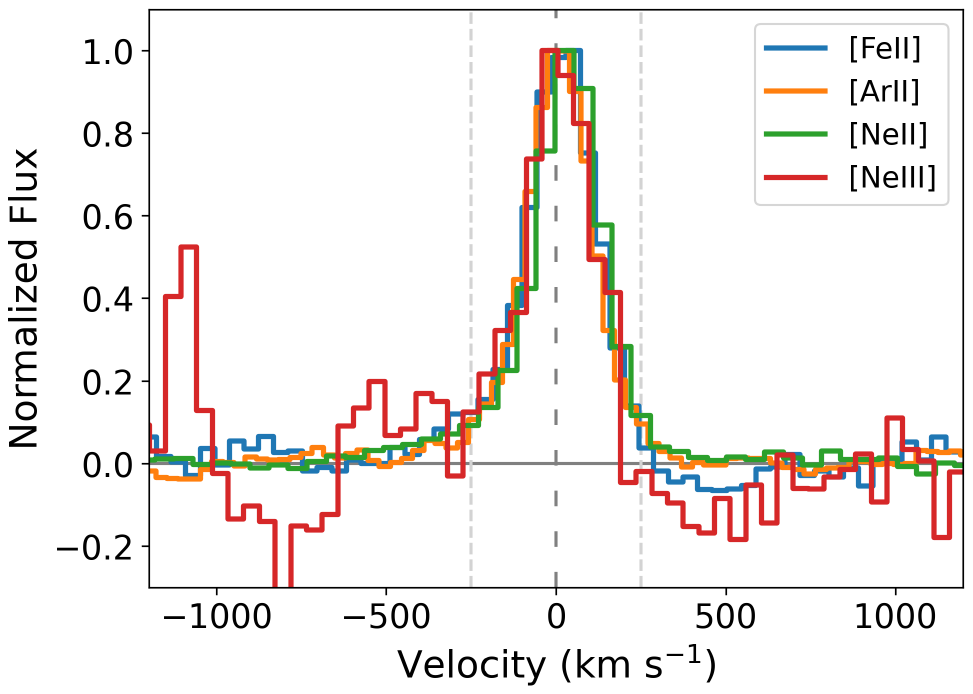}
     \includegraphics[width=5.75cm]{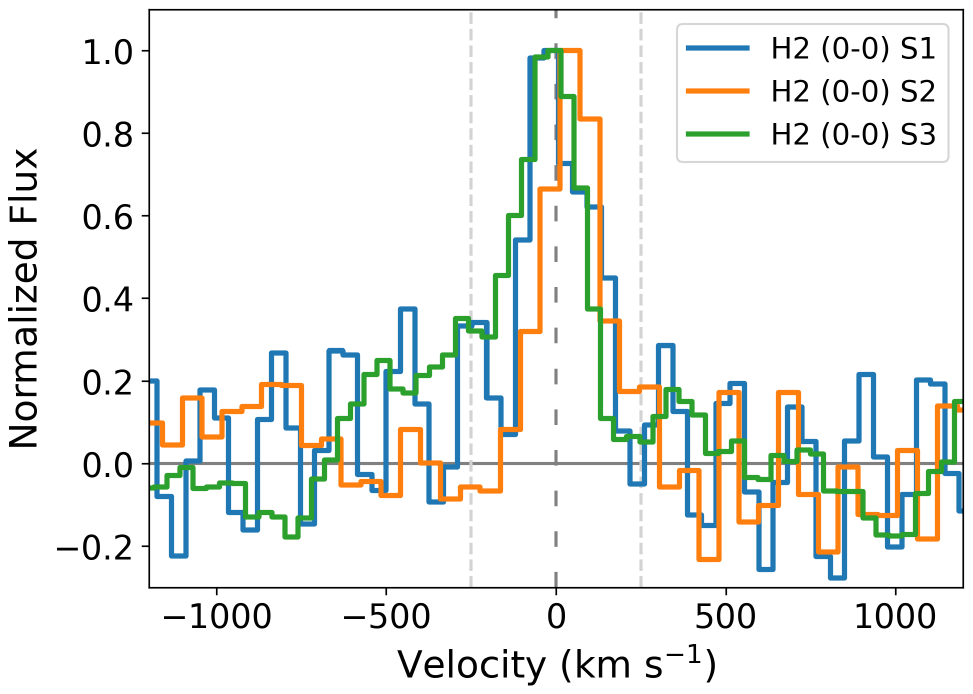}
     \includegraphics[width=5.75cm]{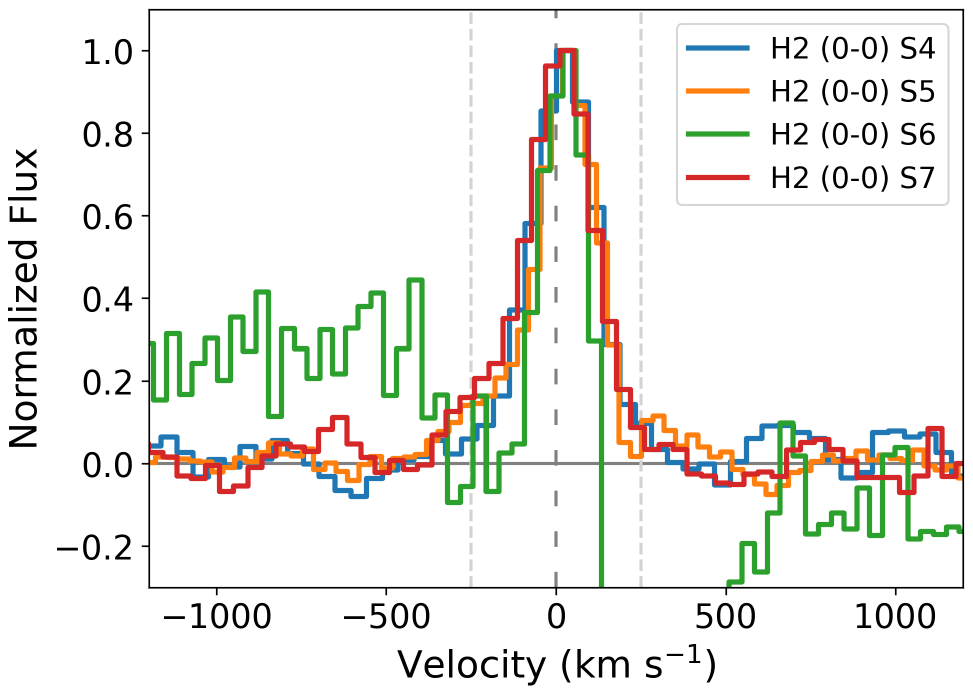}
\vspace{-2.5cm}
     \caption{Line profiles from the nuclear region ($r=1\arcsec$ aperture). {\it Left panel}: 
       fine-structure lines, and {\it middle} and {\it right panels}:
       H$_2$ 
      S(7) to S(4) and S(3) to S(1) transitions, respectively. The $0\,{\rm
        km\,s}^{-1}$ value corresponds to the 
      adopted $z=0.004217$. Note that the red part of the H$_2$ S(6)  
      line profile (green line, middle panel) is inside one of the
      H$_2$O molecular absorptions. We did not apply the 1D
       residual fringe correction.} 
     \label{fig:nuclearlineprofiles}
\end{figure*}

\subsection{Circumnuclear star-forming regions}\label{subsec:starformingregions}
The  
low-excitation fine-structure line maps have revealed the presence of
several circumnuclear star-forming regions in Mrk~231
(Sect.~\ref{sec:ionizedgas} and Fig.~\ref{fig:NeIIsketch}). That at $\simeq 1.1\arcsec$  
SE of the AGN was detected and resolved in the ch1, ch2, and ch3
line maps
(Figs.~\ref{fig:ArIImaps}, \ref{fig:NeIImaps}, and \ref{fig:NeIIImaps}). In
ch4A, this region
is partially resolved (see the [S\,{\sc iii}] maps in
Fig.~\ref{fig:SIIImaps}), since it is located slightly less than 
$2\times {\rm FWHM}_{\rm ch4A}$ from the AGN position.  We placed
a circular aperture of $r=0.3\arcsec$ on the ch1, ch2, and ch3 data
cubes to extract a 1D spectrum  
of this region. The size of the aperture encompasses well the
region at the shortest wavelengths but the long wavelength continuum
might be slightly contaminated by the bright AGN continuum
source.

In Fig.~\ref{fig:HIIregionspectrum} we show  the rest-frame
ch1 to ch3 spectra (top panel) and zoomed-in spectra (middle and
bottom panels) around several fine-structure emission lines and H$_2$  transitions. The
$15\,\mu$m continuum of this star-forming
region is more than a factor 50 fainter than the nuclear
region and the lines present larger EW.  Moreover, we 
detected two lines not identified in the nuclear spectrum,
namely [Ar\,{\sc iii}] 
     and  [S\,{\sc iv}], and tentatively the hydrogen recombination
     line H\,{\sc i}
     6-5 (Pf$\alpha$) at $\lambda_{\rm rest} = 7.46\,\mu$m. The
     central rest-frame wavelengths in these panels, which were calculated with the nominal
     redshift of $z=0.04217$, appear slightly redshifted as seen in the mean-velocity
maps (Figs.~\ref{fig:ArIImaps} and \ref{fig:NeIImaps}, middle
panels), due to rotation.

Using a single Gaussian function, we derived line ratios [Ne\,{\sc
  iii}]/[Ne\,{\sc ii}]$\simeq 0.14$ 
and   [Ar\,{\sc iii}]/[Ar\,{\sc ii}]$\simeq 0.22$. The latter is likely just
a lower limit since the [Ar\,{\sc iii}] line is inside the $9.7\,\mu$m
silicate feature. Nevertheless, these ratios can be explained by
photoionization models for relatively young ($5-6\,$Myr) stellar
populations formed in an instantaneous burst of solar metallicity with
a Salpeter initial mass function and an upper 
mass cutoff of 100\,M$_\odot$ \citep{Rigby2004}.

In the star-forming regions to the south of
the AGN, which are associated with the expanding shell S1, there is 
evidence of the presence of Wolf-Rayet
stars \citep{Lipari2009}, indicating a young SF episode.
We extracted a 1D spectrum from a region at 2.4\arcsec \, south
of the AGN with a radius of 0.8\arcsec. We measured [Ne\,{\sc
  iii}]/[Ne\,{\sc ii}]$=0.26$, which  is within 
the line ratios predicted for
this stellar phase, again for a solar
metallicity instantaneous burst \citep{Rigby2004}.

This brief analysis with the MRS observations reinforces the
conclusions from \cite{Lipari2009} about the young ages of some of the
circumnuclear star-forming regions in Mrk~ 231.

\begin{table}
  \center
  \caption{Non-parametric analysis of nuclear emission lines.}\label{tab:nonparametric} 
  \begin{tabular}{lcccccc}
    \hline
    Line & $v_{10}$ & $v_{90}$& $W_{80}$ & $v_{02}$ & $v_{\rm 98}$\\
           & \multicolumn{5}{c}{(km s$^{-1}$)} \\
    \hline
   ${\rm [Fe\,II]}$ & $-186$& 115 & 302 & $-316$ & 158 \\
    H$_2$ S(7) & $-198$ & 94 & 292 & $-324$ & 136\\
    H$_2$ S(5) & $-181$ & 119 & 300 & $-314$ & 285 \\
    ${\rm [Ar\,II]}$ & $-158$& 105 & 264 & $-290$ & 204 \\
    H$_2$ S(4) & $-139$ & 94 & 233 & $-185$ & 188 \\ 
    ${\rm [Ne\,II]}$ & $-172$ & 109 & 281 & $-396$ & 221\\ 
    \hline
  \end{tabular}
\end{table}

\section{Ionized and warm molecular gas outflows}\label{sec:outflows}
The outflows previously reported in Mrk~231 \citep[see
Sect.~\ref{sec:introduction} for references and][for a review]{Veilleux2020} were observed in the nuclear
region as well as up to a few kiloparsecs away  from the AGN.
Plausible explanations for their driving mechanism include
the quasar, strong
SF activity taking place in the galaxy, and/or the presence
of a radio jet. This jet is observed in a nearly N-S direction
but it is confined to the inner parsec.
Since the mean-velocity fields of the low-excitation emission 
line and the H$_2$ transitions
present deviations from circular motions,  on nuclear and
circumnuclear scales (Sects.~\ref{subsec:kinematics_ionized} and
\ref{subsec:kinematics_molecular}), in this section
we analyze in detail the line profiles from the nuclear region and position-velocity (p-v)
diagrams of the ionized and warm molecular gas in Mrk~231 to look for
evidence of outflows.

\subsection{Nuclear region}\label{subsec:outflows_nuclear}
We constructed emission line profiles from the nuclear spectra
($r=1\arcsec$ aperture) after subtracting the fitted local continuum
and normalizing them to the line peak.  The [Fe\,{\sc ii}], [Ar\,{\sc
  ii}], and [Ne\,{\sc ii}] profiles (Fig.~\ref{fig:nuclearlineprofiles}) show a similar overall shape with evidence of a weak blue
wing (see below). The wings of the [Ne\,{\sc iii}] profile appear noisier due to the
lower contrast of the line against the continuum. In general all the H$_2$ lines
(middle and right panels of Fig.~\ref{fig:nuclearlineprofiles}) also
show comparable shapes. Those with higher constrast over the
continuum, namely, the S(5) and S(7) lines,  present a blue wing as well.

The mid-IR line wings in Mrk~231 are not prominent and were not modeled
with the single Gaussian fits in Sect.~\ref{subsec:analysis}. To
quantify their properties,  we performed a
non-parametric analysis with a similar method to that of 
\cite{Harrison2014}. We computed
the velocities at the $10^{\rm th}$ ($v_{\rm 
  10}$)  and $90^{\rm th}$ ($v_{\rm 90}$) flux percentiles,
the line width defined as $W_{\rm 80}= {\rm abs}(v_{\rm 90}-v_{\rm 10})$, as well as the
velocities at  the $2^{\rm nd}$ and
$98^{\rm th}$  percentiles ($v_{\rm 02}$ and $v_{\rm 98}$,
respectively). The largest of these two values provides an estimate of the
maximum projected velocity  of the outflow. 
We restricted this analysis to those lines with the
highest contrast against the continuum,
which are [Fe\,{\sc ii}], [Ar\,{\sc ii}],
[Ne\,{\sc ii}], and the H$_2$ S(7), S(5), and S(4).  
Appendix~\ref{appendix:nonparametric} provides details on the
method and Fig.~\ref{fig:nonparametric} 
presents two examples of the line analysis.

\begin{figure*}
  \hspace{0.5cm}
  \includegraphics[width=8.5cm]{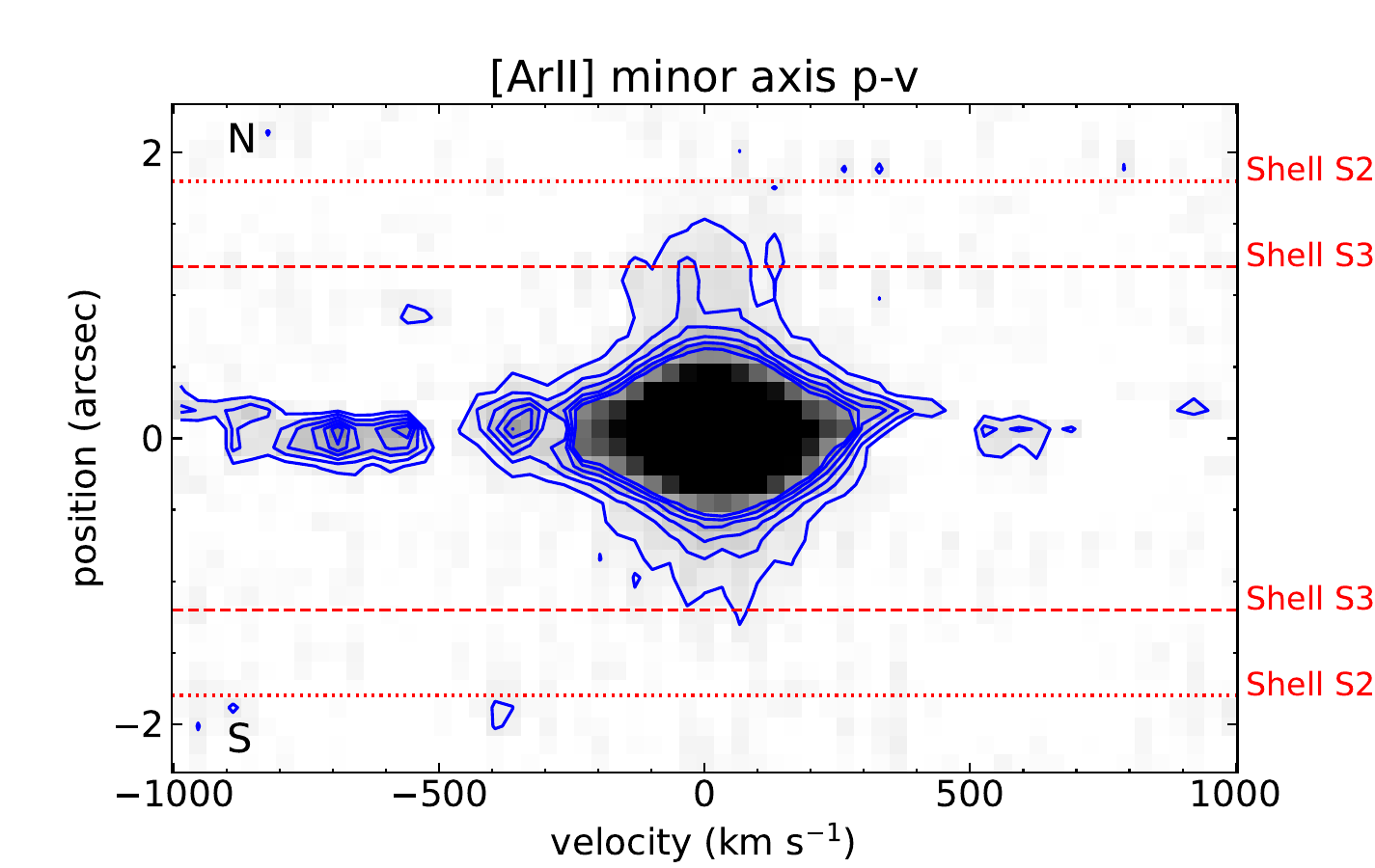}
  \hspace{0.5cm}
     \includegraphics[width=8.5cm]{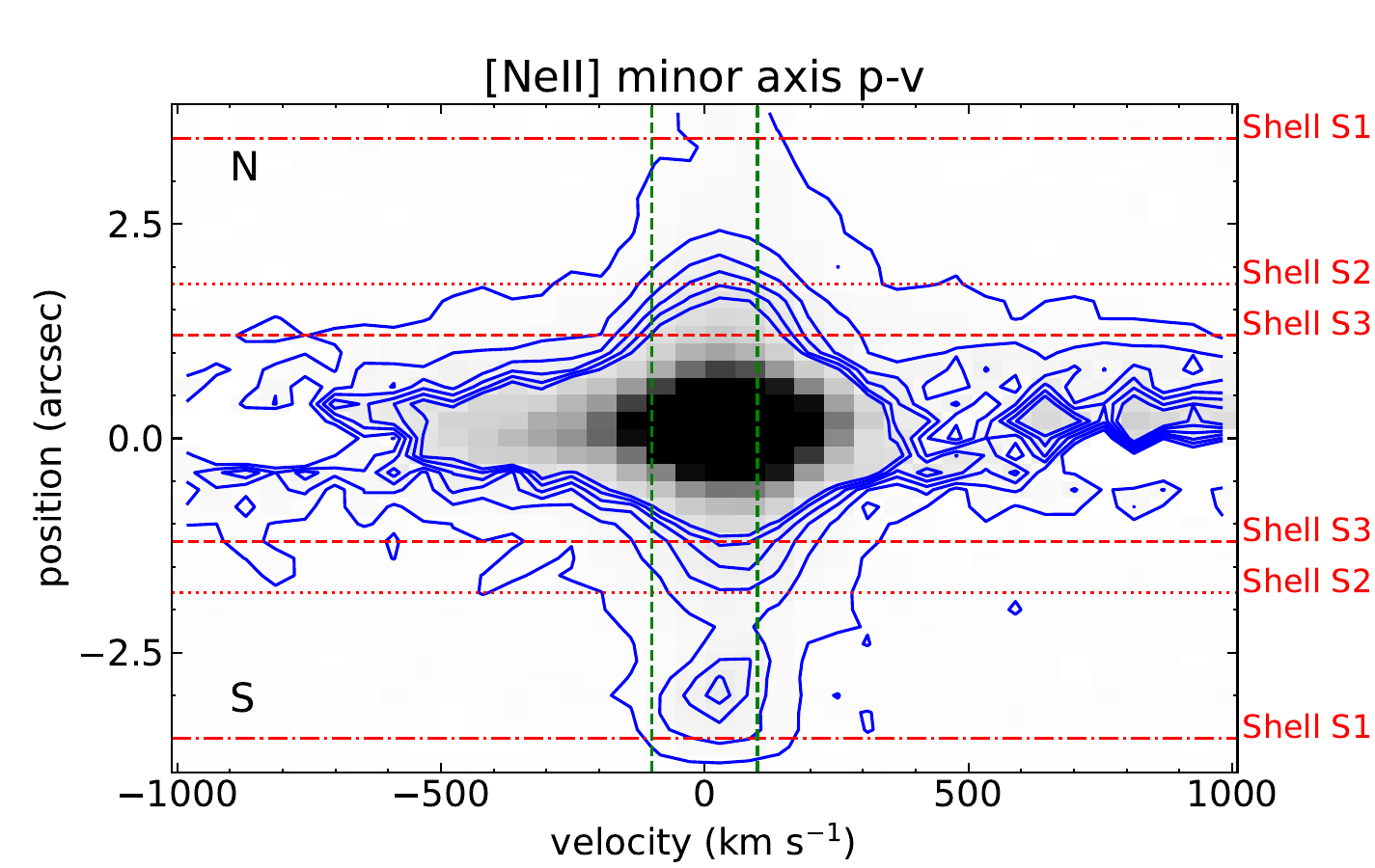}
     \caption{Kinematic minor axis p-v diagrams for [Ar\,{\sc ii}]
       ({\it left}) and  [Ne\,{\sc ii}] 
      ({\it right}). The extraction aperture is 2\arcsec. The 0,0 point in
      the spatial direction marks the AGN position. The first contour
      is drawn at three times the standard 
    deviation of 
    the (zero) continuum level. The position of the main expanding
    shells \citep{Lipari2005} are marked as horizontal lines. In
    the right panel the vertical lines denote the estimated virial range.}
     \label{fig:pvArIINeII}
\end{figure*}     

\begin{figure*}
  \hspace{0.5cm}
  \includegraphics[width=8.5cm]{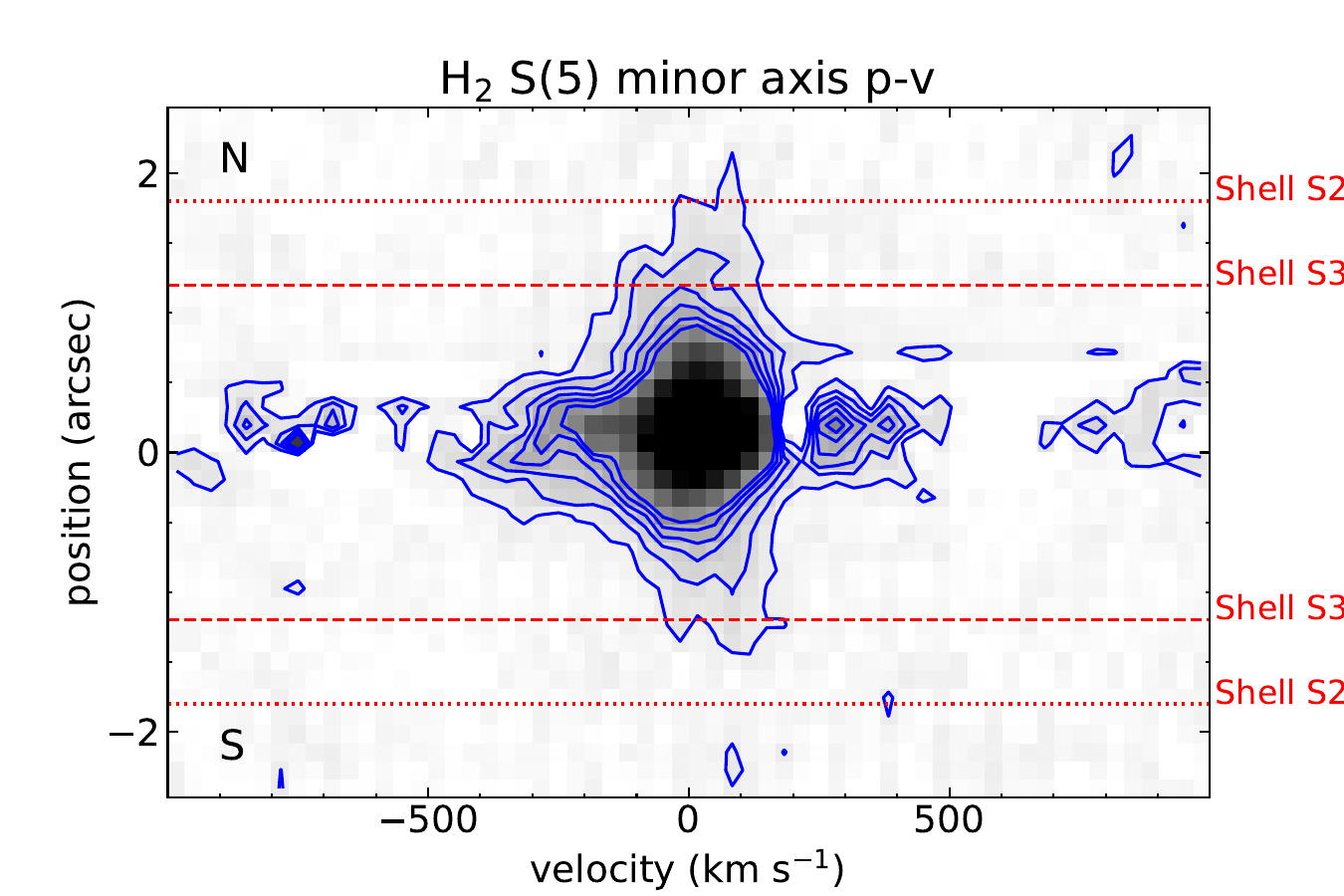}
  \hspace{0.5cm}
     \includegraphics[width=8.5cm]{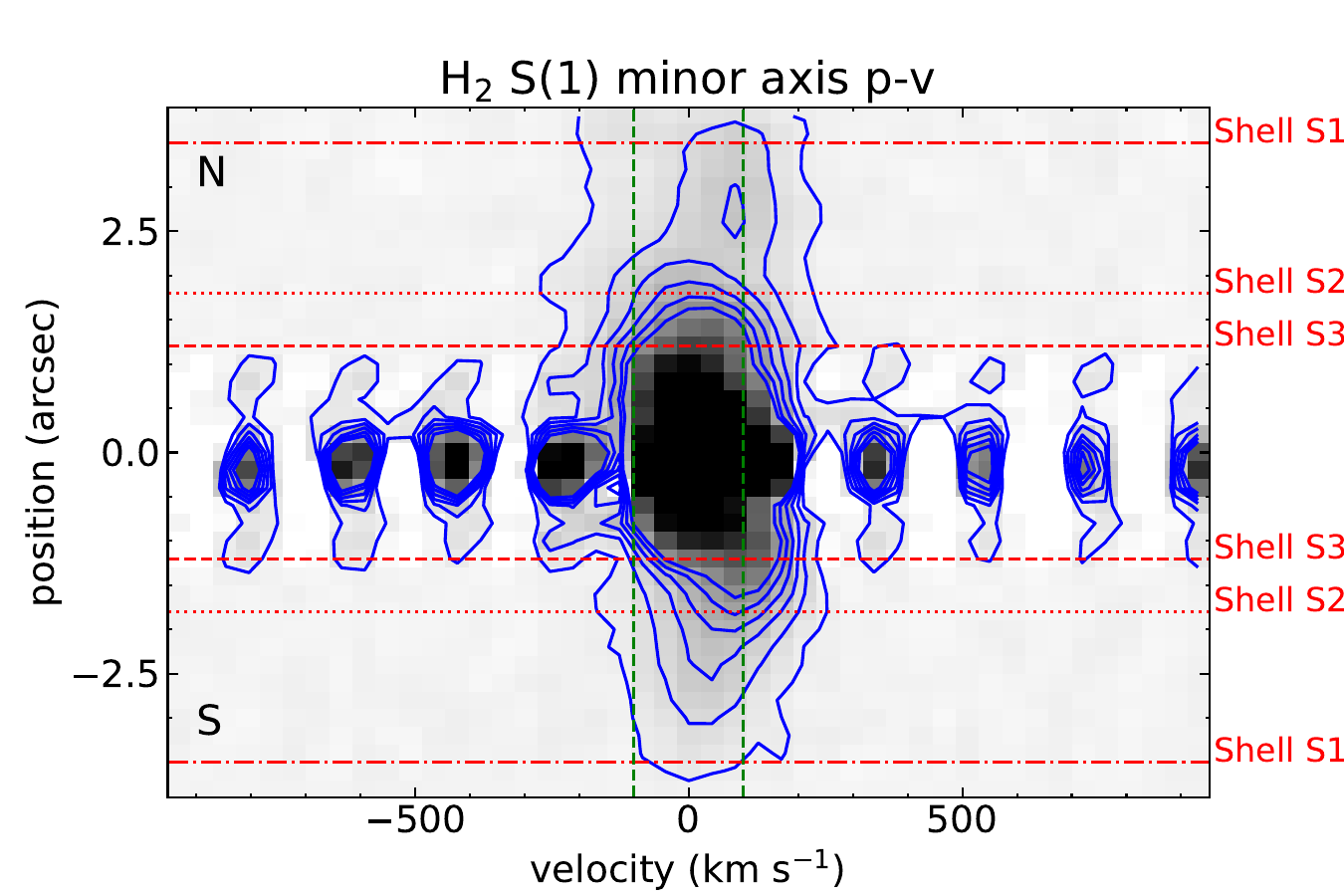}
     \caption{Kinematic minor axis p-v diagrams for 
      H$_2$ S(5) ({\it left}) and H$_2$ S(1) ({\it right})
      emission lines. Lines and symbols as
       Fig.~\ref{fig:pvArIINeII}.}
     \label{fig:pvH2S5H2S1}
\end{figure*}

All  lines show similar widths ($W_{\rm 80}\simeq 300\,{\rm
  km\,s}^{-1}$). The larger $W_{\rm 80}$ with respect to the
FWHM$_{\rm line}$ from the single Gaussian fit
(Table~\ref{tab:linefluxes}) are not solely due to the instrumental 
  broadening, which has
  FWHM$\sim 100\,{\rm km\,s}^{-1}$ for [Ne\,{\sc ii}]
  \citep{Argyriou2023}. The velocities (Table~\ref{tab:nonparametric})
  also revealed the presence of blueshifted wings being more 
extended in velocity, with [Ne\,{\sc ii}] presenting  the largest blueshifted velocities,
reaching $v_{\rm 02}=-400 \,{\rm 
  km\,s}^{-1}$ (Fig.~\ref{fig:nonparametric}, left panel). Using
  two Gaussians, we fitted this line with a narrow component
  with  FWHM$_{\rm narrow}=207\,{\rm km\,s}^{-1}$ and a blueshifted (by
  approximately $\Delta v=97\,{\rm km\,s}^{-1}$) broad component with
  FWHM$_{\rm broad}=518\pm50\,{\rm km\,s}^{-1}$. An estimate of the maximum
  outflow velocity can be computed as $v_{\rm max} = \Delta v +
  2\times \sigma_{\rm broad}$ \citep{Fiore2017}, which results in
  $v_{\rm max}\simeq 540\, {\rm
    km\,s}^{-1}$ in the nuclear region of Mrk~231.  

Additionally, the [Ne\,{\sc ii}] velocity channel maps also reveal that
the nuclear blueshifted component is slightly more prominent than the
redshifted one (see Fig.~\ref{fig:channelmaps}). 
We note that the nuclear mid-IR emission of Mrk~231 is 
 dominated by an extremely bright point source, which makes
 it  challenging to separate out
faint line wings from the residual continuum
wiggles. In the case of [Ne\,{\sc ii}], which is
the line with the highest signal-to-noise ratio, the
presence of the $12.7\,\mu$m PAH feature limits the detection of an
even more blueshifted line component, if present.

Since the observed nuclear [Ne\,{\sc
  ii}] emission in Mrk~231 is 
predominantly produced  by SF
(Sects.~\ref{subsec:lines} and ~\ref{sec:SFactivity}), the stronger
blue wings in the lines could be produced in a starburst-driven outflow
emerging perpendicular to the nearly
face-on nuclear disk. However, we cannot rule out that it is associated
with that detected in the nuclear  optical emission lines with
 faster velocities \citep[$v = -1000\,{\rm to}-750\,{\rm 
  km\,s}^{-1}$, ][]{Lipari1994, Rupke2011}. According to the latter
work, the highest velocity
component might be related to the AGN. 
 We did not detect high velocities ($|v|> 400\,{\rm
  km\,s}^{-1}$) in the warm molecular gas, which are present in 
    CO(2-1) and CO(3-2) toward the NE and SW, on
    sub-arcsecond scales \citep{Feruglio2015}. These authors explained
    these motions due to the presence of 
two molecular outflows, an isotropic one and a wide-angle biconical
outflow oriented in the NE to
SW directions. It is again possible that the high velocity motions of
the molecular gas are not detected due to the large extinction that is
affecting the nuclear region in the mid-IR (Sect.~\ref{sec:molecularhydrogen}).

\begin{figure*}
\hspace{2cm}
  \includegraphics[width=14.5cm]{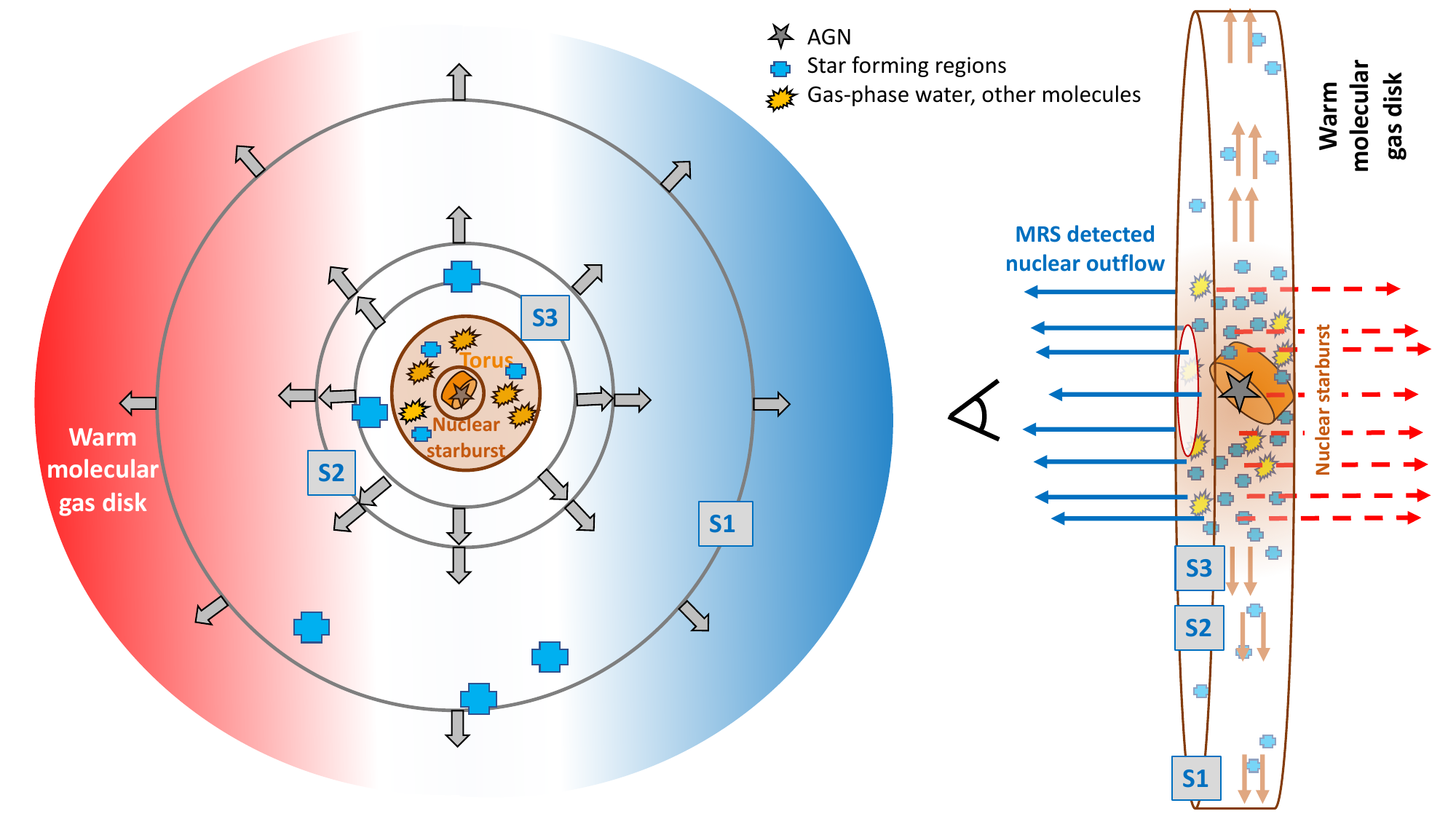}
     \caption{Cartoon summarizing the different regions (not drawn to
       scale) of Mrk~231 identified with the
       MRS observations and other literature results. {\it Left}:
       front view, that is, in the plane of the 
       sky, where the molecular gas disk is seen close to face-on. {\it
         Right}: side view to emphasize the possible scenario to
       explain some of the observed non-circular motions in the ionized and
       warm molecular gas. } 
     \label{fig:summary}
\end{figure*}

Summarizing, we favor a scenario where the nuclear outflows observed
in the mid-IR lines of Mrk~231 are driven by the powerful
starburst. However, we cannot rule 
out an AGN origin of and/or contribution to these mid-IR outflows. For
instance, some nearby Seyfert galaxies with evidence of AGN-driven outflows 
show relatively modest mid-IR line widths in their nuclear regions \citep[$W_{\rm 80} \simeq
300\,{\rm km\,s}^{-1}$, see][]{HermosaMunoz2024_NGC7172, Zhang2024}.
It is also clear that the exceptionally high  
velocities ($\simeq 1000\,{\rm km\,s}^{-1}$) detected in the
neutral gas phase of Mrk~231 and 
extending in all directions from the  center up to at least 2\,kpc \citep{Rupke2011} 
would put a starburst-driven wind origin to its limits. Indeed, based
on these observations
\citep{Rupke2013} argued that the nuclear outflow is powered by the
AGN \citep[see also][]{Leighly2014}, and 
the coupling between the nuclear wind and
the radio-jet might be responsible for accelerating the neutral gas to
these high velocities \citep[see also][]{Rupke2011}.

\subsection{Circumnuclear region}\label{subsec:extendedoutflows}
To investigate  the presence of outflows outside the nuclear region, where
the  instrumental effects due to the bright central point source are diminished, we
generated p-v diagrams along the kinematic minor axis (N-S direction)
from the data cubes of the fine-structure lines [Ar\,{\sc ii}] and  
[Ne\,{\sc ii}], and the H$_2$ S(5) and 
S(1) transitions. To do so, we placed a pseudo
slit 
of 2\arcsec \, width (as the diameter of the aperture used to
extract the nuclear 1D spectrum) on the portion of the data cube containing the line after subtracting a 
local linear continuum. At the AGN position, all four p-v diagrams
(Figs.~\ref{fig:pvArIINeII} and ~\ref{fig:pvH2S5H2S1}) show
residuals, which are due to the bright point source wiggles,
left after the 
continuum subtraction. In the case of the H$_2$ S(1) line, residual
fringes are also seen along the velocity (i.e., wavelength) axis.

In the nuclear region of Mrk~231, the  kinematic minor axis p-v diagrams of the
fine structure lines (Fig.~\ref{fig:pvArIINeII}) show velocities from
$\simeq -300\,{\rm to}\,+300\,{\rm km\,s}^{-1}$, and possibly to even higher blueshifted
velocities in [Ne\,{\sc ii}]. However, it is also clear that the high
values can also be due (in part) to the imperfect continuum
subtraction done on a spaxel-by-spaxel basis. Moving away from the AGN
continuum residuals at projected distances larger than 0.5-1\arcsec,
the ionized gas exhibits velocities centered at the systemic value
($v=0\,{\rm km\,s}^{-1}$ in these 
diagrams). The [Ne\,{\sc ii}] p-v diagram shows clearly the emission
from the star-forming region(s) located $\simeq 3\arcsec$  the south of the
AGN, while the [Ar\,{\sc ii}] p-v diagram also reveals the star-forming
region $\simeq 1\arcsec$ north of the AGN. There is, however,  another
underlying, extended, and relatively 
broad component (see below).

Out to the location of shell S3, at a projected distance of $\simeq 1.5\arcsec =
1.3\,$kpc from the AGN,  the kinematic minor axis p-v diagram of the H$_2$
S(5) line (left panel of Fig.~\ref{fig:pvH2S5H2S1})
shows a tendency for redshifted velocities to the south and blueshifted velocities to
the north. The velocities reach only a few hundred km\,s$^{-1}$ and are also
detected in the lower angular resolution p-v diagram of the H$_2$ S(1)
transition (right panel of Fig.~\ref{fig:pvH2S5H2S1}). A comparable
behavior was observed in the near-IR line 
$2.12\,\mu$m line on physical scales even closer to
the AGN \citep{Davies2004}. There are some similarities between the H$_2$
S(5) kinematic minor axis p-v diagram and those of the CO(2-1)
transition \citep[see middle panel of Fig.~7
of][]{Feruglio2015}. However, CO(2-1)  shows redshifted
velocities in excess 500\,km\,s$^{-1}$ in the nuclear region that are not observed in this
H$_2$ line. In the innermost region of Mrk~231, these  deviations from rotation in the
hot and cold molecular gas kinematics have
been attributed to a warp of the inner disk \citep{Davies2004,
  Feruglio2015}. They could also indicate the presence of a molecular outflow
in the disk of the galaxy.

On larger scales, at projected distances of a few arcseconds (several
kpc) from the AGN,
the H$_2$ S(1) and [Ne\,{\sc ii}] minor axis p-v diagrams (right panels of
Figs.~\ref{fig:pvH2S5H2S1} and ~\ref{fig:pvArIINeII}) show a  
range of velocities. At $\simeq 2\arcsec$ ($\simeq 1.6\,$kpc)  of the
AGN, they go from approximately $\simeq -200\, {\rm to}\,-300\,{\rm km\,s}^{-1}$ to
$\simeq +200\, {\rm to}\,+300\,{\rm km\,s}^{-1}$.
Following \cite{GarciaBurillo2015}, we
estimated a virial range  of $\pm 100\,{\rm km\,s}^{-1}$ for Mrk~231, by
adding the contributions from turbulence as half the FWHM of the
narrow component of [Ne\,{\sc ii}] in the southern star-forming region 
(observed FWHM$=130-135\,{\rm km\,s}^{-1}$), and in-plane
non-circular motions approximated as 1/2$\times v_{\rm rot}$, where the (projected) rotation velocity 
is $v_{\rm rot}=60\,{\rm km\,s}^{-1}$ \citep{DownesSolomon1998}. We can
conclude that the MRS high velocities on these scales are not likely be attributed to the expected
virial range of the rotating disk and could be related to extended
outflows and/or expanding shells.


\section{Summary}\label{sec:conclusions}
We presented JWST/MIRI MRS $4.9-27.9\,\mu$m observations of the
ULIRG and BAL quasar Mrk~231
covering the central 
$\simeq 5\arcsec \times 6\arcsec$ ($4.2\,{\rm kpc} \times
5\,{\rm kpc}$) in ch1 to
$\simeq 11\arcsec \times 11\arcsec$ ($9.2\,{\rm kpc} \times
9.2\,{\rm kpc}$) in ch4. The observations are part of the MIRI
European Consortium GTO program MICONIC.  Our main results are
summarized as follows  and in the
cartoon of Fig.~\ref{fig:summary}. 

\begin{itemize}
\item
 High excitation  (IP$\ge 95$\,eV) emission lines,
such as  [Mg\,{\sc v}]  and [Ne\,{\sc v}],  remain
undetected in these new and sensitive MRS observations. This is likely
due to a combination of the X-ray weak nature and the bright mid-IR continuum emission 
of  Mrk~231.
The mid-IR evidence for the presence of an AGN  comes from the bright
unresolved nuclear continuum and low EW
  of PAH features. This bright continuum likely prevents the detection
  of [Ar\,{\sc iii}] and
[S\,{\sc iv}]  (IP of 28 and 35\,eV, respectively) in
the nuclear region, whereas they are clearly observed in a 
  star-forming region $\simeq 920\,$pc SE from the AGN position. 

 \item
   A bright nuclear starburst with a size of $\simeq 400\,$pc (FWHM)
   is resolved for the first time in
the mid-IR  in the [Ne\,{\sc ii}]  and [Ar\,{\sc ii}] lines as well as
rotational H$_2$ transitions
and PAH emission.  The starburst extent is consistent with estimates 
derived at other wavelengths \citep{Carilli1998, Davies2004}. The  low
[Ne\,{\sc
  iii}]/[Ne\,{\sc ii}] ratio measured in the nuclear 
region is more typical of star-forming galaxies  than AGN.
Numerous absorption features associated with the gas-phase of
water are present in the $\sim 5-7\,\mu$m spectral range. Assuming
that they have similar origin as the water features observed in the far-IR, 
then they are formed within the nuclear starburst \citep{GonzalezAlfonso2010}.  
All this  supports the scenario where most of the mid-IR
low-excitation line emission in the central region of Mrk~231 is produced by SF
activity.  

\item
Several pieces of evidence, including the
neon-based SFR, EW of PAH features, and the H$_2$ excitation
diagram, indicate that the nuclear starburst in Mrk~231 is partly obscured, even in the
mid-IR. The extinction is higher  than what would be derived  
from the apparent depth of the $9.7\,\mu$m silicate feature
($S_{9.7\mu{\rm m}}=-0.75$), because it is partly filled in by the strong
mid-IR continuum produced by the AGN.

\item
  Using the H$_2$ S(1) to S(8) transition fluxes corrected for
    extinction, we estimated a mass of warm molecular gas in  
the central $\simeq 1.7\,$kpc of $9\times 10^6\,{\rm M}_\odot$
which is  between 
the hot and cold molecular gas masses \citep{DownesSolomon1998,
  Krabbe1997}.

\item
  The nuclear line profiles show weak wings, which
reach velocities of up to $v_{\rm 02} \simeq -400\,{\rm km\,s}^{-1}$ in [Ne\,{\sc
  ii}]. We interpreted these blueshifted velocities, as produced by a
 starburst-driven outflow. The more prominent blueshifted wings can be explained if
the outflow is driven by the nearly face-on nuclear 
starburst. Similar blueshifted components are
identified from optical observations, although with higher velocities \citep{Lipari2009,
  Rupke2011}. We note however that the MRS observations cannot
  rule out an AGN origin for the nuclear outflow. Regardless of its
  origin, the  lack of high velocity components and redshifted
  wings in 
  the mid-IR lines could be due to a 
combination of a strong mid-IR continuum and obscuration.

\item
  The extended [Ne\,{\sc ii}], H$_2$ S(1), and  $11.3\,\mu{\rm m}$
PAH emissions  trace the  ionized
and warm molecular gas phases of the large-scale disk of Mrk~231 over the
entire mapped region of  $\simeq 5-8\,$kpc. We detected several star-forming
regions and complexes located up to a few kpc away from the AGN,
which were  already
known from previous observations \citep{Surace1998,  Lipari2009,
  Rupke2011}. The emission from these regions is  
superimposed on a  more extended and diffuse component.

\item
The large-scale kinematics of the ionized and warm molecular gas show
circular motions in a disk oriented in the 
approximate E-W direction, similar to that detected in cold molecular
gas \citep{DownesSolomon1998, Feruglio2015}. There are also some deviations from
rotation, seen as
blue- and red-shifted motions with respect to the systemic velocity in 
the $\simeq 200-300\,{\rm km\,s}^{-1}$ range. 
These non-circular motions are likely
connected with  the expanding  shells and super-bubbles identified from
optical imaging and IFU 
observations \citep{Lipari2009} and/or the isotropic
cold molecular gas outflow \citep{Feruglio2015}.

\end{itemize}

Summarizing, the unprecedented combination of sensitivity, high angular resolution, and
spectral resolving power together with the broad spectral coverage afforded
by MIRI MRS has enabled, for the first time, a detailed, spatially resolved 
study of Mrk~231 in the
mid-IR. The picture that emerges is that a large fraction of the
mid-IR line emission is produced in a spatially resolved powerful and obscured nuclear
starburst. It is located in the central regions of a large ($\sim 8\,$kpc) and
massive and rotating disk where  circumnuclear SF
regions and widely-spread (low velocity) outflows in both the
ionized and molecular gas are present.  
The Mrk~231 AGN reveals itself
as a bright mid-IR continuum which, at the same time,  might prevent the
detection of relatively faint high-excitation lines produced by this  X-ray weak
quasar. 


    \begin{acknowledgements}
We thank the anonymous referee for comments that helped improve
  the manuscript. We are grateful to I. Garc\'{\i}a-Bernete, M. Pereira-Santaella,
S. Garc\'{\i}a-Burillo, and F. Donnan for insightful discussions
about the MRS observations and scientific interpretation.

AAH and LHM acknowledge
support from grant PID2021-124665NB-I00 funded by the Spanish
Ministry of Science and Innovation and the State Agency of Research
MCIN/AEI/10.13039/501100011033  and ERDF A way of making Europe.

MB acknowledges funding from the Belgian Science Policy Office
(BELSPO) through the PRODEX project "JWST/MIRI Science exploitation"
(C4000142239).  

JAM and LC acknowledge support by grant PIB2021-127718NB-100 from the Spanish
Ministry of Science and Innovation/State Agency of Research MCIN/AEI/10.13039/50110001103.

G\"O acknowledges support from the Swedish National Space Administration (SNSA).

MJW acknowledges support from a Leverhulme Emeritus Fellowship, EM-2021-064.

MIRI draws on the scientific and technical expertise of the following organisations:
Ames Research Center, USA; Airbus Defence and Space, UK; CEA-Irfu, Saclay,
France; Centre Spatial de Liége, Belgium; Consejo Superior de Investigaciones
Científicas, Spain; Carl Zeiss Optronics, Germany; Chalmers University of Technology, Sweden; Danish Space Research Institute, Denmark; Dublin Institute
for Advanced Studies, Ireland; European Space Agency, Netherlands; ETCA,
Belgium; ETH Zurich, Switzerland; Goddard Space Flight Center, USA; Institute d’Astrophysique Spatiale, France; Instituto Nacional de Técnica Aeroespacial, Spain; Institute for Astronomy, Edinburgh, UK; Jet Propulsion Laboratory,
USA; Laboratoire d’Astrophysique de Marseille (LAM), France; Leiden University, Netherlands; Lockheed Advanced Technology Center (USA); NOVA OptIR group at Dwingeloo, Netherlands; Northrop Grumman, USA; Max-Planck
Institut für Astronomie (MPIA), Heidelberg, Germany; Laboratoire d’Etudes
Spatiales et d’Instrumentation en Astrophysique (LESIA), France; Paul Scherrer Institut, Switzerland; Raytheon Vision Systems, USA; RUAG Aerospace,
Switzerland; Rutherford Appleton Laboratory (RAL Space), UK; Space Telescope Science Institute, USA; Toegepast- Natuurwetenschappelijk Onderzoek
(TNO-TPD), Netherlands; UK Astronomy Technology Centre, UK; University
College London, UK; University of Amsterdam, Netherlands; University of
Arizona, USA; University of Cardiff, UK; University of Cologne, Germany;
University of Ghent; University of Groningen, Netherlands; University of Leicester, UK; University of Leuven, Belgium; University of Stockholm, Sweden; Utah State University, USA. A portion of this work was carried out at
the Jet Propulsion Laboratory, California Institute of Technology, under a contract with the National Aeronautics and Space Administration. We would like
to thank the following National and International Funding Agencies for their
support of the MIRI development: NASA; ESA; Belgian Science Policy
Office; Centre Nationale D’Etudes Spatiales (CNES); Danish National Space Centre;
Deutsches Zentrum fur Luft-und Raumfahrt (DLR); Enterprise Ireland; Ministerio De Economía y Competitividad; Netherlands Research School for Astronomy (NOVA); Netherlands Organisation for Scientific Research (NWO);
Science and Technology Facilities Council; Swiss Space Office; Swedish National Space Board; UK Space Agency. This work is based on observations
made with the NASA/ESA/CSA James Webb Space Telescope. The data were
obtained from the Mikulski Archive for Space Telescopes at the Space Telescope Science Institute, which is operated by the Association of Universities for
Research in Astronomy, Inc., under NASA contract NAS 5-03127 for JWST;
and from the European JWST archive (eJWST) operated by the ESDC.

This research has made use of the NASA/IPAC Extragalactic Database (NED),
which is operated by the Jet Propulsion Laboratory, California Institute of Technology,
under contract with the National Aeronautics and Space
Administration.

This research made use of NumPy \citep{Harris2020}, Matplotlib \citep{Hunter2007} and Astropy \citep{Astropy2013, Astropy2018}.

\end{acknowledgements}

%
%

   \bibliographystyle{aa} 
   \bibliography{bibliography} 

\begin{appendix}
\section{Continuum maps}\label{appendix:continuummaps}
\begin{figure}[h]

       \includegraphics[width=4.5cm]{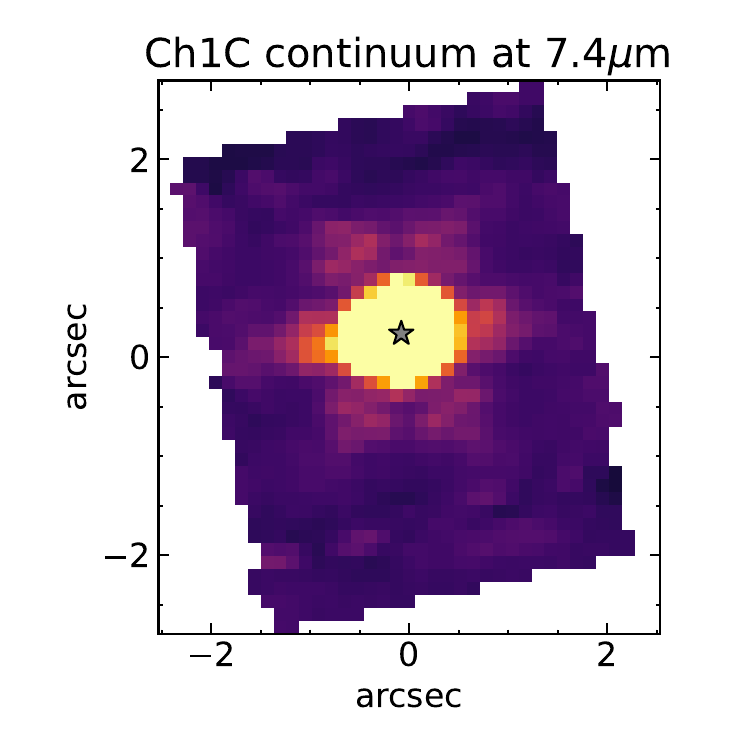}
      \hspace{-0.5cm}
     \includegraphics[width=4.5cm]{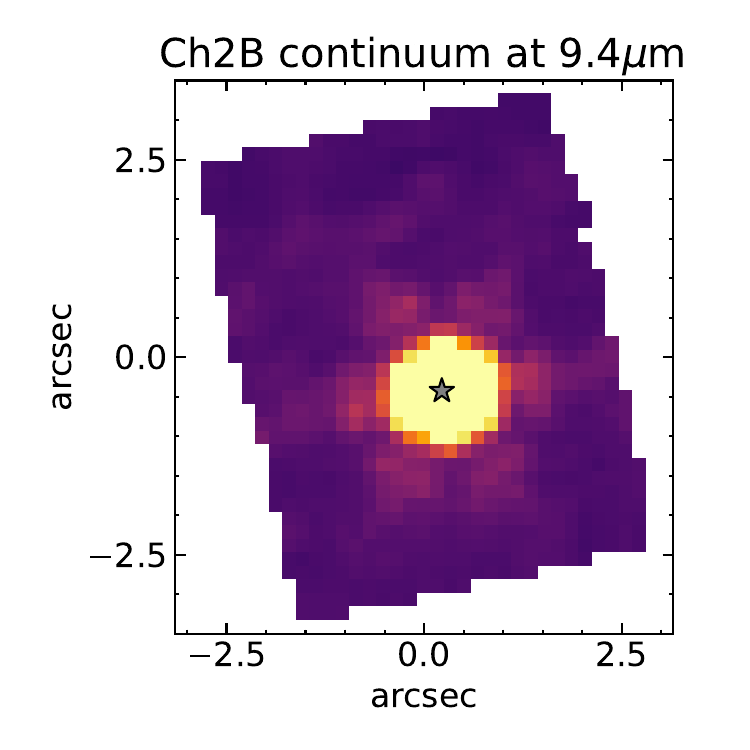}

     \includegraphics[width=4.5cm]{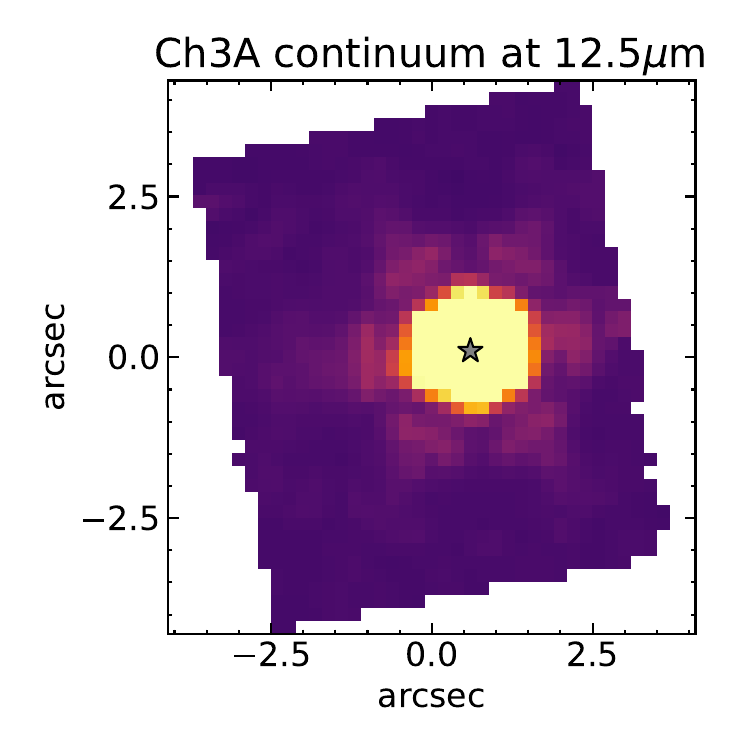}
      \hspace{-0.4cm}
     \includegraphics[width=4.5cm]{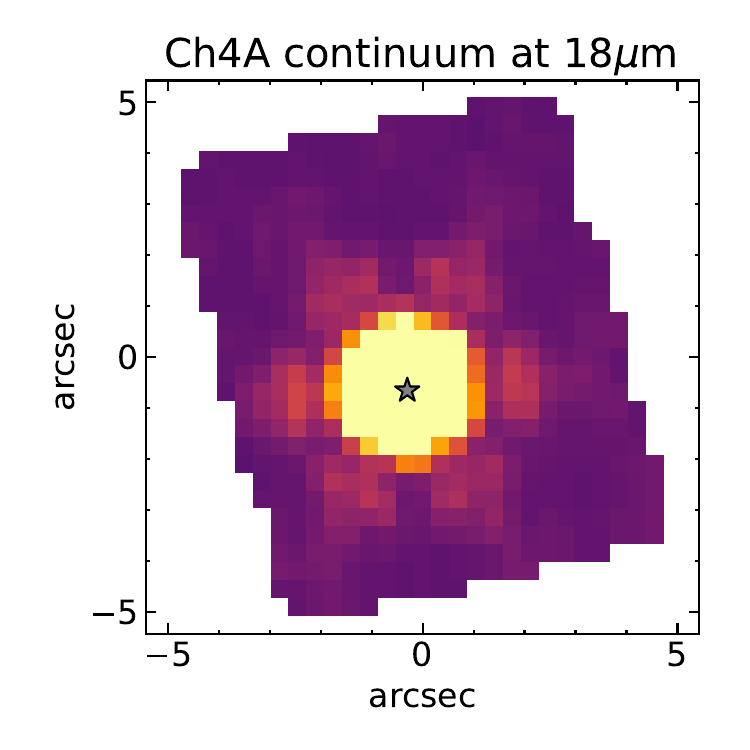}
     \caption{MRS continuum maps of Mrk~231 in ch1, ch2, ch3, and ch4 at the
     wavelengths specified at the top of each subpanel. The images are
     shown in a linear scale (arbitrary units) with cuts chosen to
     emphasize the low level structure of the point source The star symbol marks the peak of
     the continuum. The 0,0 point on the axes of each map refers to the center of
     the corresponding sub-channel array, after rotation.}
     \label{fig:continuummaps}
\end{figure}

        Figure~\ref{fig:continuummaps} presents MRS continuum maps
     of Mrk~231 in  sub-channels of ch1,
     ch2, ch3, and ch4, at selected observed wavelengths of 7.4
     (ch1C), 9.4 (ch2B),
     12.5 (ch3A), and $18\,\mu$m (ch4A). As can be seen from the figures, the
     continuum emission of Mrk~231
     is dominated by the presence of a bright nuclear point source. We
     note that the position of this source (the continuum peak is marked with the star
     symbol) on the array is different for 
     each sub-channel.

\section{Spitzer/IRS spectra}\label{appendix:spitzerirs}

\begin{figure}

 \includegraphics[width=9cm]{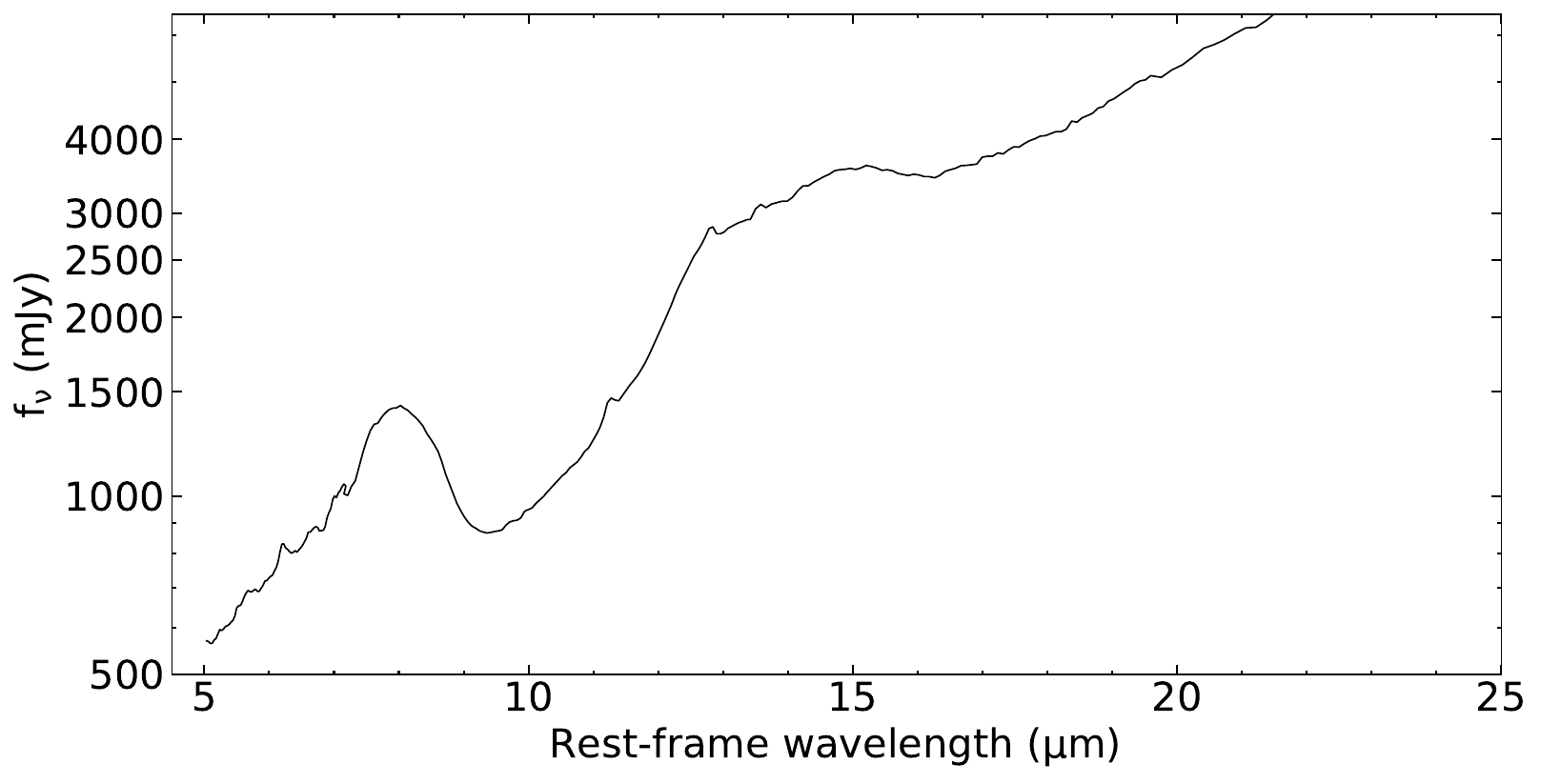}

 \includegraphics[width=9cm]{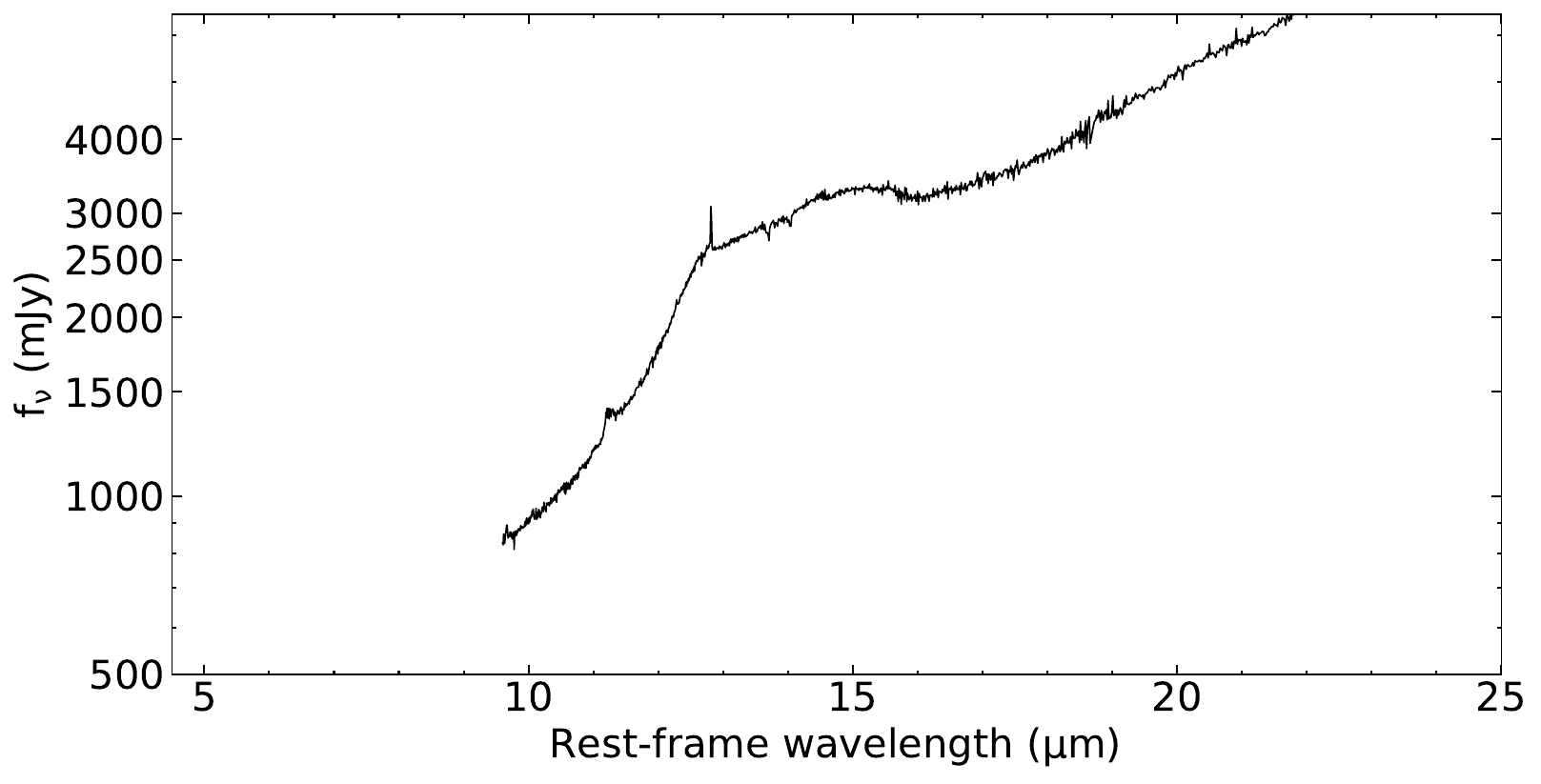}
  \caption{{\it Spitzer}/IRS spectra of Mrk~231. Those obtained with the
      low-resolution modules SL and LL are in the top panel  and those
      of the high-resolution
      modules SH and LH in the bottom. Wavelength and flux density ranges
      are as in Fig.~\ref{fig:nuclearspectrum}. }
     \label{fig:SpitzerIRS}
\end{figure}

 We retrieved {\it Spitzer}/IRS fully reduced spectra of Mrk~231 from the Combined
  Atlas of Sources with {\it Spitzer} IRS Spectra \citep[CASSIS,][]{Lebouteiller2011,
    Lebouteiller2015}. There are several observations taken in
  different cycles for this galaxy. Figure~\ref{fig:SpitzerIRS} shows those from program ID: 1459
  (P.I.: L. Armus) using the low-resolution modules (top panel): short-low (SL)
  with a slit width of 3.6--3.7\arcsec \, and long-low (LL) with a slit width of 10.5-10.7\arcsec,
and the high-resolution modules (bottom panel): short-high (SH) with a slit width of
4.7\arcsec \, and long-high (LH) with a slit width of 11.1\arcsec. CASSIS identified
this source as point-like and used the optimal extraction option.

 \section{Emission line maps}\label{appendix:extramaps}
\begin{figure*}
  \includegraphics[width=19cm]{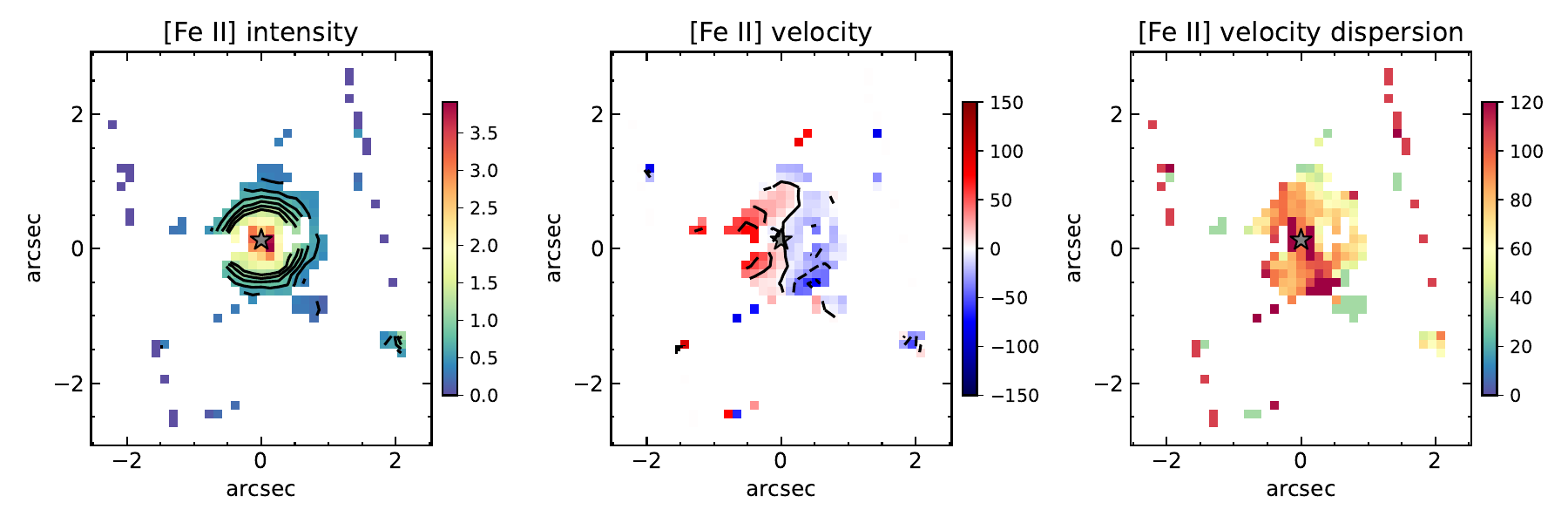}
  \vspace{-0.75cm}
  \caption{MRS maps of [Fe\,{\sc ii}] at $\lambda_{\rm
         rest}  = 5.34\,\mu$m. We smoothed this sub-channel data cube with a $2\times2$
    pixel box in the spatial directions prior to fitting the emission
    line on a spaxel-by-spaxel basis.  Panels and 
       symbol are as in
       Fig.~\ref{fig:ArIImaps}.  The
       0,0 point on the axes refers to the center of this sub-channel
       array, after rotation. In 
       all maps, north is up and east to the left.}
     \label{fig:FeIImaps}
\end{figure*}

\begin{figure*}
  \includegraphics[width=19cm]{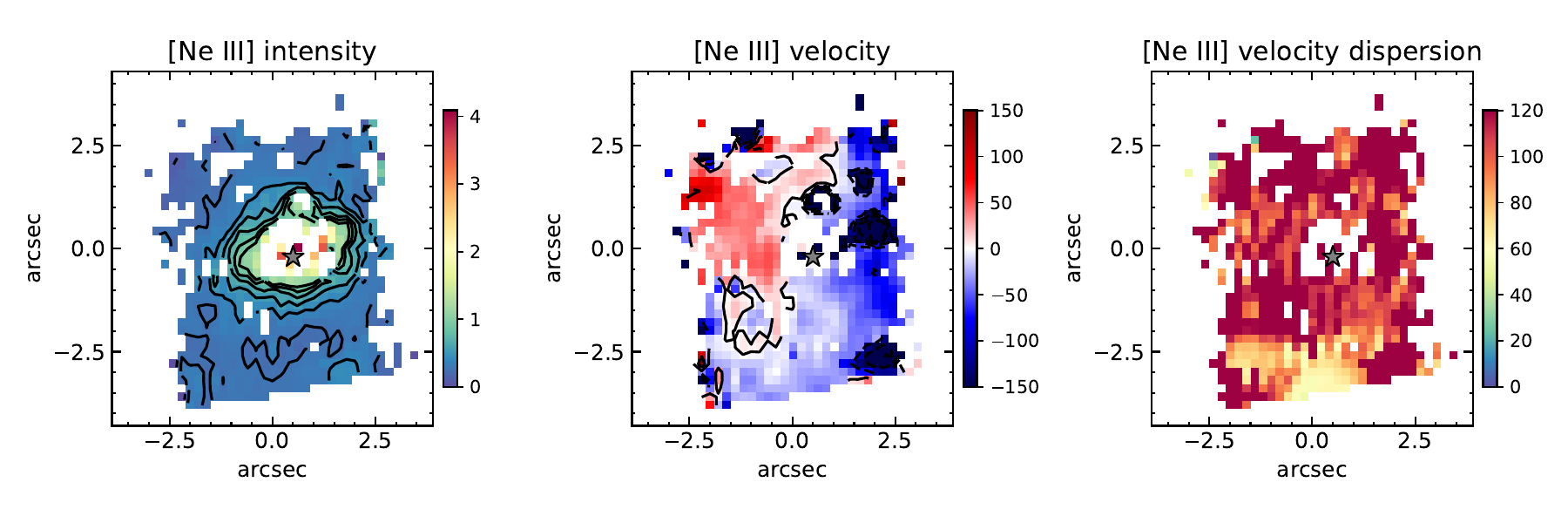}
  \vspace{-0.75cm}
  \caption{MRS maps of   [Ne\,{\sc iii}] at $\lambda_{\rm
         rest} =15.56\,\mu$m. We smoothed this sub-channel data cube with a $2\times2$
    pixel box in the spatial directions prior to fitting the emission
    line on a spaxel-by-spaxel basis. Panels and 
       symbol are as in
       Fig.~\ref{fig:ArIImaps}.  The
       0,0 point on the axes refers to the center of this sub-channel
       array, after rotation. In 
       all maps, north is up and east to the left.}
     \label{fig:NeIIImaps}
\end{figure*}

\begin{figure*}
  \includegraphics[width=19cm]{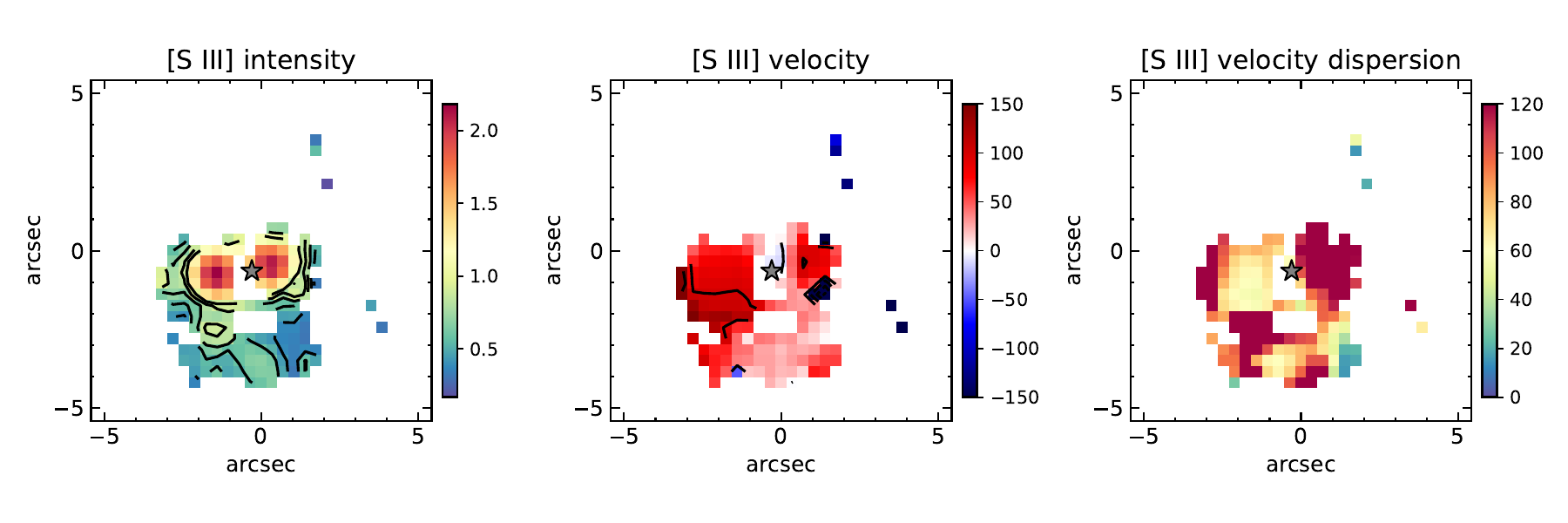}
  \vspace{-0.75cm}
  \caption{MRS maps of   [S\,{\sc iii}]  at $\lambda_{\rm
         rest} =18.71\,\mu$m. We smoothed sub-channel data cube with a $2\times2$
    pixel box in the spatial directions prior to fitting the emission
    line on a spaxel-by-spaxel basis. Panels and 
       symbol are as in
       Fig.~\ref{fig:ArIImaps}.  The
       0,0 point on the axes refers to the center of this sub-channel
       array, after rotation. In 
       all maps, north is up and east to the left.}
     \label{fig:SIIImaps}
\end{figure*}

\begin{figure*}
  \includegraphics[width=19cm]{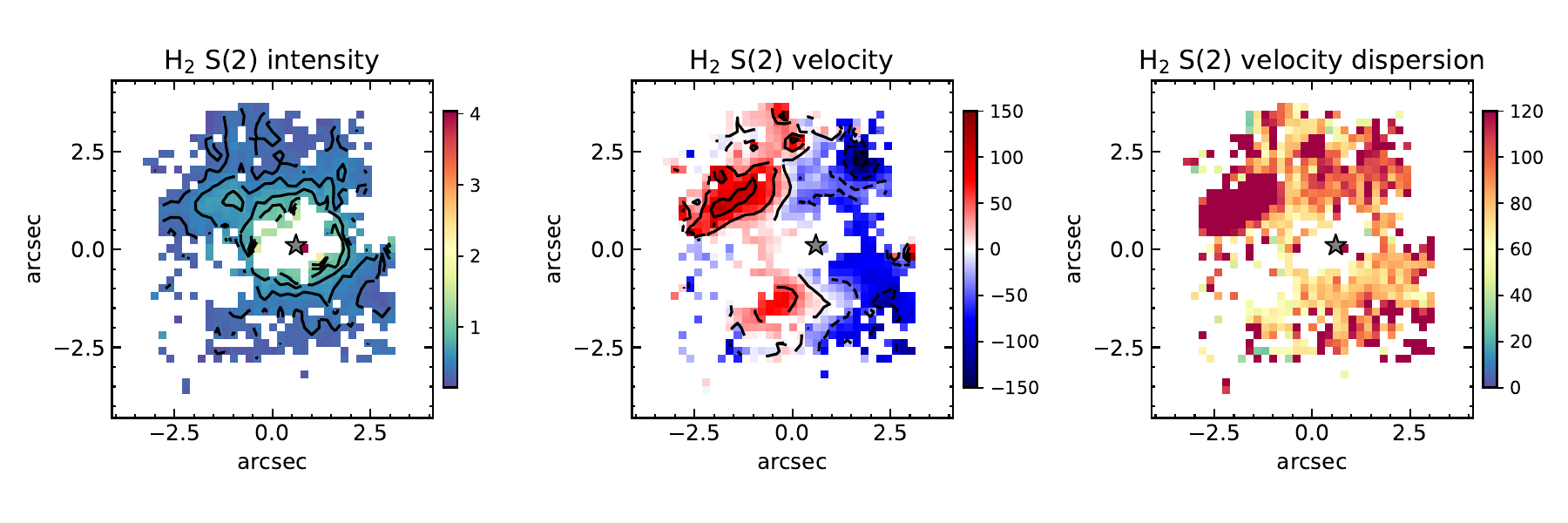}
  \vspace{-0.75cm}
  \caption{MRS maps of  H$_2$ S(2) line  at $\lambda_{\rm
         rest} = 12.28\,\mu$m. Panels and 
       symbol are as in
       Fig.~\ref{fig:ArIImaps}.  The
       0,0 point on the axes refers to the center of this sub-channel
       array, after rotation. In 
       all maps, north is up and east to the left.}
     \label{fig:H2S2maps}
   \end{figure*}

\begin{figure*}
  \includegraphics[width=19cm]{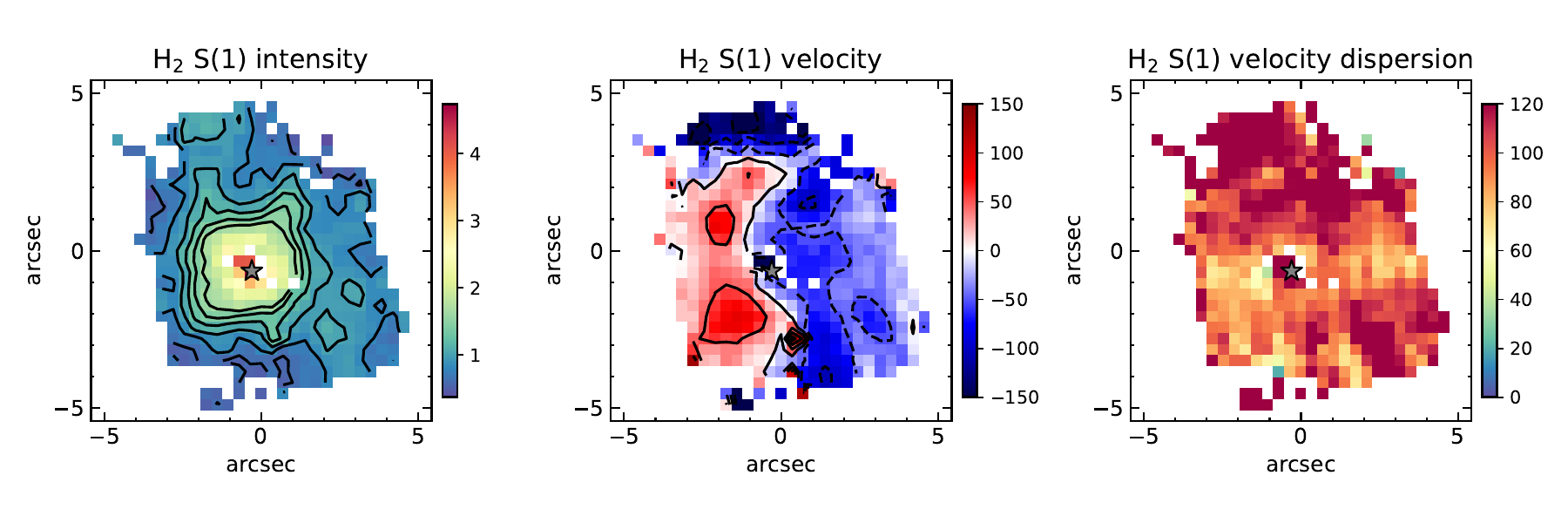}
  \vspace{-0.75cm}
  \caption{MRS maps of H$_2$ S(1) at $\lambda_{\rm
         rest} =17.03\,\mu$m observed in ch4. Panels and 
       symbol are as in
       Fig.~\ref{fig:ArIImaps}.  The
       0,0 point on the axes refers to the center of this sub-channel
       array, after rotation. In 
       all maps, north is up and east to the left.}
     \label{fig:H2S1ch4maps}
\end{figure*}

     In this appendix, Figs.~\ref{fig:FeIImaps}, ~\ref{fig:NeIIImaps} and \ref{fig:SIIImaps} present maps of the line intensity, mean-velocity field, and velocity dispersion for the low-excitation
   lines [Fe\,{\sc ii}], 
     [Ne\,{\sc iii}], and [S\,{\sc iii}], while
Figs.~\ref{fig:H2S2maps} and \ref{fig:H2S1ch4maps}
for  H$_2$ S(2) and  
H$_2$ S(1) observed in ch4.

Since the [Ne\,{\sc iii}] and [S\,{\sc iii}] emission lines show a
relatively low contrast with respect to the  
local continuum, especially in the nuclear region (see
Figs.~\ref{fig:nuclearspectrum} and \ref{fig:nuclearspectrum_channels}),
we smoothed the data cube spaxels along the 
spatial directions with a $2\times2$ average box before fitting
Gaussians to the
lines and  constructing the maps. Despite this,
there are still
artificial holes of line emission in the central few
spaxels due to the strong continuum. Nevertheless,  both line flux maps reveal
emission similar to that of [Ne\,{\sc ii}] (see Fig.~\ref{fig:NeIImaps}).

\begin{figure}[h]
  \hspace{-1.5cm}
  \includegraphics[width=12cm]{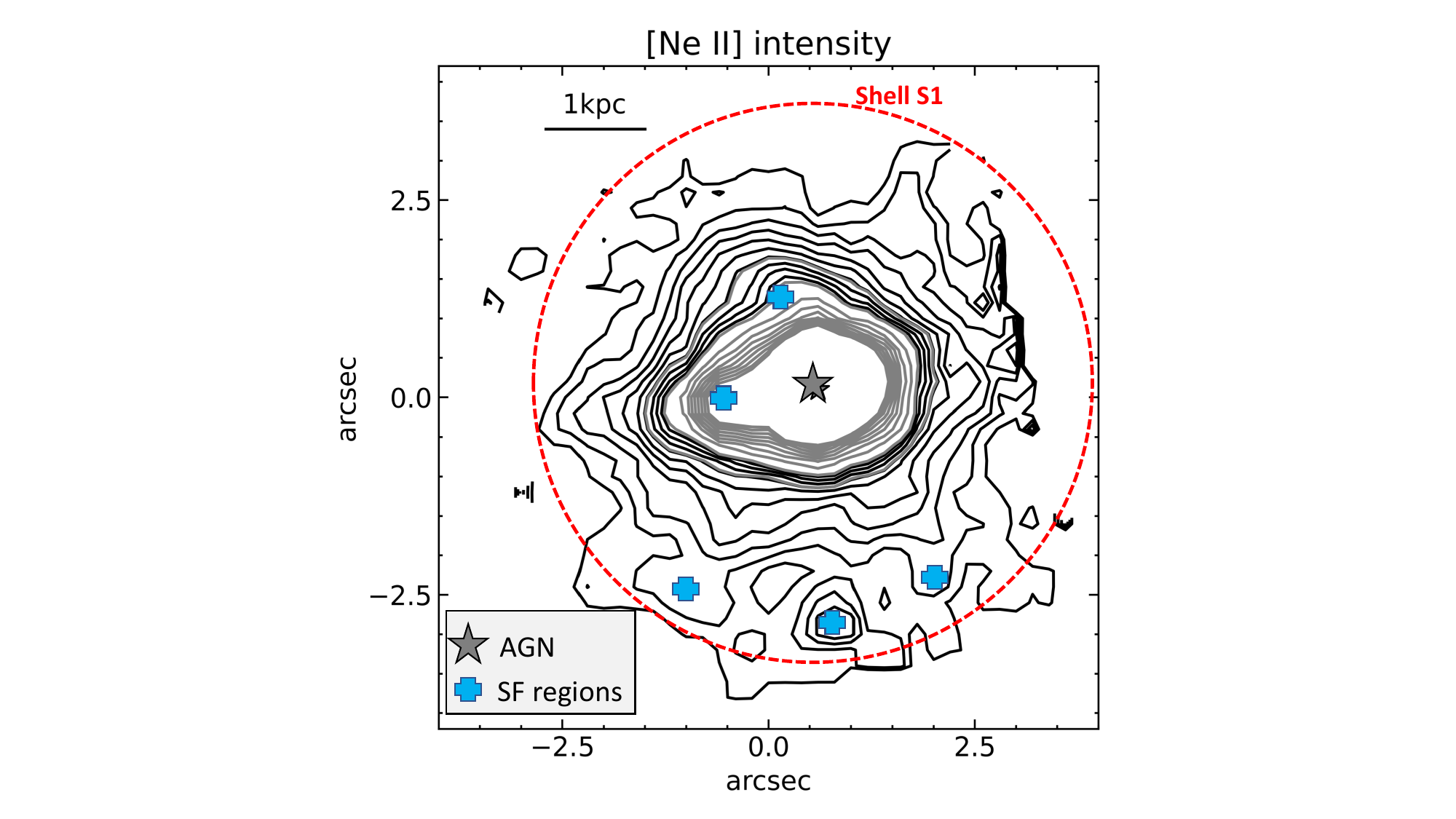}
  \vspace{-0.5cm}
     \caption{Contour map of the MRS [Ne\,{\sc ii}] emission. We mark some
       of the SF regions discussed in the text. We generated this [Ne\,{\sc ii}] map
       by integrating all the line emission and
       subtracting a local 
       continuum measured on both sides of the line. The black
       contours are shown in a square root scale, while the gray ones
       are in a linear scale.}
     \label{fig:NeIIsketch}
\end{figure}

To help identify the SF regions discussed in the text, we marked them in
the contour plot of the [Ne\,{\sc ii}] emission of
Fig.~\ref{fig:NeIIsketch}. In this case, this intensity map was
constructed by integrating the line and subtracting the local adjacent
continuum from both sides of the line, instead of fitting a single
Gaussian.

\section{[Ne\,{\sc ii]} velocity channel maps}\label{appendix:channelmaps}
\begin{figure}
  \includegraphics[width=9cm]{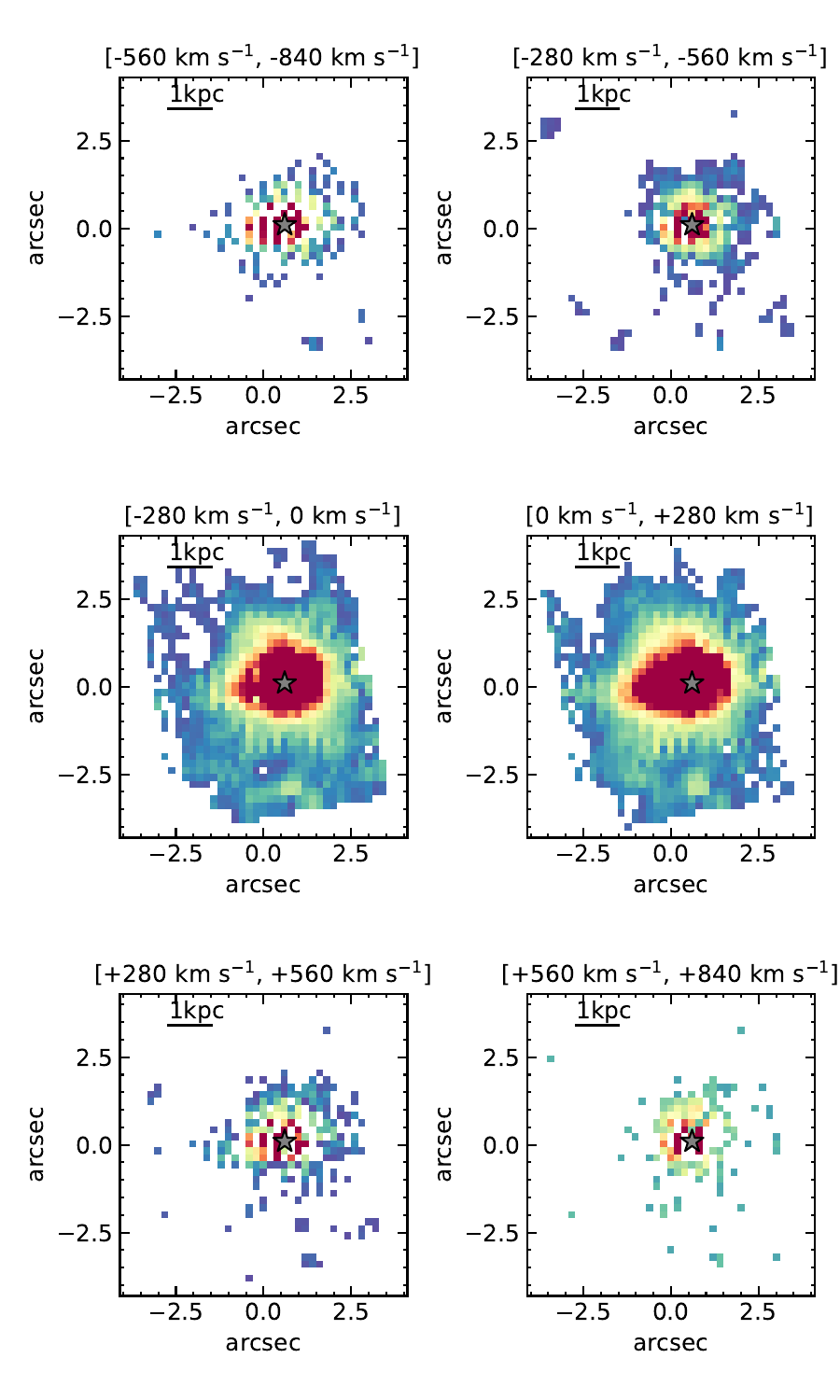}
  \vspace{-1cm}
  \caption{Velocity (relative to the systemic value) channel maps for
    [Ne\,{\sc ii}]. Fluxes are plotted in a 
    square root scale, which is the same in all panels.}
     \label{fig:channelmaps}
\end{figure}

     In this appendix, we show in Fig.~\ref{fig:channelmaps}]  the
     velocity (relative to the systemic value) 
     channel maps for the [Ne\,{\sc ii}]. We generated them from an extracted data cube
     around the line after subtracting the local continuum on a
     spaxel-by-spaxel basis. For each velocity range, the channel maps
     are plotted for pixels with fluxes three times the standard
     deviation above the mean
     value  measured in regions with no line emission. The exception
     is for the
     channels including the systemic velocity, where there is line
     emission practically all over the ch3 FoV and the fluxes are
     those above the mean value in a small region with no line emission. 
     
\section{Non-parametric emission line
       analysis}\label{appendix:nonparametric} 
In this appendix, we describe the method followed to perform the
non-parametric analysis of the 
 most prominent nuclear emission lines in
the nuclear region of Mrk~231.

A non-parametric analysis allows to
obtain the kinematic properties of the emission lines without assuming
any prior conditions for the line profile, as is the case with a
parametric modeling (i.e., Gaussian profiles). This method is
exclusively based on measuring the flux of the lines. We used a
modified  version of the method put forward by \cite{Harrison2014}. Instead
of using fitted Gaussians to isolate the line, we defined the line with
respect to the noise of the zero level continuum.
To perform the analysis,  we selected a region ranging from $-1200$ to
1200\,km\,s$^{-1}$ from the redshifted-corrected central wavelength of
the line, and normalised the whole range to the maximum flux
(that is, the line peak) after the continuum subtraction. Then we
estimated the standard deviation in 
the continuum near the emission line, and defined the line as all
those spectral elements where
the flux is larger than two times that value. Then we summed up all
the flux of the line within that range and estimated the
velocities. The derived velocities are more
conservative as this method takes into account the continuum variations, which can
include real structure (e.g., broad features) as well as
noise. This is especially relevant for lines which do not present a
strong contrast against the continuum.

As mentioned in
Sect.~\ref{subsec:outflows_nuclear}, we measured the following parameters
$v_{\rm 02}$,
$v_{\rm 10}$, $v_{\rm 50}$, $v_{\rm 90}$, and 
$v_{\rm 98}$, which are the velocities at which we found 2\%, 10\%, 50\%, 90\%,
and 98\% of the line fluxes, and $W_{\rm 80}$, which is a measurement of the
line width, defined as abs($v_{\rm 90}-v_{\rm 10}$). The line width $W_{\rm 80}$ is equivalent to 
1.088$\times$FWHM for a single Gaussian-like profile \citep{Harrison2014}.  
We note that the
W80 parameter is not corrected from the instrumental broadening.  We also note that the
analysis was done on the spectra before the 1D residual fringe
correction to avoid removing real spectral features.

\begin{figure}

\hspace{0.75cm}
  \includegraphics[width=7cm]{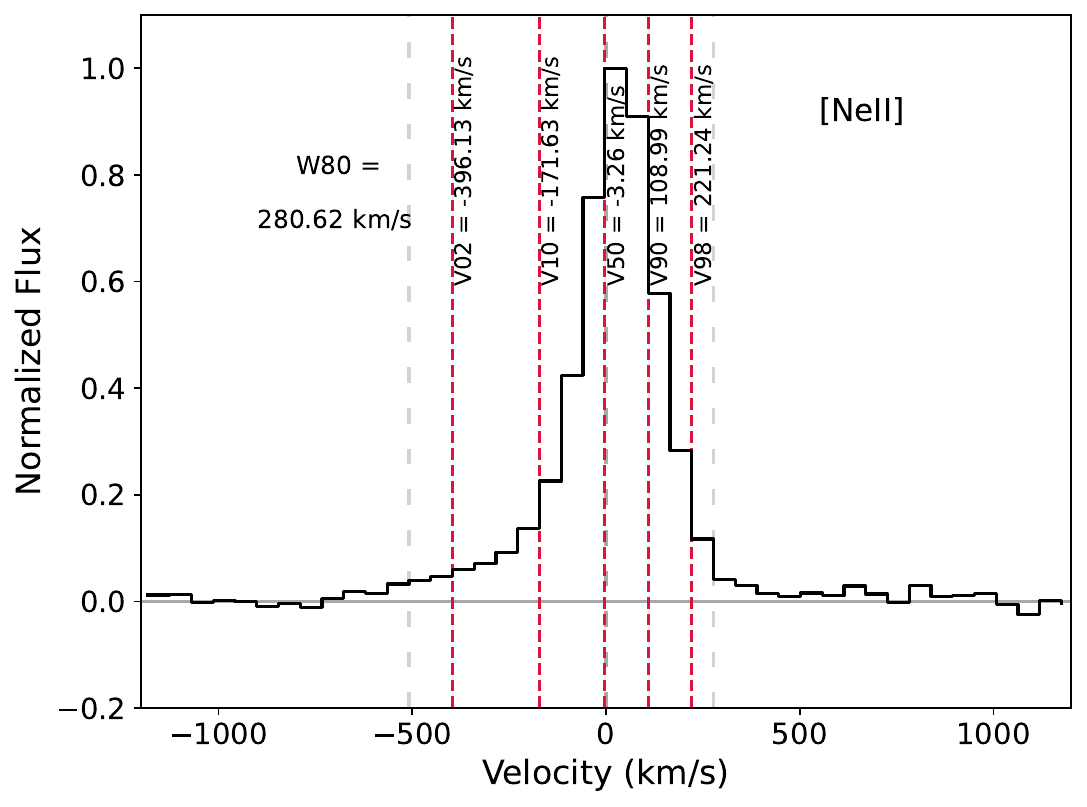}

 \hspace{0.75cm}
  \includegraphics[width=7cm]{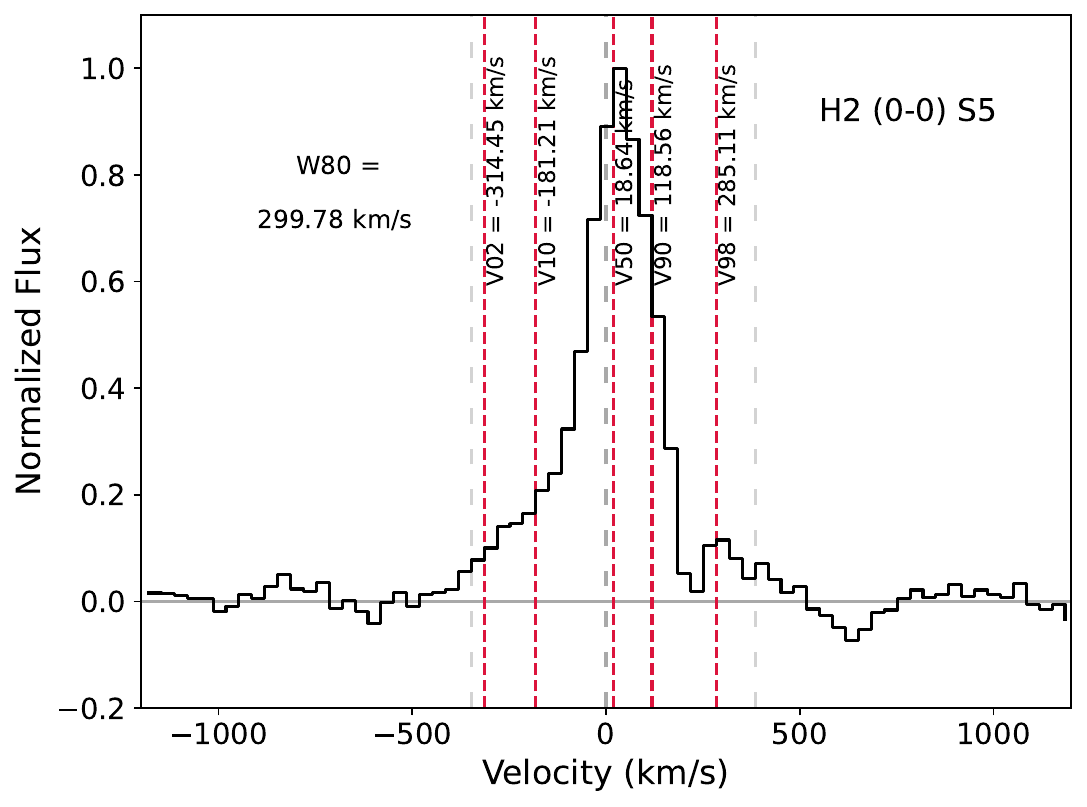}
  \caption{Non-parametric analysis of two nuclear line profiles: [Ne\,{\sc
      ii}] ({\it top panel}) and H$_2$ S(5) ({\it bottom
      panel}). The derived velocities (see text for details) are
    marked with red
    vertical lines.  The gray lines indicate the limits of the
    emission line defined as fluxes above two times the standard
    deviation of the (zero) continuum.}
     \label{fig:nonparametric}
\end{figure}

As an illustration, Fig.~\ref{fig:nonparametric} 
presents the analysis for [Ne\,{\sc ii}] and H$_2$ S(5).
The uncertainties of the measured velocities
only  associated with the spectral step of the different sub-channels are of the order of $43\,{\rm
    km\,s}^{-1}$ for [Fe\,{\sc ii}], $33\,{\rm
    km\,s}^{-1}$ for [Ar\,{\sc ii}] and H$_2$ S(5), $47\,{\rm
    km\,s}^{-1}$ for H$_2$ S(4), and $56\,{\rm
    km\,s}^{-1}$ for [Ne\,{\sc ii}].

\end{appendix}

  \end{document}